\documentclass{aa} 

\usepackage{hyperref}
\graphicspath{.}
\usepackage{graphicx}
\usepackage[mode=buildnew]{standalone}
\usepackage[varg]{txfonts}
\usepackage{bm}
\usepackage{physics}
\usepackage{multicol}
\usepackage{nccmath}
\usepackage[version=4]{mhchem}
\bibpunct{(}{)}{;}{a}{}{,}             

\hypersetup{
    colorlinks=true,
    urlcolor=blue,
    linkcolor=blue,
    citecolor=blue,
}

\defcitealias{2001PhRvD..65b4001S}{S01}
\defcitealias{1997ApJ...491..839L}{LS97}
\defcitealias{2019A&A...627A..64P}{P19}
\defcitealias{1978ApJ...221..937F}{FS78}
\defcitealias{1984ApJ...279..394B}{BG84}
\defcitealias{1994A&A...291..481G}{GB94}
\defcitealias{2012MNRAS.420.2387L}{L12}
\defcitealias{2021A&A...655A..59V}{V21}
\defcitealias{1982AcA....32..147D}{D82}

\newcommand{\mymatmod}[2]{\mathrm{mod}\left(#1\,,\,#2\right)}

\newcommand{\myintset}[1]{\llbracket#1\rrbracket}
\newcommand{\myintsetrange}[2]{\llbracket#1\,,\,#2\rrbracket}

\newcommand{\imp}[1]{{\rm Im}\left[#1\right]}
\newcommand{\rp}[1]{{\rm Re}\left[#1\right]}
\newcommand{\ccoef}[1]{c_{#1}\left(t\right)}
\newcommand{\ccoefconj}[1]{c^*_{#1}\left(t\right)}
\newcommand{\spatmode}[1]{\vec{\xi}_{#1}\left(\bm{x}\right)}
\newcommand{\spatmodeconj}[1]{\vec{\xi}^*_{#1}\left(\bm{x}\right)}
\newcommand{\BOP}[1]{\bm{B}\left(#1\right)}

\newcommand{\triadquant}[3]{\left(#1,\,#2,\,#3\right)}
\newcommand{\triadquantrad}[3]{\left(-#1,\,-#2,\,-#3\right)}

\newcommand{\absfraktxt}[2]{\lvert#1\rvert\,\displaystyle/\,\lvert#2\rvert}

\newcommand{\orcit}[1]{\protect\href{https://orcid.org/#1}{\protect\includegraphics[width=3mm]{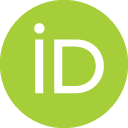}}}

\begin{document}

   \title{Non-linear three-mode coupling of gravity modes in rotating slowly pulsating B stars}
   \titlerunning{Non-linear three-mode coupling of $g$ modes in rotating SPB stars}

   \subtitle{Stationary solutions and modeling potential}

   \author{J. Van Beeck\inst{\ref{kul}, \ref{Caltech},\ref{Kavli}\, \orcit{0000-0002-5082-3887}} \and
   T. Van Hoolst \inst{\ref{kul},\ref{rob}\, \orcit{0000-0002-9820-8584}}
   \and C. Aerts \inst{\ref{kul},\ref{nijm},\ref{MPIA},
  \ref{CCA}\, \orcit{0000-0003-1822-7126}}
    \and J. Fuller \inst{\ref{Caltech}\, \orcit{0000-0002-4544-0750}}}

\institute{Institute of Astronomy, KU Leuven, Celestijnenlaan 200D, 3001 Leuven, Belgium\\ e-mail: \texttt{jordan.vanbeeck@kuleuven.be} \label{kul} \and TAPIR, Mailcode 350-17, California Institute of Technology, Pasadena, CA 91125, USA \label{Caltech} \and Kavli Institute for Theoretical Physics, University of California, Santa Barbara, CA 93106, USA \label{Kavli} \and Royal Observatory of Belgium, Ringlaan 3, Brussels, Belgium \label{rob} \and Dept. of Astrophysics, IMAPP, Radboud University Nijmegen, 6500 GL, Nijmegen, The Netherlands \label{nijm} 
\and Max Planck Institute for Astronomy, Koenigstuhl 17, 69117 Heidelberg, Germany \label{MPIA}
\and Guest Researcher, Center for Computational Astrophysics, Flatiron Institute, 162 Fifth Ave, New York, NY 10010, USA \label{CCA}
}

   \date{Received 24 October 2023 / Accepted 21 December 2023}

\abstract{Slowly pulsating B (SPB) stars display multi-periodic variability in the gravito-inertial mode regime with indications of non-linear resonances between modes. Several have undergone asteroseismic modeling in the past few years to infer their internal properties, but only in a linear setting. These stars rotate fast, so that rotation is typically included in the modeling by means of the traditional approximation of rotation (TAR).}{We aim to extend the set of tools available for asteroseismology, by describing time-independent (stationary) resonant non-linear coupling among three gravito-inertial modes within the TAR. Such coupling offers the opportunity to use mode amplitude ratios in the asteroseismic modeling process, instead of only relying on frequencies of linear eigenmodes, as has been done so far.}{Following observational detections, we derive expressions for the resonant stationary non-linear coupling between three gravito-inertial modes in rotating stars. We assess selection rules and stability domains for stationary solutions. We also predict non-linear frequencies and amplitude ratio observables that can be compared with their observed counterparts.}{The non-linear frequency shifts of stationary couplings are negligible compared to typical frequency errors derived from observations. The theoretically predicted amplitude ratios of combination frequencies match with some of their observational counterparts in the SPB targets. Other, unexplained observed ratios could be linked to other saturation mechanisms, to interactions between different modes, or to different opacity gradients in the driving zone.}{For the purpose of asteroseismic modeling, our non-linear mode coupling formalism can explain some of the stationary amplitude ratios of observed resonant mode couplings in single SPB stars monitored during 4 years by the {\em Kepler} space telescope.}


   \maketitle
%
%
\section{Introduction}

Slowly pulsating B (SPB) stars are mid-to-late-B variable stars on the main sequence that display a variety of low-frequency oscillations, and have masses ranging from $\sim3$ to $\sim9$ $M_\sun$ (e.g., \citealt{1991A&A...246..453W,2002A&A...393..965D, 2021NatAs...5..715P,2021MNRAS.503.5894S}).
Much of their multi-periodic variability is attributed to gravity modes excited by the $\kappa$ mechanism associated with the \ce{Fe} opacity bump in their envelope \citep{1993MNRAS.262..213G,1993MNRAS.265..588D,1999AcA....49..119P}.
Their rotation rates vary from $\sim1$\% of critical to nearly critical velocity.
The Coriolis force is a significant restoring force for most SPB oscillations, in addition to buoyancy (e.g., \citealt{2012MNRAS.420.2387L, 2021NatAs...5..715P}).  
We refer to such gravito-inertial modes whenever we mention $g$ modes in this work.
The large number of oscillations identified in space photometry of SPB stars have made them the subject of many asteroseismic modeling studies in the past few years (e.g., \citealt{2010Natur.464..259D,2018MNRAS.478.2243S,2019MNRAS.485.3544W,2019ApJ...881...86W,2020ApJ...899...38W,2021MNRAS.503.5894S,2021NatAs...5..715P,2022MNRAS.511.1529S}).

The broad general review of asteroseismology by \citet[][and references therein]{2021RvMP...93a5001A} makes it clear that stellar modeling is currently mainly done in a linear framework. 
Signals with frequencies approximately equal to linear combinations of frequencies of other detected signals, termed combination frequencies, are often detected in frequency lists generated by the harmonic analyses of stellar variability commonly used in asteroseismology \citep{2010aste.book.....A}. 
Some of these combination frequencies can be explained by a non-linear response of the stellar medium to the pulsation wave (see e.g., \citealt{2016MNRAS.460.1970B}), which is referred to as non-linear distortion by \citet{2009A&A...506..111D}.
In this work, however, we focus on combination frequencies that are explained by non-linear coupling among oscillation modes (e.g., \citealt{1984ApJ...279..394B} and \citealt{1994A&A...286..879V}).
The amplitudes of heat-driven non-radial oscillations in SPB stars cannot be explained in the linear approximation.
Observed amplitudes are therefore currently not used in asteroseismic inference.
Non-linear mode interactions that exchange energy among (coupled) modes must be taken into account to describe $g$ mode amplitude limitation.
Such interactions also change the mode frequencies.
Instead of resorting to resource-costly numerical integration of the non-linear hydrodynamical equations that govern the oscillation dynamics, we consider weakly non-linear effects of the lowest order.
We therefore consider isolated weak non-linear mode coupling among three $g$ modes, and followed and extended the approach of \citet{2012MNRAS.420.2387L}, hereafter referred to as \citetalias{2012MNRAS.420.2387L}.
Our approach is guided by the detected properties of SPB stars in {\em Kepler} observations, which are summarized in the sample studies by \citet{2021NatAs...5..715P} and \citet{2021MNRAS.503.5894S}.

Weak non-linear mode coupling among three modes has been a topic of interest for stellar pulsation modes since the 1970s, with several seminal papers written long before space photometry was available (limiting ourselves to mode coupling among non-radial pulsations, see e.g., \citealt{1982AcA....32..147D,1983A&A...123..331B,1984MNRAS.206..833A,1985AcA....35..229M,1985AcA....35....5D,1985A&A...151..174D,1988AcA....38...61D,1993ASPC...40..521D,1993A&A...279..417V,1993Ap&SS.210....9B,1993Ap&SS.210.....T,1994A&A...292..471V,1994A&A...286..879V,1994A&A...291..481G,1995A&A...296..405B,1995A&A...295..371V,1996A&A...308...66V,1997A&A...321..159B,1998BaltA...7...21G, 2001ApJ...546..469W}).
Most of these formalisms focused on the description of mode coupling in non-rotating stars. 
A notable exception is the formalism developed by Friedman and Schutz \citep{1978ApJ...221..937F,1978ApJ...222..281F,1979ApJ...232..874S}, on which \citet{2001PhRvD..65b4001S}, hereafter referred to as \citetalias{2001PhRvD..65b4001S}, based their treatment of non-linear three-mode coupling in rotating stars. 
\citetalias{2001PhRvD..65b4001S} included the effects of the Coriolis force perturbatively, but their framework is generic, allowing for the derivation of formalisms that do not treat the Coriolis force as a perturbation.
Several studies are based on the \citetalias{2001PhRvD..65b4001S} formalism, modeling non-linear tides in multiple-star systems (e.g., \citealt{2012MNRAS.420.3126F,2012MNRAS.421..983B,2013MNRAS.429.2425F,2013MNRAS.433..332B,2014MNRAS.440.3036O,2014MNRAS.443.2957B,2016ApJ...819..109W,2017MNRAS.472.1538F,2020ApJ...896..161G,2020MNRAS.496.5482Y,2021AJ....161..263Z,2021MNRAS.501.1836Y}) and in star-exoplanet systems \citep{2016ApJ...816...18E,2019MNRAS.484.5645V,2021ApJ...917...31Y,2022ApJ...928..140Y}, non-linear interactions among modes in neutron stars (e.g., \citealt{2002ApJ...571..435M,2003ApJ...591.1129A,2006PhRvD..74b4007L,2008MNRAS.387L..64W,2013ApJ...769..121W}), tidal migration in the moon systems of Jupiter and Saturn \citep{2016MNRAS.458.3867F}, non-linear interactions among mixed modes in red giant stars \citep{2019ApJ...873...67W,2021ApJ...918...70W}, or resonant mode coupling in $\delta$ Sct stars \citep{2023ApJ...950....6M}.

\citetalias{2012MNRAS.420.2387L} used the \citetalias{2001PhRvD..65b4001S} formalism as a basis for an extension that described rapidly rotating stars in which the Coriolis force cannot be treated as a perturbation.
To do so, they adopted the so-called traditional approximation for rotation (TAR), in which the latitudinal component of the rotation vector in a spherical coordinate system was neglected, assuming spherical symmetry.
This decoupled the radial and horizontal components of the pulsation equations (e.g., \citealt{1968RSPTA.262..511L,1997ApJ...491..839L,2003MNRAS.340.1020T,2013LNP...865...23M}).
For high-radial order $g$ modes it is justified to ignore the centrifugal force within the TAR, because the dominant contribution to the mode energy occurs deep inside the star, close to the convective core, where rotational deformation is small and spherical symmetry is a good approximation \citep{2019A&A...631A..26M,2021A&A...648A..97H,2021A&A...652A.154D,2021A&A...656A.122D}.
There is a marked difference in scale of the rotation frequency $\Omega$ and the Brunt-V\"ais\"al\"a frequency $N$ in those deep near-core regions.
The assumption that the Coriolis force is weaker than the buoyancy force in the direction of stable entropy or chemical stratification is therefore fulfilled near the core for low-frequency (Poincar\'e) modes.\footnote{The strength of the Coriolis force is determined by the Coriolis frequency $2\,\Omega$. The TAR frequency hierarchies $2\,|\Omega| \ll |N|$ and $|\Omega_\varphi| \ll |N|$, where $\Omega_\varphi$ is the real-valued frequency of pulsation mode $\varphi$ in the co-rotating reference frame, impose strong stratification in specific regions of the stellar interior, through which the low-frequency waves described within the TAR propagate (see e.g., \citealt{2021A&A...652A.154D}).}
Their horizontal velocities are also greater than the vertical velocities (see \citealt{2019A&A...631A..26M}), justifying the neglect of the latitudinal rotation vector component within the TAR.

\citetalias{2012MNRAS.420.2387L} used their quadratic non-linear mode coupling formalism within the TAR (based on the \citetalias{2001PhRvD..65b4001S} formalism) to provide a numerical mode coupling example for a specific SPB star model near the zero-age main sequence.
The computed mode properties of that example were not compared to observed SPB mode properties.
In this work we extend and correct the \citetalias{2012MNRAS.420.2387L} formalism with the aim of creating a modeling framework that can be used to model non-linear three-mode coupling of $g$ modes in some of the $38$ SPB stars considered by \citet{2021A&A...655A..59V}, hereafter referred to as \citetalias{2021A&A...655A..59V}.
We specifically focus on deriving the conditions for which the amplitudes of modes in coupled mode triads, and their combination phase, do not vary over time.
The mode parameters inferred by \citetalias{2021A&A...655A..59V} allowed for the discovery of many such potentially `locked' mode couplings.
We therefore contrast the theoretically predicted observables computed by our formalism for mode couplings obtained from models typical for the ensemble of SPB stars analyzed by \citetalias{2021A&A...655A..59V} with their detected observational counterparts.
We first provide a rigorous overview of our theoretical oscillation model in Sect.~\ref{sect:theoretical+model}, followed by Sect.~\ref{sect:theoretical+observables}, which describes the non-linear theoretical observables.
In Sect.~\ref{sect:num+results} we show the numerical results for resonant mode couplings typical for SPB stars, while Sect.~\ref{sect:asteroseismic+modeling+impact} discusses the potential of our theoretical framework for future asteroseismic modeling.
Finally, Sect.~\ref{sect:conclusions} outlines our conclusions and prospects.

%
%

\section{Theoretical oscillation model}\label{sect:theoretical+model}

\subsection{Linear free oscillations within the TAR}\label{sect:linear+free+oscillations+TAR}

The linearized momentum equation governing linear stellar oscillations in uniformly rotating stars is expressed in a co-rotating reference frame as (e.g., \citealt{1960RvMP...32..898F}, \citealt{1967MNRAS.136..293L} or \citetalias{2001PhRvD..65b4001S})
\begin{equation}\label{eq:general+linear+eqs+motion}
    \ddot{\vec{\xi}} + \bm{B}(\dot{\vec{\xi}}) + \bm{C}(\vec{\xi}) = \bm{a}_{\rm ext}\,,
\end{equation}
where $\vec{\xi}$ denotes the Lagrangian displacement, the superscripted dot indicates a partial time derivative, $\bm{B}(\dot{\vec{\xi}})\equiv 2\,\bm{\Omega}\times\dot{\vec{\xi}}$ is the Coriolis term, with $\vec{\Omega} = \Omega\, \bm{e}_z$ the (uniform) rotation vector and $\bm{e}_z$ the unit vector along the rotation axis, $\bm{C}(\vec{\xi})$ is the term that describes forces not depending on the oscillation frequency, and $\bm{a}_{\rm ext}$ is any acceleration due to external forces.
For the free oscillations used in linear asteroseismology, $\bm{a}_{\rm ext} = \vec{0}$.
The operators $\bm{B}$ and $\bm{C}$ are anti-Hermitian and Hermitian, respectively \citep{1967MNRAS.136..293L}.
An equivalent tensor representation of the linearized momentum equation is described in Appendix~\ref{app:alternative+representation+oscillation+model} and used in Sect~\ref{sect:coupled+mode+equations}. 

With time dependence {\em Ansatz}
\begin{equation}\label{eq:time+Ansatz}
    \vec{\xi}(\bm{x},t) = \vec{\xi}(\bm{x}) \, e^{-i \omega t}\,,
\end{equation}
we can rewrite Eq.~\eqref{eq:general+linear+eqs+motion} for free oscillations as
\begin{equation}\label{eq:general+linear+eqs+motion+time+dependence}
    - \,\omega^2 \vec{\xi} - i\, \omega\, \bm{B}(\vec{\xi}) + \bm{C}(\vec{\xi}) = 0\,,
\end{equation}
where $\omega$ is the (real- or complex-valued) angular frequency in the co-rotating frame (e.g., \citealt{1978ApJ...221..937F}, \citetalias{2001PhRvD..65b4001S} and \citealt{2019A&A...627A..64P}).
The Hermiticity of the operators $i\bm{B}$ and $\bm{C}$ allow one to define a generic orthogonality relation valid for two distinct (ordinary) eigenmodes $\vec{\xi}_\varphi$ and $\vec{\xi}_\beta$ of Eq.~\eqref{eq:general+linear+eqs+motion+time+dependence} if $\omega_\beta \neq \omega_\varphi^*$ and $\vec{\Omega}$ is time-independent, where we use the superscript `*' to indicate a complex conjugated quantity,
\begin{equation}\label{eq:complex+orthogonality+relation}
    \left(\omega_\beta + \omega_\varphi^*\right)\left<\vec{\xi}_\varphi,\vec{\xi}_\beta\right> + \left<\vec{\xi}_\varphi, i\bm{B}(\vec{\xi}_\beta)\right> = 0\,.
\end{equation} 
In Eq.~\eqref{eq:complex+orthogonality+relation}, the inner product of the Hilbert space $\mathcal{H}$ spanned by the complex eigenvectors $\vec{\xi}$ and $\vec{\xi}'$ of Eq.~\eqref{eq:general+linear+eqs+motion+time+dependence} is defined as
\begin{equation}\label{eq:inner+product}
    \left<\vec{\xi},\vec{\xi}'\right> = \int \rho\, \vec{\xi}^* \cdot \vec{\xi}' dV\,.
\end{equation}
A proof of the orthogonality condition implied in Eq.~\eqref{eq:complex+orthogonality+relation} for the two distinct modes $\vec{\xi}_\varphi$ and $\vec{\xi}_\beta$ is given in Appendix~\ref{app:orthogonality}.
For a mode $\varphi$ with complex-valued $\omega_\varphi$ described by Eq.~\eqref{eq:general+linear+eqs+motion+time+dependence}, the linear heat-driven growth (damping) rate $\gamma_\varphi$ is defined as $\gamma_\varphi \equiv \imp{\omega_\varphi}$.
Because of {\em Ansatz}~\eqref{eq:time+Ansatz}, linear growth (damping) occurs when $\gamma_\varphi$ is positive (negative).

We describe the coupling among non-degenerate modes using their adiabatic eigenfunctions (in Sect.~\ref{sect:coupled+mode+equations}), for which the generic orthogonality relation implied in Eq.~\eqref{eq:complex+orthogonality+relation} for the modes $\varphi$ and $\beta$ can be written as (see Appendix~\ref{app:orthogonality} and \citetalias{2001PhRvD..65b4001S})
\begin{equation}\label{eq:real+orthogonality+relation}
    \left(\Omega_\varphi + \Omega_\beta\right)\left<\vec{\xi}_\varphi,\vec{\xi}_\beta\right> + \left<\vec{\xi}_\varphi,i\bm{B}\left(\vec{\xi}_\beta\right)\right> = \delta^{\varphi}_\beta\, b_\varphi\,,
\end{equation}
where $\delta^\varphi_\beta$ denotes a Kronecker delta, $\rp{\omega_{\varphi}} = \Omega_{\varphi}$ (and similar for mode $\beta$), and the real-valued constant $b_\varphi$ is given by
\begin{equation}\label{eq:b+A}
    b_\varphi = \left<\vec{\xi}_\varphi, i \bm{B}\left(\vec{\xi}_\varphi\right)\right> + 2\, \Omega_\varphi \left<\vec{\xi}_\varphi,\vec{\xi}_\varphi\right>\,.
\end{equation}
The constant $b_\varphi$ defined in Eq.~\eqref{eq:b+A} is related to the rotating-frame mode energy $\epsilon_\varphi$ at unit complex amplitude, if $\Omega_\varphi \neq 0$ \citepalias{2001PhRvD..65b4001S}:
\begin{equation}\label{eq:mode+energy+rot+frame}
    \epsilon_\varphi = \Omega_\varphi\, b_\varphi\,.
\end{equation}

Because we describe uniformly rotating non-magnetic stars and use the Cowling approximation in which the Eulerian perturbation of the gravitational potential is set to zero, 
\begin{equation}\label{eq:C+operator+def}
    \bm{C}\left(\vec{\xi}\right) = \dfrac{\bm{\nabla}\delta P}{\rho} - \dfrac{\bm{\nabla}P}{\rho^2}\delta\rho\,,
\end{equation}
with $\delta P$ and $\delta \rho$ being the Eulerian perturbations of the pressure and density around their equilibrium values $P$ and $\rho$ (e.g., \citetalias{2019A&A...627A..64P}). The explicit dependence of $\bm{C}\left(\vec{\xi}\right)$ on $\vec{\xi}$ is defined in, for example, \citet{1967MNRAS.136..293L}.

We use the TAR, which ignores the latitudinal component $-\,\Omega\,\sin\theta\,\bm{e}_\theta$ of the rotation vector (e.g., \citealt{1997ApJ...491..839L}).
The Lagrangian displacement of a $g$ mode $\varphi$ can then be expressed in spherical coordinates ($r$,$\,\theta$,$\,\phi$) as \citep{2019A&A...627A..64P}
\begin{equation}\label{eq:lag+disp+TAR}
    \vec{\xi}_\varphi = [\xi_r^\varphi(r)\,H_r(\theta),\,\xi_h^\varphi(r)\,H_{\theta}(\theta),\,i\,\xi_h^\varphi(r)\,H_{\phi}(\theta)]\,e^{i\,(m_\varphi\,\phi_\varphi\,-\,\Omega_\varphi\,t)}\,,
\end{equation}
where $m_\varphi$ is the mode azimuthal order, $\phi_\varphi$ is the mode phase, $\xi_r^\varphi(r)$ and $\xi_h^\varphi(r)$ are the radial and horizontal mode displacement components, and $H_r(\theta)$, $H_{\theta}(\theta)$ and $H_{\phi}(\theta)$ are the real-valued radial, latitudinal and azimuthal Hough functions of a mode $\varphi$ (see Appendix A of \citealt{2019A&A...627A..64P} for their definitions).
Following \citetalias{2012MNRAS.420.2387L}, we normalize radial Hough function $H_r(\theta)$ as
\begin{equation}\label{eq:normalization+radial+hough}
    2\,\pi\,\int_{-1}^{1} \lvert H_r\,(\theta)\,\rvert^2\, d\mu = 1\,,
\end{equation}
where $\mu = \cos\theta$.
The normalization of $H_r(\theta)$ also determines the normalization of the latitudinal and azimuthal Hough functions $H_\theta(\theta)$ and $H_\phi(\theta)$.

\subsection{Non-linear oscillations within the TAR}

Non-linear mode interactions add acceleration (i.e., forcing) terms to the governing equation of motion, so that the higher-order equation of motion becomes the following quadratic eigenfunction problem (see e.g., \citealt{1960RvMP...32..898F})
\begin{equation}\label{eq:nonlinear+eqs+motion}
    -\, \omega^2 \vec{\xi} - i\, \omega\, \bm{B}\left(\vec{\xi}\right) + \bm{C}\left(\vec{\xi}\right) = \bm{a}[\vec{\xi}]\,,   
\end{equation}
under the {\em Ansatz} given by Eq.~\eqref{eq:time+Ansatz}: $ \vec{\xi}(\bm{x},t) = \vec{\xi}(\bm{x}) \, e^{-i \omega t}$.
In Eq.~\eqref{eq:nonlinear+eqs+motion}, the first term describes the acceleration.
That acceleration is changed by the non-linear coupling term $\bm{a}[\vec{\xi}]$, which can be split up into contributions from pressure gradients ($\bm{a}_P[\vec{\xi}]$) and gravity ($\bm{a}_G[\vec{\xi}]$).

Because we limit ourselves to non-linear three-mode interactions, it is sufficient to expand $\bm{a}[\vec{\xi}]$ to second order (e.g., \citetalias{2001PhRvD..65b4001S}),\vspace{-0.3cm}
\begin{equation}\label{eq:nonlinear+acceleration+quadratic}
    \bm{a}[\vec{\xi}] = \bm{a}^{(2)}[\vec{\xi},\vec{\xi}'] + O\left(\vec{\xi}^3\right) = \bm{a}^{(2)}_P[\vec{\xi},\vec{\xi}'] + \bm{a}^{(2)}_G[\vec{\xi},\vec{\xi}'] + O\left(\vec{\xi}^3\right)\,,
\end{equation}
where $\bm{a}^{(2)}[\vec{\xi},\vec{\xi}']$, the second order pressure term $\bm{a}^{(2)}_P[\vec{\xi},\vec{\xi}']$, and the second order gravitational acceleration term $\bm{a}^{(2)}_G[\vec{\xi},\vec{\xi}']$ are symmetric bilinear functions of the Lagrangian displacements associated with the interacting modes.
The nth components of $\bm{a}^{(2)}_P[\vec{\xi},\vec{\xi}']$ and $\bm{a}^{(2)}_G[\vec{\xi},\vec{\xi}']$ are expressed as (e.g., \citetalias{2001PhRvD..65b4001S})
\begin{subequations}\label{eq:a+G+expressions}
\begin{align}
    \bm{a}^{(2)}_{G,\,n}[\vec{\xi},\vec{\xi}'] =&\, -\dfrac{1}{2}\,\xi^k\,\xi^l\,\nabla_k\,\nabla_l\,\nabla_n\,\Phi\,,\\
    \bm{a}^{(2)}_{P,\,n}[\vec{\xi},\vec{\xi}'] =&\, -\dfrac{1}{\rho}\,\nabla_j\left[p\,\left(\Gamma_1 - 1\right)\Theta^j_n + p\,\Xi^j_n + \Psi\,\delta_n^j\right]\,,
\end{align}
\end{subequations}
where we use the Cowling approximation and the Einstein summation convention, $\nabla_\varphi$ denotes the covariant derivative with respect to coordinate $x^\varphi$, the adiabatic exponent $\Gamma_1 = (\partial \ln p\,\displaystyle/\, \partial \ln \rho)_S$ with subscript $S$ denoting adiabatic conditions, and $\Phi$ is the gravitational potential. 
We also use the following definitions of the tensors
\begin{subequations}\label{eq:tensors+nonlinear+acceleration}
\begin{align}
    \Theta^j_i =&\, \left(\nabla_i\,\xi^j\right)\left(\nabla_k\,\xi^k\right) = \left(\nabla_i\,\xi^j\right)\left(\nabla\cdot\vec{\xi}\right)\,,\\
    \Xi^j_i =&\, \left(\nabla_k\,\xi^j\right)\left(\nabla_i\,\xi^k\right)\,,
\end{align}
\end{subequations}
and of the quantities
\begin{subequations}\label{eq:traces+scalar+nonlinear+acceleration}
\begin{align}
    \Psi =&\, \dfrac{p}{2}\left\{\Theta \left[\left(\Gamma_1 - 1\right)^2 + \left(\pdv{\Gamma_1}{\ln \rho}\right)_S\right] + \Xi\left(\Gamma_1 - 1\right)\right\}\,,\\
    \Theta =&\, \delta^i_j\,\Theta^j_i = \left(\nabla\cdot\vec{\xi}\right)^2\,,\\
    \Xi =&\, \delta^i_j\,\Xi^j_i = \left(\nabla_k\,\xi^i\right)\left(\nabla_i\,\xi^k\right)\,.
\end{align}
\end{subequations}
We provide additional information on the expressions for the quantities in Eqs.~\eqref{eq:tensors+nonlinear+acceleration} and \eqref{eq:traces+scalar+nonlinear+acceleration} in Appendix~\ref{app:expressions+qcc} when discussing the explicit terms of the mode coupling coefficient defined in Eq.~\eqref{eq:coupling+coefficient} of Sect.~\ref{sect:coupled+mode+equations}.
Appendix~\ref{app:expressions+qcc} also contains corrections and simplifications for some of the expressions in \citetalias{2012MNRAS.420.2387L}.

\subsection{Coupled-mode equations}\label{sect:coupled+mode+equations}

Eigenvalue equation~\eqref{eq:nonlinear+eqs+motion} needs to be solved to retrieve the eigenfunctions $\vec{\xi}(\bm{x})$ and eigenfrequencies $\omega$ for interacting modes in rotating stars.
Each pulsation mode $\varphi$ of the star is characterized by a pair ($\vec{\xi}_\varphi, \omega_\varphi$).
The linear eigenvalue equation~\eqref{eq:general+linear+eqs+motion+time+dependence} is used to calculate the linear mode eigenfunctions, which are subsequently employed in the mode coupling computations.

We follow the procedure of \citetalias{2001PhRvD..65b4001S} to derive a set of equations that describe the couplings among all eigenmodes, which we refer to as the coupled-mode equations.
This procedure solves the phase space equivalent of the eigenvalue equations~\eqref{eq:nonlinear+eqs+motion}, defined in Eq.~\eqref{eq:phase+space+eigenvalue+problem}, by employing the phase space mode decomposition for a physically relevant real Lagrangrian displacement $\vec{\xi}\left(\bm{x},t\right)$ of the set of coupled modes in a rotating star, which is given by \citepalias{2001PhRvD..65b4001S}
\begin{equation}\label{eq:real+mode+expansion}
    \begingroup\renewcommand*{\arraystretch}{1.2}\begin{bmatrix}
    \vec{\xi}(t)\\
    \dot{\vec{\xi}}(t)
    \end{bmatrix}\endgroup = \sum_\varphi c_\varphi(t) \begingroup\renewcommand*{\arraystretch}{1.2}\begin{bmatrix}
    \vec{\xi}_\varphi\left(\bm{x}\right)\\
    -i\, \Omega_\varphi\, \vec{\xi}_\varphi\left(\bm{x}\right)
    \end{bmatrix}\endgroup + c_\varphi^*(t) \begingroup\renewcommand*{\arraystretch}{1.2}\begin{bmatrix}
    \vec{\xi}_\varphi^*\left(\bm{x}\right)\\
    i\, \Omega_\varphi\, \vec{\xi}_\varphi^*\left(\bm{x}\right)
    \end{bmatrix}\endgroup\,,
\end{equation}
under the assumption that mode $\varphi$ is not degenerate and has a real-valued frequency $\Omega_\varphi$.
This is justified because the linear mode eigenfrequencies associated with the (linear) adiabatic eigenfunctions and used to compute the coefficient relevant for mode coupling (see Eq.~\eqref{eq:real+mode+expansion}) are real. The real-valuedness of the eigenfrequencies further justifies the use of eigenmode orthogonality relation~\eqref{eq:real+orthogonality+relation}.
In principle, the expansion~\eqref{eq:real+mode+expansion} should also include modes with non-ordinary eigenvectors (i.e., modes with Jordan chains of length greater than zero, as was explained in e.g., \citealt{1979ApJ...232..874S}, \citealt{1980MNRAS.190...21S,1980MNRAS.190....7S} and \citetalias{2001PhRvD..65b4001S}).
Similar to the assumption made for $r$-modes in \citetalias{2001PhRvD..65b4001S}, we assume that such modes with non-ordinary eigenvectors are not important dynamically for the saturation of $g$ modes in SPB stars and therefore do not include these modes in the sum in Eq.~\eqref{eq:real+mode+expansion}.
A formalism that includes these non-ordinary eigenvectors can for example be found in Appendix A of \citetalias{2001PhRvD..65b4001S}.

The complex mode amplitudes $c_\varphi (t)$ in expansion~\eqref{eq:real+mode+expansion} and their complex conjugates $c_\varphi^* (t)$ are given by (see Appendix~\ref{app:c_coefficients})
\begin{subequations}\label{eq:real+inverse+mode+expansions}
\begin{align}
    c_\varphi(t) =&\, \dfrac{1}{b_\varphi}\left<\vec{\xi}_\varphi,\Omega_\varphi \,\vec{\xi}(t) + i \dot{\vec{\xi}}(t) + i \bm{B}\left(\vec{\xi}(t)\right)\right>\,, \label{eq:real+mode+coefficient}\\
    c_\varphi^*(t) =&\, \dfrac{1}{b_\varphi}\left<\vec{\xi}_\varphi^*,\Omega_\varphi \,\vec{\xi}(t) - i \dot{\vec{\xi}}(t) - i \bm{B}\left(\vec{\xi}(t)\right)\right>\,, \label{eq:imaginary+mode+coefficient}
\end{align}
\end{subequations}
for non-zero $b_\varphi$ and $\Omega_\varphi \neq -\Omega_\beta$, where the subscript $\beta$ refers to a different non-degenerate mode in expansion~\eqref{eq:real+mode+expansion}.
The rotating-frame mode energy $E_\varphi$ for mode $\varphi$ with rotating-frame frequency $\Omega_\varphi$ and (complex) amplitude $c_\varphi$ is (see Appendix K in \citetalias{2001PhRvD..65b4001S})
\begin{equation}\label{eq:mode+energy+rot+frame_non_unit_amplitude}
    E_\varphi = \epsilon_\varphi\,\lvert c_\varphi\rvert^2 = \Omega_\varphi\,b_\varphi\,\lvert c_\varphi\rvert^2\,\,.
\end{equation}

We disregard non-linear coupling of $g$ modes with toroidal $r$ modes (i.e., $r$-$g$ mode coupling) in this work because $r$ modes have not been detected in the SPB stars analyzed in \citetalias{2021A&A...655A..59V}. Instead we focus on inter-$g$ mode coupling.
For faster-rotating SPB stars, such $r$-$g$ mode couplings can become significant (see e.g., \citetalias{2012MNRAS.420.2387L}), but only if their coupling initiation threshold is similar to or smaller than that of the inter-$g$ mode couplings.
As discussed by \citet{1995A&A...296..405B}, stable (constant), periodic or irregular time-dependent behavior of the oscillation amplitudes (and frequencies) may be observed.
In this work, we focus on the stable (time-independent) amplitude and frequency solutions of the amplitude equations (see Sect.~\ref{sect:AEs}) because the observations of mode couplings indicate that this is the dominant observed behavior in many of the SPB stars (see Sect.~\ref{sect:stationary+AEs}).
We therefore do not discuss periodic modulation or irregular behavior of amplitudes or frequencies.\footnote{Such periodic frequency and/or amplitude modulation (or even irregular behavior) can be simulated (numerically) using the amplitude equations or coupled-mode equations derived in this work.}

By substituting the phase space mode expansion~\eqref{eq:real+mode+expansion} into the phase space equivalent of the equation of motion, defined in Eq.~\eqref{eq:tensor+equations+of+motion}, and accounting for mode orthogonality, we derive the coupled-mode equations in the quadratic approximation (see appendix A of \citetalias{2001PhRvD..65b4001S} for the technical details),
\begin{equation}\label{eq:real+coupled+mode+b+alpha}
    \dot{c}_\varphi + i\, \Omega_\varphi\,c_\varphi = \dfrac{i}{b_\varphi} \left(\kappa_{\varphi}^{\beta\gamma} c_\beta c_\gamma + \kappa_{\varphi}^{\overline{\beta}\gamma} c_\beta^* c_\gamma + \kappa_{\varphi}^{\beta\overline{\gamma}} c_\beta c_\gamma^* + \kappa_{\varphi}^{\overline{\beta}\overline{\gamma}} c_\beta^* c_\gamma^*\right)\,,
\end{equation}
in which we employ the Einstein summation convention.
If we combine Eqs.~\eqref{eq:mode+energy+rot+frame} and \eqref{eq:real+coupled+mode+b+alpha} and assume non-zero $\Omega_\varphi$, we get
\begin{equation}\label{eq:real+coupled+mode}
    \dfrac{\dot{c}_\varphi}{i} + \Omega_\varphi\,c_\varphi = \Omega_\varphi \left(\eta_{\varphi}^{\beta\gamma} c_\beta c_\gamma + \eta_{\varphi}^{\overline{\beta}\gamma} c_\beta^* c_\gamma + \eta_{\varphi}^{\beta\overline{\gamma}} c_\beta c_\gamma^* + \eta_{\varphi}^{\overline{\beta}\overline{\gamma}} c_\beta^* c_\gamma^*\right)\,.
\end{equation}
The implicit sums over indices $\beta$ and $\gamma$ in Eqs.~\eqref{eq:real+coupled+mode+b+alpha} and \eqref{eq:real+coupled+mode} run over all modes with ordinary eigenvectors.
Coupling coefficient $\kappa_{\varphi}^{\beta\gamma}$ and energy-scaled coupling coefficient $\eta_{\varphi}^{\beta\gamma}$ involved in the coupling of the three modes $\varphi$, $\beta$ and $\gamma$ are defined by
\begin{equation}
    \label{eq:coupling+coefficient}
    \kappa_{\varphi}^{\beta\gamma} = \epsilon_\varphi\, \eta_{\varphi}^{\beta\gamma} = \left<\vec{\xi}_\varphi,\,\bm{a}^{(2)}[\vec{\xi}_\beta,\,\vec{\xi}_\gamma]\right>\,,
\end{equation}
which deviates from the definition in \citetalias{2001PhRvD..65b4001S}: $\vec{\xi}_\varphi$ is used instead of its complex conjugate.
The (real-valued) total displacement is thus expanded in terms of non-conjugated products of mode amplitudes and eigenvectors, instead of their complex conjugated analogues used in \citetalias{2001PhRvD..65b4001S}.
The resulting expression~\eqref{eq:coupling+coefficient} is equivalent, and directly yields the specific coupling coefficient selection rules derived in Sect.~\ref{sect:selection+rules}.
A bar over a subscript or superscript of $\kappa_{\varphi}^{\beta\gamma}$ or $\eta_{\varphi}^{\beta\gamma}$ indicates that a complex conjugate of the corresponding mode eigenfunction is used in the definition in Eq.~\eqref{eq:coupling+coefficient}.
The expressions for the coupling coefficients $\kappa_{\varphi}^{\beta\gamma}$ and $\eta_{\varphi}^{\beta\gamma}$ are symmetric under permutation of indices, due to the presence of the symmetric bilinear function $\bm{a}^{(2)}[\vec{\xi}_\beta,\,\vec{\xi}_\gamma]$.
We normalize the eigenfunctions $\vec{\xi}_\varphi$, $\vec{\xi}_\beta$ and $\vec{\xi}_\gamma$ such that $\epsilon_\varphi = \epsilon_\beta = \epsilon_\gamma = GM^2/R$, following \citetalias{2012MNRAS.420.2387L}.
Additional information on the explicit expressions used in Eq.~\eqref{eq:coupling+coefficient} can be found in Appendix~\ref{app:expressions+qcc}.

The coupled-mode equations~\eqref{eq:real+coupled+mode+b+alpha} and \eqref{eq:real+coupled+mode} thus describe the temporal dynamics of the amplitudes of all (coupled) modes in a coupling network.
To study specific three-mode non-linear interactions among $g$ modes in SPB stars, we derive amplitude equations (AEs) in Sect.~\ref{sect:AEs}, based on the coupled-mode equations.
These AEs simplify the study of mode resonances.

\subsection{Coupling coefficient selection rules}\label{sect:selection+rules}

For the coupling coefficients defined in Eq.~\eqref{eq:coupling+coefficient} to be non-zero, selection rules need to be satisfied (see e.g., \citetalias{2001PhRvD..65b4001S}).
Considering the integration of coupling coefficient $\kappa_{\varphi}^{\beta\gamma}$ over longitude $\phi$,
\begin{equation}\label{eq:longitudinal+integral}
    \kappa_{\varphi}^{\beta\gamma}\propto\int_0^{2\,\pi} \exp\left({i\,\left(-H_\varphi\, m_\varphi + H_\beta\, m_\beta + H_\gamma\, m_\gamma\right)\,\phi}\right)d\phi\,.
\end{equation}
The same longitudinal dependence is obtained for $\eta_{\varphi}^{\beta\gamma}$, which yields the generic azimuthal selection rule
\begin{equation}\label{eq:azimuthal+selection+rule}
    -H_\varphi\, m_\varphi + H_\beta\, m_\beta + H_\gamma\, m_\gamma = 0\,,
\end{equation}
in which we introduce the sign factor $H_\varphi$ that has the value $-1$ for complex conjugated modes in the coupling coefficient expression (i.e., modes that have a bar over their index) and $1$ otherwise.
Equation~\eqref{eq:azimuthal+selection+rule} for example leads to the azimuthal selection rule $m_\varphi = m_\beta + m_\gamma$ for coupling coefficients $\kappa_\varphi^{\beta\gamma}$ and $\eta_{\varphi}^{\beta\gamma}$.

The other selection rule takes into account the symmetry of the modes across the stellar equator.
Whether $g$ mode $\varphi$ is odd or even is determined by $\left(-1\right)^{|k_\varphi|}$. If $\mymatmod{\lvert k_\varphi \rvert}{2} = 1$, mode $\varphi$ is odd, if it is zero, the mode is even (e.g., \citealt{1997ApJ...491..839L}).
This selection rule requires that two or no odd modes partake in the mode coupling (see Appendix~\ref{app:meridional+selection+rule}), which can also be written as a selection rule based on the mode ordering number $k$,
\begin{equation}\label{eq:meridional+selection+rule}
    \mymatmod{\, \lvert k_\varphi\rvert + \lvert k_\beta\rvert + \lvert k_\gamma\rvert\,}{2\,} = 0\,,
\end{equation}
where, for a $g$ mode $\varphi$, $k_\varphi \equiv l_\varphi - \lvert m_\varphi \rvert$ \citep{1997ApJ...491..839L}.
We refer to Eq.~\eqref{eq:meridional+selection+rule} as the meridional selection rule because \citet{2022A&A...662A..58V} called $k$ the meridional degree.

\subsection{Resonant coupling networks}\label{sect:three+mode+coupling+scenarios}
\begin{figure}
\centering
\includegraphics[]{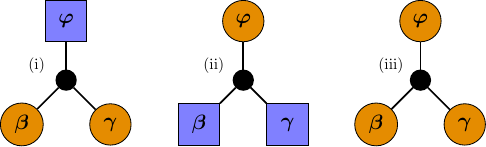}
\caption{Representations of isolated resonant three-mode coupling networks for modes $\varphi$, $\beta$ and $\gamma$.
They represent direct resonances (i), parametric resonances (ii), and driven resonances (iii). A linearly unstable (i.e., excited) mode is pictured as an orange circle, whereas a linearly stable (i.e., damped) mode is pictured as a blue square.}
\label{fig:all+three+mode+resonances}
\end{figure}

The three potential scenarios for resonant coupling among three modes, hereafter referred to as the isolated resonant three-mode coupling networks, are pictured in Fig.~\ref{fig:all+three+mode+resonances}. 
A direct resonant coupling is an interaction between a linearly damped daughter mode $\varphi$ and the linearly driven parent modes $\beta$ and $\gamma$, and is depicted as scenario (i) in Fig.~\ref{fig:all+three+mode+resonances}.
In this scenario, $\gamma_\varphi < 0$, and $\gamma_\beta,\,\gamma_\gamma > 0$.
A parametric resonant coupling is an interaction between a linearly driven parent mode $\varphi$ and the linearly damped daughter modes $\beta$ and $\gamma$, and is depicted as scenario (ii) in Fig.~\ref{fig:all+three+mode+resonances}.
Hence, in this case, $\gamma_\varphi > 0$, and $\gamma_\beta,\,\gamma_\gamma < 0$.
We call the resonant interaction between three linearly driven modes $\varphi$, $\beta$ and $\gamma$ a driven resonant coupling, which is depicted as scenario (iii) in Fig.~\ref{fig:all+three+mode+resonances}, for which $\gamma_\varphi,\, \gamma_\beta,\, \gamma_\gamma > 0$.

In addition to these three isolated resonant coupling networks, non-isolated resonant coupling networks exist, in which a daughter or parent mode is shared among different resonantly coupled mode triads.
\citet{2014MNRAS.440.3036O} called this multiple-mode or multi-mode coupling (see also e.g., \citealt{2021FrASS...8...67G}).
The selection rules derived in Sect.~\ref{sect:selection+rules} remain valid in such networks, because coupling occurs at the same quadratic order.
Networks with granddaughter and great-granddaughter modes have also been constructed. The reader is referred to \citet{1996ApJ...466..946K} and \citet{2021ApJ...918...70W} for additional information on such networks.

Higher-order coupling terms, such as those that appeared in the cubic four-wave coupling networks developed to describe four interacting modes in non-rotating stars (by for example \citealt{1993A&A...279..417V} and \citealt{1994A&A...292..471V}) do not adhere to the selection rules in Sect.~\ref{sect:selection+rules}.
The four-wave selection rules can however be determined from similar arguments.
For example, the four-wave azimuthal selection rule was derived in \citet{2016ApJ...819..109W}.
They however used the complex conjugate of mode $\varphi$ in their definition of the mode coupling coefficient and did not expand in pairs of complex conjugates (as in Eq.~\eqref{eq:real+mode+expansion}).
This yielded a slightly different formulation of the azimuthal selection rule than what would be obtained for a coupling coefficient defined in a similar way as in Eq.~\eqref{eq:coupling+coefficient}.

A priori, it is not clear whether multi-mode or higher-order coupling terms are necessary to explain the (stationary) amplitudes and phases of the three-mode resonance candidate couplings detected among the modes of SPB stars analyzed in \citetalias{2021A&A...655A..59V}.
The most parsimonious models for these identified candidate couplings would be isolated three-mode sum-frequency coupling networks that produce stable stationary solutions, if those solutions effectively describe observed (time-independent) amplitudes, phases and frequencies.
We limit ourselves in the remainder of this work to describing such coupling networks.

\subsection{Amplitude equations for $\Omega_1 \simeq \Omega_2 + \Omega_3$}\label{sect:AEs}

Under the assumption that the linear driving or linear damping rate $\gamma_\varphi$ of a mode $\varphi$ fulfills $\lvert \gamma_\varphi/\Omega_\varphi\rvert \ll 1$ for the modes that are weakly non-linearly coupled, we can derive the AEs.
In this section we use the coupled-mode equations \eqref{eq:real+coupled+mode+b+alpha} and \eqref{eq:real+coupled+mode} valid for quadratically non-linear oscillations in rotating stars to derive the specific AEs for the sum-frequency resonant interaction $\Omega_1 \simeq \Omega_2 \,+\, \Omega_3$ between three distinct modes.
Similar AEs were derived for the resonant harmonic interaction $\Omega_{1_h} \simeq \Omega_{2_h}$ (with subscript $h$ denoting the harmonic nature of the resonance) in Appendix B.2 of \citet{PhDThesisVanBeeck}.
The AEs derived in this Section also describe the temporal evolution of the coupled mode's properties of difference-frequency resonances $\Omega_{1_d} = \Omega_{2_d} - \Omega_{3_d}$, as we show in Appendix~\ref{app:diff+freq}.

Following \citet{1995A&A...295..371V}, we apply the multiple time scales perturbation method (e.g., \citealt{1973pm.book.....N,1981itpt.book.....N,1979noos.book.....N}) to compute the mode amplitudes due to non-linear coupling at the quadratic level.
We introduce time variables
\begin{equation}\label{eq:times}
    t_0 = t,\hspace{3mm}t_1 = \mathfrak{J}\,t, \ldots\,,
\end{equation}
where $\mathfrak{J}$ is a small, dimensionless ordering parameter.
The time derivative in Eq.~\eqref{eq:real+coupled+mode+b+alpha} becomes
\begin{equation}\label{eq:time+derivatives}
    \pdv{}{t} = \pdv{}{t_0} + \mathfrak{J}\,\pdv{}{t_1} + O(\mathfrak{J}^2)\,.
\end{equation}
The multiple time scales method then searches for an approximate solution for the complex amplitudes $c_\varphi$ of mode $\varphi$ as
\begin{equation}\label{eq:uniform+amplitude+expansion}
    c_\varphi = \mathfrak{J}\, c_{\varphi 1}\left(t_0, t_1, \ldots\right) + \mathfrak{J}^2\, c_{\varphi 2}\left(t_0, t_1, \ldots\right) + O(\mathfrak{J}^3)\,. 
\end{equation}
Perturbation series~\eqref{eq:uniform+amplitude+expansion} needs to be uniform: each of the higher-order terms should be a small correction to lower-order terms.

Substituting expansions \eqref{eq:time+derivatives} and \eqref{eq:uniform+amplitude+expansion} into the coupled-mode equations~\eqref{eq:real+coupled+mode+b+alpha}, subsequently dividing by the non-zero parameter $\mathfrak{J}$ and keeping terms up to first order in $\mathfrak{J}$, yields
\begin{equation}\label{eq:expanded+coupled+mode}
\begin{aligned}
    &\pdv{c_{\varphi 1}}{t_0} + \mathfrak{J} \pdv{c_{\varphi 1}}{t_1} + \mathfrak{J} \pdv{c_{\varphi 2}}{t_0} + i\,\Omega_\varphi \left( c_{\varphi 1} + \mathfrak{J} c_{\varphi_2}\right)\\
    &= \mathfrak{J} \dfrac{i}{b_\varphi} \left(\kappa_{\varphi}^{\beta\gamma} c_{\beta 1} c_{\gamma 1} + \kappa_{\varphi}^{\overline{\beta}\gamma} c_{\beta 1}^* c_{\gamma 1} + \kappa_{\varphi}^{\beta\overline{\gamma}} c_{\beta 1} c_{\gamma 1}^* + \kappa_{\varphi}^{\overline{\beta}\overline{\gamma}} c_{\beta 1}^* c_{\gamma 1}^*\right)\\
    &\hspace{4mm}+\, O(\mathfrak{J}^2)\,\,.
\end{aligned}
\end{equation}
We examine this equation order by order in $\mathfrak{J}$ to determine the complex amplitude variation of the eigenmode $\varphi$ in the form of AEs.
The (linear) equation at order $\mathfrak{J}^0$ is
\begin{equation}\label{eq:first+order}
    \pdv{c_{\varphi 1}}{t_0} + i\,\Omega_{\varphi}\, c_{\varphi 1} \equiv L_\varphi(c_{\varphi 1}) = 0\,,
\end{equation}
in which we introduce the linear operator $L_\varphi$ as the mapping $L_\varphi\colon \mathbb{C} \to \mathbb{C}$ defined as $L_\varphi(c) = \left(\pdv{}{t_0} + i\, \Omega_\varphi\right)\left[c\right]$ for a complex amplitude $c$.
The solution of Eq.~\eqref{eq:first+order} is
\begin{equation}\label{eq:first+order+solution}
    c_{\varphi 1} = a_{\varphi} \left(t_1, \ldots\right) e^{-i\,\Omega_\varphi\,t_0}\,,
\end{equation}
where $a_\varphi$ is a complex amplitude factor that varies only on time scales slower than the time scale $t_0$ for each mode $\varphi$.

The equation at order $\mathfrak{J}$ is
\begin{equation}\label{eq:second+order}
\begin{aligned}
    L_\varphi(c_{\varphi 2}) = - \pdv{c_{\varphi 1}}{t_1} + \dfrac{i}{b_\varphi} &\left(\kappa_{\varphi}^{\beta\gamma} c_{\beta 1} c_{\gamma 1} + \kappa_{\varphi}^{\overline{\beta}\gamma} c_{\beta 1}^* c_{\gamma 1} \right.\\
    &\left.\hspace{2mm}+\, \kappa_{\varphi}^{\beta\overline{\gamma}} c_{\beta 1} c_{\gamma 1}^*
    + \kappa_{\varphi}^{\overline{\beta}\overline{\gamma}} c_{\beta 1}^* c_{\gamma 1}^*\right)\,.
\end{aligned}
\end{equation}
Substituting the linear solution~\eqref{eq:first+order+solution} into Eq.~\eqref{eq:second+order} yields
\begin{equation}\label{eq:second+order+subst}
\begin{aligned}
    &L_\varphi(c_{\varphi 2}) = - \pdv{a_{\varphi}}{t_1}\,e^{-i\,\Omega_\varphi\,t_0}\\
    &+ \dfrac{i}{b_\varphi} \left(a_\beta\,a_\gamma\, \kappa_{\varphi}^{\beta\gamma} e^{-i(\Omega_\beta + \Omega_\gamma)t_0} + a_\beta^*\,a_\gamma\,\kappa_{\varphi}^{\overline{\beta}\gamma} e^{-i(-\Omega_\beta + \Omega_\gamma)t_0} \right.\\
    &\hspace{10mm}\left.+\, a_\beta\,a_\gamma^*\,\kappa_{\varphi}^{\beta\overline{\gamma}} e^{-i(\Omega_\beta - \Omega_\gamma)t_0} + a_\beta^*\,a_\gamma^*\,\kappa_{\varphi}^{\overline{\beta}\overline{\gamma}} e^{i(\Omega_\beta + \Omega_\gamma)t_0} \right)\,.
\end{aligned}
\end{equation}
Some of the terms in the solution of Eq.~\eqref{eq:second+order+subst} increase linearly in time.
Such terms are called secular terms (e.g., \citealp{1973pm.book.....N,1981itpt.book.....N} and \citealp{1979noos.book.....N}) and must vanish to ensure we obtain a uniformly valid perturbation series.

Setting the terms that generate the secular terms in the solution of Eq.~\eqref{eq:second+order+subst} equal to zero yields the AEs for the mode amplitudes and phases.
These secular-term-generating terms have a multiplying factor $\exp \left(-i\Omega_\varphi t_0\right)$ in Eq.~\eqref{eq:second+order+subst}, as shown in Appendix~B.1 of \citet{PhDThesisVanBeeck}.
The first term on the right side of Eq.~\eqref{eq:second+order+subst} always generates a secular term.
We obtain additional terms that generate secular terms by substituting the resonance condition $\Omega_1 \simeq \Omega_2 + \Omega_3$, defined in terms of the detuning parameters $\delta\omega$ and $\Delta\Omega^l$,
\begin{equation}\label{eq:frequency+detuning}
    \mathfrak{J}\, \delta\omega = \Omega_1 - \Omega_2 - \Omega_3 \equiv \Delta\Omega^l\,,
\end{equation}
in Eq.~\eqref{eq:second+order+subst}.
The condition that ensures that the sum of these secular-term-generating terms vanishes leads to the (extended complex) AEs for the sum-frequency resonance $\Omega_1 \simeq \Omega_2 + \Omega_3$:
\begin{equation}\label{eq:final+form+extended+AEs}
    \pdv{\vec{a}}{t_1} = \vec{\gamma}\circ\vec{a} + 2\,i\,\left(\vec{\eta}\circ\vec{a}_M\circ\vec{\Omega}_M\circ\vec{e}\right)\,,
\end{equation}
in which we introduce the linear growth or linear damping rates $\gamma_\varphi$ for mode $\varphi$ (e.g., \citealt{1995A&A...295..371V}) and use Eq.~\eqref{eq:coupling+coefficient} to convert $\kappa_{\varphi}^{\beta\gamma}$ into $\eta_{\varphi}^{\beta\gamma}$ (and similar).
In Eq.~\eqref{eq:final+form+extended+AEs}, we also introduce the symbol `$\circ$', which denotes a Hadamard-Schur (i.e., element-wise, \citealt{Bhatia+2007,Bernstein+2009}) product.
We use the vectors
\begin{equation}\label{eq:defined+vectors+1}
    \vec{a} = \begingroup\renewcommand*{\arraystretch}{1.2}\begin{bmatrix}
    a_1 \\ a_2 \\ a_3
    \end{bmatrix}\endgroup\,,\hspace{0.3cm}\begingroup\renewcommand*{\arraystretch}{1.2}\vec{a}_M = \begin{bmatrix}
    a_2\,a_3 \\ a_1\,a_3^* \\ a_1\,a_2^*
    \end{bmatrix}\endgroup\,,\hspace{0.3cm}\begingroup
    \renewcommand*{\arraystretch}{1.2}\vec{e} =  \begin{bmatrix}
    \exp(i\,\delta\omega\,t_1) \\ \exp(-i\,\delta\omega\,t_1) \\ \exp(-i\,\delta\omega\,t_1)
    \end{bmatrix}\endgroup\,,
\end{equation}
and
\begin{equation}\label{eq:defined+vectors+2}
    \vec{\gamma} = \begingroup\renewcommand*{\arraystretch}{1.2}\begin{bmatrix}
    \gamma_1 \\ \gamma_2 \\ \gamma_3
    \end{bmatrix}\endgroup\,,\hspace{0.3cm} \begingroup\renewcommand*{\arraystretch}{1.2} \vec{\eta} = \begin{bmatrix}
    \eta_1 \\ \eta_1^* \\ \eta_1^*
    \end{bmatrix}\endgroup
    \,,\hspace{0.3cm} \vec{\Omega}_M = \begingroup\renewcommand*{\arraystretch}{1.2}\begin{bmatrix}
    \Omega_1 \\ \Omega_2 \\ \Omega_3
    \end{bmatrix}\endgroup\,,
\end{equation}
in Eq.~\eqref{eq:final+form+extended+AEs}.
In Eq.~\eqref{eq:defined+vectors+2}, we define the isolated three-mode coupling coefficient $\eta_1$:
\begin{equation}\label{eq:symmetry+coupling+coefficients}
    \eta_1^{23} = \eta_1^{32} = \left(\eta_2^{1\overline{3}}\right)^* = \left(\eta_2^{\overline{3}1}\right)^* = \left(\eta_3^{1\overline{2}}\right)^* = \left(\eta_3^{\overline{2}1}\right)^* \equiv \eta_1 \,,
\end{equation}
which respects the symmetry of coupling coefficient~\eqref{eq:coupling+coefficient} under permutation of its indices.
For exact resonances (i.e., $\delta\omega = 0$), the AEs reduce to a set of trivial equations.
In such situations, efficient non-linear energy transfer is expected, and one enters the regime of intermediate to strong non-linear coupling, whereas the AEs are valid for weak coupling only.

By introducing real amplitudes $A_\varphi$ and phases $\phi_\varphi$ as $a_\varphi = A_\varphi \exp(i\, \phi_\varphi)$ with $\varphi \in \myintset{3}$, where $\myintset{u}$ denotes the set of integers $\left\{\, x \in \mathbb{N}_0\,|\, x\leq u \,\right\}$, and separating the real and imaginary parts of the extended complex AEs~\eqref{eq:final+form+extended+AEs}, we obtain
\begin{subequations}\label{eq:AEs}
\begin{align}
    \pdv{\vec{A}}{t_1} &= \vec{\gamma} \circ \vec{A} + 2\, \lvert \eta_1\rvert\, \sin\left(\Upsilon\right)\, \left(\vec{A}_N\circ \vec{\Omega}_M\right)\,, \label{eq:AE+A}\\
    \vec{A}\circ\pdv{\vec{\phi}}{t_1} &= 2\, \lvert \eta_1\rvert\, \cos\left(\Upsilon\right)\, \left(\vec{A}_P\circ \vec{\Omega}_M\right)\,,\label{eq:AE+P}
\end{align}
\end{subequations}
where $\eta_1 = |\eta_1|\,e^{-i\,\delta_1}$, and
\begin{equation}\label{eq:amplitude+products}
    \vec{A} = \begingroup\renewcommand*{\arraystretch}{1.2}\begin{bmatrix}
    A_1 \\ A_2 \\ A_3
    \end{bmatrix}\endgroup\,,\hspace{0.2cm} \vec{\phi} = \begingroup\renewcommand*{\arraystretch}{1.2}\begin{bmatrix}
    \phi_1 \\ \phi_2 \\ \phi_3
    \end{bmatrix}\endgroup\,,\hspace{0.2cm} \vec{A}_N = \begingroup\renewcommand*{\arraystretch}{1.2}\begin{bmatrix}
    A_2\, A_3 \\ -A_1\, A_3 \\ -A_1 \, A_2
    \end{bmatrix}\endgroup\,,\hspace{0.2cm} \vec{A}_P = \begingroup\renewcommand*{\arraystretch}{1.2}\begin{bmatrix}
    A_2\, A_3 \\ A_1\, A_3 \\ A_1 \, A_2
    \end{bmatrix}\endgroup\,.
\end{equation}
In the AEs~\eqref{eq:AEs} we also introduce the generic phase coordinate
\begin{equation} \label{eq:upsilon+parameter}
    \Upsilon \equiv -\delta\omega\,t_1 + \phi_1 - \phi_2 - \phi_3 + \delta_1\,,
\end{equation}
which contains the combination phase $\Phi = \phi_1 - \phi_2 - \phi_3$.
Because the coupling coefficients~\eqref{eq:coupling+coefficient} do not depend on time, we have
\begin{equation}\label{eq:AE+4+no+zero+A}
    \pdv{\Upsilon}{t_1} = -\delta\omega + \cot\left(\Upsilon\right)\left(-\gamma_\boxplus + \pdv{\ln A_1}{t_1} + \pdv{\ln A_2}{t_1} + \pdv{\ln A_3}{t_1}\right)\,,
\end{equation}
for non-zero resonant mode amplitudes, because of Eq.~\eqref{eq:AE+A} and the definition $\gamma_\boxplus \equiv \gamma_1 + \gamma_2 + \gamma_3$.
Equations~\eqref{eq:AE+A} and \eqref{eq:AE+4+no+zero+A} thus form an autonomous four-dimensional system equivalent to the six-dimensional system in Eqs.~\eqref{eq:AE+A} and \eqref{eq:AE+P}, in which the individual mode phases $\phi_1$, $\phi_2$ and $\phi_3$ do not explicitly appear.

Non-linear interactions lead to non-linear frequency shifts, which can be derived from Eq.~\eqref{eq:AE+P} for the individual mode phases.
By integrating these equations, the first order solution~\eqref{eq:first+order+solution} can be expressed as
\begin{equation}\label{eq:explicit+time+nonlinear+non+stationary+first+order}
\begin{aligned}
    &\vec{c}_1 = \vec{A}\left(t_1,\ldots\right)\\
    &\exp\left\{i\left(\vec{\phi}_0 - \vec{\Omega}_M\,t_0 + 2\int\left(\vec{A}_P\oslash\vec{A}\circ\vec{\Omega}_M\right) \lvert \eta_1\rvert\cos\Upsilon d t_1\right)\right\}\,,
\end{aligned}
\end{equation}
in which the symbol `$\oslash$' denotes a Hadamard-Schur (element-wise) division, and where
\begin{equation}
    \vec{\phi}_0 = \begingroup\renewcommand*{\arraystretch}{1.2}\begin{bmatrix}
    \left(\phi_1\right)_0 \\ \left(\phi_2\right)_0 \\ \left(\phi_3\right)_0
    \end{bmatrix}\endgroup\,,\hspace{0.3cm}\vec{c}_1 = \begingroup\renewcommand*{\arraystretch}{1.2}\begin{bmatrix}
    c_{11} \\ c_{21} \\ c_{31}\\
    \end{bmatrix}\endgroup\,.
\end{equation}
The third term in the exponential factor of Eq.~\eqref{eq:explicit+time+nonlinear+non+stationary+first+order} defines the quadratic non-linear frequency shift multiplied with the elapsed time.
If the linear angular mode frequency in the co-rotating frame -- the second term in that factor -- is added to the non-linear frequency shift, we obtain a shifted frequency, hereafter referred to as the non-linear frequency, in the same reference frame.
The expression for the (second-order) non-linear correction to the complex amplitude, $c_{\varphi 2}$, can be found in Appendix~B.1 of \citet{PhDThesisVanBeeck}. 
The harmonic resonance equivalents of the non-linear frequency shifts and second-order non-linear corrections to the complex amplitudes can be found in Appendix~B.2 of that work.

\subsection{Stationary solutions of the amplitude equations}\label{sect:stationary+AEs}

The candidate resonant three-mode couplings identified by \citetalias{2021A&A...655A..59V} are time-independent, with the modes in a non-linear frequency (and phase) lock (i.e., they are synchronized).
We therefore derive the time-independent or stationary solutions of the AEs (i.e., the fixed points of Eqs.~\eqref{eq:AE+A} and \eqref{eq:AE+4+no+zero+A}) in this section.

The trivial stationary solution $\vec{A}^s = \vec{0}$, in which the superscript `s' denotes the stationarity of a quantity, is not oscillatory, and therefore cannot explain the observed couplings.
From Eq.~\eqref{eq:AE+A}, we obtain a oscillatory stationary solution,
\begin{equation}\label{eq:stat+amp+relations}
    \vec{\gamma}\circ\vec{A}^s = - 2\, \lvert \eta_1\rvert\, \sin \left(\Upsilon^s\right)\,\left(\vec{A}_N^s \circ \vec{\Omega}_M\right)\,,
\end{equation}
in which we define the stationary phase coordinate $\Upsilon^s$ as
\begin{equation}\label{eq:stationary+Gamma}
    \Upsilon^s = -\delta\omega\,t_1 + \phi_1^s - \phi_2^s - \phi_3^s + \delta_1\,,
\end{equation}
with $\pdv{\Upsilon^s}{t_1} = 0$.
The stationary equivalent of Eq.~\eqref{eq:AE+4+no+zero+A} yields an expression for the detuning parameter of stationary solutions,
\begin{equation}\label{eq:stat+phase+rewrite+2}
    \delta\omega = -\cot\left(\Upsilon^s\right)\, \gamma_\boxplus\,,
\end{equation}
by setting $\pdv{\vec{A}^s}{t_1}=\vec{0}$ and $\pdv{\Upsilon^s}{t_1}=0$.
To derive this equation, we assume that $\Upsilon^s \neq p\,\pi$ ($p\in \mathbb{Z}$).
If $\Upsilon^s = p\,\pi$, we retrieve the trivial solution, due to Eq.~\eqref{eq:stat+amp+relations}.

By considering the products of two equations of the three in Eq.~\eqref{eq:stat+amp+relations}, and using Eq.~\eqref{eq:stat+phase+rewrite+2}, we write the squared stationary amplitudes as
\begin{equation}\label{eq:stat+amps+q}
    \left(\vec{A}^s\right)^{\circ 2} = \dfrac{\vec{Q}}{4\,\lvert \eta_1\rvert^2}\left[1 + q^2 \right]\,,
\end{equation}
where the superscript `$\circ n$' indicates a Hadamard-Schur (element-wise) nth power. 
In Eq.~\eqref{eq:stat+amps+q}, we use the detuning-damping ratio $q$ and quality factor vector $\vec{Q}$, which we define as
\begin{equation}\label{eq:definition+q}
    \vec{Q} = \begingroup
    \renewcommand*{\arraystretch}{1.2}\begin{bmatrix}
    \tfrac{\gamma_2\,\gamma_3}{\Omega_2\,\Omega_3}\\
    -\tfrac{\gamma_1\,\gamma_3}{\Omega_1\,\Omega_3}\\
    -\tfrac{\gamma_1\,\gamma_2}{\Omega_1\,\Omega_2}\\
    \end{bmatrix}\endgroup = \begingroup\renewcommand*{\arraystretch}{1.2}\begin{bmatrix}
    \tfrac{1}{Q_2\,Q_3} \\ -\tfrac{1}{Q_1\, Q_3} \\ -\tfrac{1}{Q_1\, Q_2}
    \end{bmatrix}\endgroup\,,\hspace{0.3cm}\,q \equiv \dfrac{\delta\omega}{\gamma_\boxplus} = - \cot\Upsilon^s\,,
\end{equation}
with quality factor $Q_\varphi \equiv \Omega_\varphi\,/\,\gamma_\varphi$ for $\varphi \in \myintset{3}$.
The resonant stationary amplitudes are real if
\begin{equation}\label{eq:reality+conditions+driven+case}
    \vec{\Omega}_N >_\circ 0 \,\lor\, \vec{\Omega}_N <_\circ 0 \Rightarrow \vec{A}^s \in_\circ \mathbb{R}\,,
\end{equation}
for a driven resonance scenario, or if
\begin{equation}\label{eq:reality+conditions+direct+parametric+case}
    \vec{\Omega}_P >_\circ 0 \,\lor\, \vec{\Omega}_P <_\circ 0 \Rightarrow \vec{A}^s \in_\circ \mathbb{R}\,,
\end{equation}
for direct and parametric resonance scenarios.
In Eqs.~\eqref{eq:reality+conditions+driven+case} and \eqref{eq:reality+conditions+direct+parametric+case}, $\vec{\Omega}_N$ and $\vec{\Omega}_P$ are the mode frequency variants of the vectors $\vec{A}_N$ and $\vec{A}_P$ defined in Eq.~\eqref{eq:amplitude+products}.
The operators $>_\circ$ and $\in_\circ$ are the element-wise equivalents of the operators $>$ and $\in$.
Equivalent expressions describing stationary solutions for harmonic resonances can be found in Appendix~B.2 of \citet{PhDThesisVanBeeck}.

Three quantities determine the stationary mode amplitudes: the non-linear interaction described by coupled-mode equations~\eqref{eq:real+coupled+mode+b+alpha} and \eqref{eq:real+coupled+mode}, the linear eigenmode properties given by $\vec{Q}$, and the detuning of the resonance measured by $q$.
For decreasing values of the non-linear coupling coefficients $|\eta_1|$, the efficiency for non-linear energy transfer between modes is lower, hence, the stationary amplitudes~\eqref{eq:stat+amps+q} become larger, because a larger mode energy (and thus mode amplitude) is required to transfer enough energy in order to have a significant non-linear effect leading to amplitude saturation.
Quality factor $Q_\varphi$ expresses the ratio of the mode $\varphi$'s e-folding damping or driving time ($\gamma_\varphi^{-1}$) relative to its angular period ($\Omega_\varphi^{-1}$).
With increasing $\lvert Q_\varphi\rvert$, more mode periods are needed for the mode energy (normalized to $GM^2/R$) to damp or grow by a factor of $e$ due to linear heat-driven damping or driving.
The stationary amplitude of a certain mode therefore decreases if the other modes involved in the triad have larger values of $\lvert Q_\varphi \rvert$, because then either a linearly (heat-)driven mode $\varphi$ gains less energy per cycle or a linearly (heat-)damped mode $\varphi$ loses energy more slowly per cycle.
For example, in the parametric resonance scenario, the stationary amplitude of the daughter mode $2$ or $3$ is increased when the parent mode is driven faster (by the $\kappa$ mechanism) and when the other daughter mode $3$ or $2$ is damped faster.
The stationary amplitude of the parent mode $1$ is larger when it is coupled to daughter modes that are more difficult to non-linearly excite (i.e., a larger parent mode energy is required to have a measurable non-linear effect).

Increasing values of $\lvert q\rvert$ lead to increasing values of the stationary amplitudes.
Two factors affect the value of the detuning-damping ratio $q$: the detuning $\delta\omega$ and $\gamma_\boxplus\, (\equiv \gamma_1 + \gamma_2 + \gamma_3)$.
For larger detunings the stationary amplitudes increase, because the increased detuning reduces the efficiency of the non-linear mode coupling (similar to decreasing $|\eta_1|$), requiring a larger amplitude for a non-linear effect.
For smaller values of $\lvert\gamma_\boxplus\rvert$, that is, for the case where the parent and daughter mode linear heat-driven growth and damping rates almost balance out, there is very weak overall linear excitation or damping, requiring a very close resonance (i.e., small detuning) and large amplitudes to have a non-linear effect.

The relative stationary energies of the modes in the triad are related by the ratios of their quality factors, assuming conditions~\eqref{eq:reality+conditions+driven+case} or \eqref{eq:reality+conditions+direct+parametric+case} hold:
\begin{equation}\label{eq:theoretical+amplitude+ratios}
    \left(\dfrac{A_2^s}{A_1^s}\right)^2 = -\dfrac{Q_2}{Q_1}\,,\hspace{4mm}\left(\dfrac{A_3^s}{A_1^s}\right)^2 = -\dfrac{Q_3}{Q_1}\,,\hspace{4mm}\left(\dfrac{A_3^s}{A_2^s}\right)^2 = \dfrac{Q_3}{Q_2}\,.
\end{equation}
The energy ratios~\eqref{eq:theoretical+amplitude+ratios} do not depend on the non-linear coupling coefficients because of the symmetry of these coefficients, and are determined by linear properties of the modes only.
Therefore, of two linearly damped modes, the one with the longest damping time per cycle (largest $\lvert Q_\varphi \rvert$) reaches the largest stationary amplitude through non-linear energy transfer from the linearly excited mode, because it loses energy more slowly.

Non-linear stationary frequencies $\Omega^{nl,s}$ for modes $\varphi \in \myintset{3}$ are obtained from Eq.~\eqref{eq:explicit+time+nonlinear+non+stationary+first+order} in the form
\begin{equation}\label{eq:general+frequency+shift}
    \Omega_\varphi^{nl,s} = \Omega_\varphi + \delta\Omega_\varphi^s\,,
\end{equation}
with the non-linear stationary frequency shifts $\delta\Omega_\varphi^s$ given by
\begin{equation}\label{eq:explicit+frequency+shift}
    \begingroup
    \renewcommand*{\arraystretch}{1.2}\begin{bmatrix}
    \delta\Omega_1^s\\
    \delta\Omega_2^s\\
    \delta\Omega_3^s\\
    \end{bmatrix}\endgroup = \mathfrak{J}\,q\,\begingroup\renewcommand*{\arraystretch}{1.2}\begin{bmatrix}
    -\gamma_1 \\
    \gamma_2 \\
    \gamma_3 \\
    \end{bmatrix}\endgroup = \Delta\Omega^l\,\begingroup\renewcommand*{\arraystretch}{1.2}\begin{bmatrix}
    -\gamma_1 / \gamma_\boxplus \\
    \gamma_2 / \gamma_\boxplus \\
    \gamma_3 / \gamma_\boxplus \\
    \end{bmatrix}\endgroup\,,
\end{equation}
to first order.
For parametric and direct resonance scenarios, the sign of the frequency shift is thus determined by the sign of $\Delta\Omega^l$.
The non-linear stationary frequencies $\Omega_\varphi^{nl,s}$ are frequency-locked, meaning that they satisfy the resonance condition~\eqref{eq:frequency+detuning} exactly:
\begin{equation}\label{eq:locked+condition}
    \Omega_1^{nl,s} - \Omega_2^{nl,s} - \Omega_3^{nl,s} = 0\,,
\end{equation}
which follows from Eq.~\eqref{eq:stationary+Gamma} and $\pdv{\Upsilon^s}{t_1} = 0$.
Explicit expressions for harmonic resonance frequency shifts can be found in Appendix~B.2 of \citet{PhDThesisVanBeeck}.
As stated in Sect.~\ref{sect:AEs}, the derived stationary quantities in this Section also describe the stationary properties of modes in difference-frequency resonances.

\subsection{Stability of the stationary amplitude equation solution}\label{sect:AE+stability}

From a mathematical perspective, the system of amplitude equations defined by Eqs.~\eqref{eq:AE+A} and \eqref{eq:AE+4+no+zero+A} is an autonomous dynamical system, and its stationary solutions are fixed points of that dynamical system.
One of the fundamental results of dynamical systems theory is the Hartman-Grobman or linearization theorem, which states that if a fixed point is hyperbolic, a linearization of the dynamical system can be used to trace the asymptotic behavior of dynamical system solutions near the fixed point \citep{1983guckenheimerbook,2010betounesbook}.
The stability of a hyperbolic fixed point can thus be determined from the linearization of the system around that point.
Two conditions need to be fulfilled for a fixed point $\overline{\tens{X}}$ of a dynamical system $\pdv{\tens{X}}{t} = f(\tens{X})$ to be hyperbolic: both the Jacobian and the real parts of the eigenvalues of the Jacobian matrix of $f(\overline{\tens{X}})$ cannot be equal to zero \citep{1983guckenheimerbook,2010betounesbook}.
Hereafter, we collectively refer to these two conditions as the hyperbolicity condition.
The Jacobian matrix of a dynamical system of amplitude equations depends on the values of the respective non-linear coupling coefficients because the hyperbolicity condition considers the geometry of the non-linear solution around the fixed point (see Appendices~B.1 and B.2 of \citealt{PhDThesisVanBeeck} for the expressions).

In the remainder of this section we assume that the hyperbolicity condition is fulfilled for the stationary solution, so that a linearization can be used to guarantee the stability of that solution.
Let us now examine the stability of the stationary solutions for each of the three isolated resonant coupling scenarios discussed in Sect.~\ref{sect:three+mode+coupling+scenarios}.
Physically, the stability of the (hyperbolic) stationary solutions of the AEs is governed by the reaction of the system (i.e., the pulsating star) to small disturbances away from the stationary state (in this case, away from the stationary pulsation solution).
Such disturbances can be damped or get amplified over time.
If they are damped, the stationary solution is stable.
The solution is unstable if they get amplified.

To derive the linearized dynamical system, we introduce small perturbations of the stationary amplitudes ($\delta A_\varphi$ for mode $\varphi$) and phases ($\delta \phi_\varphi$ for mode $\varphi$), so that the complex amplitudes take the form \citep{1995A&A...295..371V}
\begin{equation}\label{eq:linearization+subst}
    a_{\varphi} = \left(A^s_{\varphi} \,+\, \delta A_{\varphi}\right) \exp\left[i\left(\phi_{\varphi}^s \,+\, \delta\phi_{\varphi}\right)\right],\hspace{4mm} \forall\, \varphi \in \myintset{3}\,.
\end{equation}
Substituting the perturbed complex amplitude factor~\eqref{eq:linearization+subst} into the complex-valued AEs~\eqref{eq:final+form+extended+AEs}, subsequently linearizing and separating the real-valued and complex-valued parts, yields the tensor equation that describes the time evolution of the perturbations,
\begin{equation}\label{eq:tensor+linearization+case}
    \pdv{\tens{Z}}{t} = \tens{M}\,\tens{Z}\,.
\end{equation}
Here, $\tens{M}$ is the stability matrix, defined as 
\begin{equation}\label{eq:stability+matrix}
    \tens{M} \equiv \begingroup\renewcommand*{\arraystretch}{1.2}\begin{pmatrix}
    \gamma_1 & -\gamma_1 & -\gamma_1 & q \,\gamma_1\\
    -\gamma_2 & \gamma_2 & -\gamma_2 & q\, \gamma_2 \\
     -\gamma_3 & -\gamma_3 & \gamma_3 & q \,\gamma_3\\
    q\,\gamma_{\boxplus_1} & q\,\gamma_{\boxplus_2} & q\,\gamma_{\boxplus_3} & \gamma_\boxplus \\
    \end{pmatrix}\endgroup\,,
\end{equation}
and $\tens{Z}$ is the perturbation tensor defined as
\begin{equation}
    \tens{Z} = \left(\delta A_1/A_1^s\,,\, \delta A_2/A_2^s\,,\, \delta A_3/A_3^s\,,\, \delta\Upsilon\right)^{\rm T}\,,
\end{equation}
for which
\begin{equation}
    \delta \Upsilon = \delta\phi_1 - \delta\phi_2 - \delta\phi_3\,,\hspace{4mm}
    \gamma_{\boxplus_\varphi} \equiv \gamma_\boxplus - 2\,\gamma_\varphi, \hspace{4mm} \forall \varphi \in \myintset{3}\,.
\end{equation}

We infer an analytical stability domain of the stationary solutions of the AEs based on the Routh-Hurwitz stability criterion (e.g., \citealt{Hahn1967_book,1971_cesari_book,1983cmbt.book.....K}).
Necessary but not sufficient conditions for stability are that the Hurwitz determinants $H_u$ are positive (see Appendix~B.4 in \citealt{PhDThesisVanBeeck}):
\begin{subequations}\label{eq:hurwitz+determinant+criteria}
\begin{align}
    H_{1} =&\, w_1 = -2\,\gamma_\boxplus > 0 \Leftrightarrow \gamma_\boxplus < 0\,,\label{eq:first+hurwitz}\\
    H_{2} =&\, w_1\,w_2 - w_3 > 0\,,\label{eq:second+hurwitz}\\
    H_{3} =&\, w_3\,H_2 - w_1^2\,w_4 > 0\label{eq:third+hurwitz}\,,\\
    H_{4} =&\, w_4 \, H_{3} > 0 \label{eq:fourth+hurwitz}\,,
\end{align}
\end{subequations}
in which the coefficients $w_u\,\,(u \in \myintset{4}_0$, with $\myintset{n}_0 \equiv \left\{0\right\}\cup\myintset{n} = \left\{\, x \in \mathbb{N}\,|\, x\leq u \,\right\}$) are the coefficients of the characteristic polynomial $p_{\tens{Z}}(\lambda)$ of the stability matrix~\eqref{eq:stability+matrix},
\begin{equation}\label{eq:characteristic+equation+case+4D}
    p_{\tens{Z}}(\lambda) \equiv \sum_{u=0}^{4} w_{4-u}\,\lambda^u = w_4 + w_3\, \lambda + w_2\, \lambda^2 + w_1\, \lambda^3 + w_0\, \lambda^4 = 0\,,
\end{equation}
In Eq.~\eqref{eq:characteristic+equation+case+4D}, $\lambda$ denotes an eigenvalue of the tensor equation~\eqref{eq:tensor+linearization+case}, and the coefficients $w_u$ are
\begin{subequations}\label{eq:characteristic+equation+case+4D+coefficients}
\begin{align}
    w_0 =&\, 1\,,\\
    w_1 =&\, -2\,\gamma_\boxplus\,,\\
    w_2 =&\, \gamma^2_\boxplus\left[1 + q^2\right] -4\,q^2\,\left[\gamma_1\,\gamma_2 + \gamma_1\,\gamma_3 + \gamma_2\,\gamma_3\right]\,,\\
    w_3 =&\, 4\,\gamma_1\,\gamma_2\,\gamma_3\, \left[1 + 3\,q^2\right]\,,\\
    w_4 =&\, -4\, \gamma_\boxplus\, \gamma_1\,\gamma_2\,\gamma_3\, \left[1 + q^2\right] \,.
\end{align}
\end{subequations}
These are the same coefficients as the ones that were determined by \citet{1982AcA....32..147D} in his formalism for quadratic non-linear mode coupling of three distinct modes in non-rotating stars.
Because $H_3 > 0$ due to the stability condition~\eqref{eq:third+hurwitz}, stability condition~\eqref{eq:fourth+hurwitz} becomes $w_4 > 0$.

The necessary but not sufficient stability conditions in Eq.~\eqref{eq:hurwitz+determinant+criteria} show that a stationary solution at the quadratic coupling level is always unstable for the driven resonant coupling scenario, because Eq.~\eqref{eq:first+hurwitz} and $\vec{\gamma} >_\circ 0$ are contradictory.
This is a logical consequence of only considering modes that are linearly excited and can exchange energy between them, and not including third-order couplings that can limit the amplitude.
The direct resonant three-mode coupling scenario also always yields unstable stationary solutions for distinct modes $2$ and $3$, similar to what was found by \citet{1982AcA....32..147D}, because Eqs.~\eqref{eq:first+hurwitz} and \eqref{eq:fourth+hurwitz} contradict each other if the necessary stability condition~\eqref{eq:third+hurwitz} is to be satisfied.
Hence, driven and direct resonance scenarios lead to time-variable amplitudes for their coupled modes. Modes in such coupling scenarios can for example lead to limit cycle behavior of the amplitudes (see e.g., \citealt{2009SeydelBook}).
For the parametric resonant three-mode coupling scenario, Eq.~\eqref{eq:fourth+hurwitz} is trivially fulfilled if Eqs.~\eqref{eq:first+hurwitz} and \eqref{eq:third+hurwitz} are fulfilled.
The stability of the fixed points of harmonic resonance analogues is discussed in Appendix~B.2 of \citet{PhDThesisVanBeeck}.

\begin{figure*}
  \includegraphics[width=17cm]{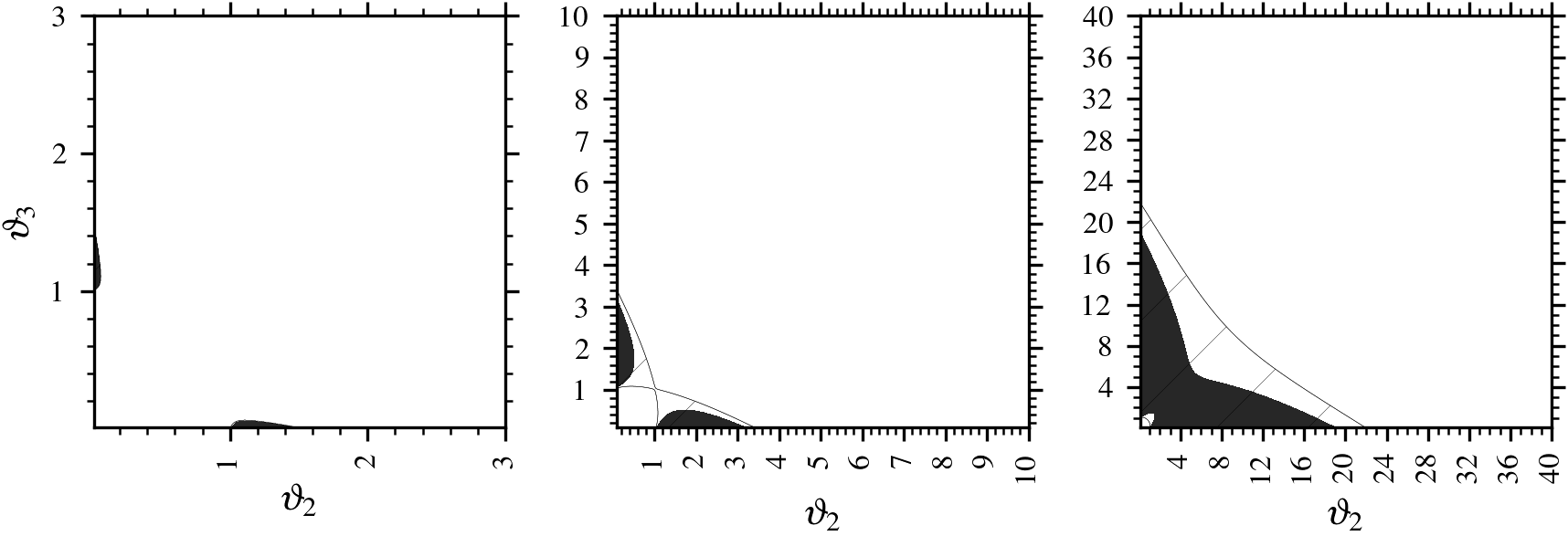}\\
  \includegraphics[width=17cm]{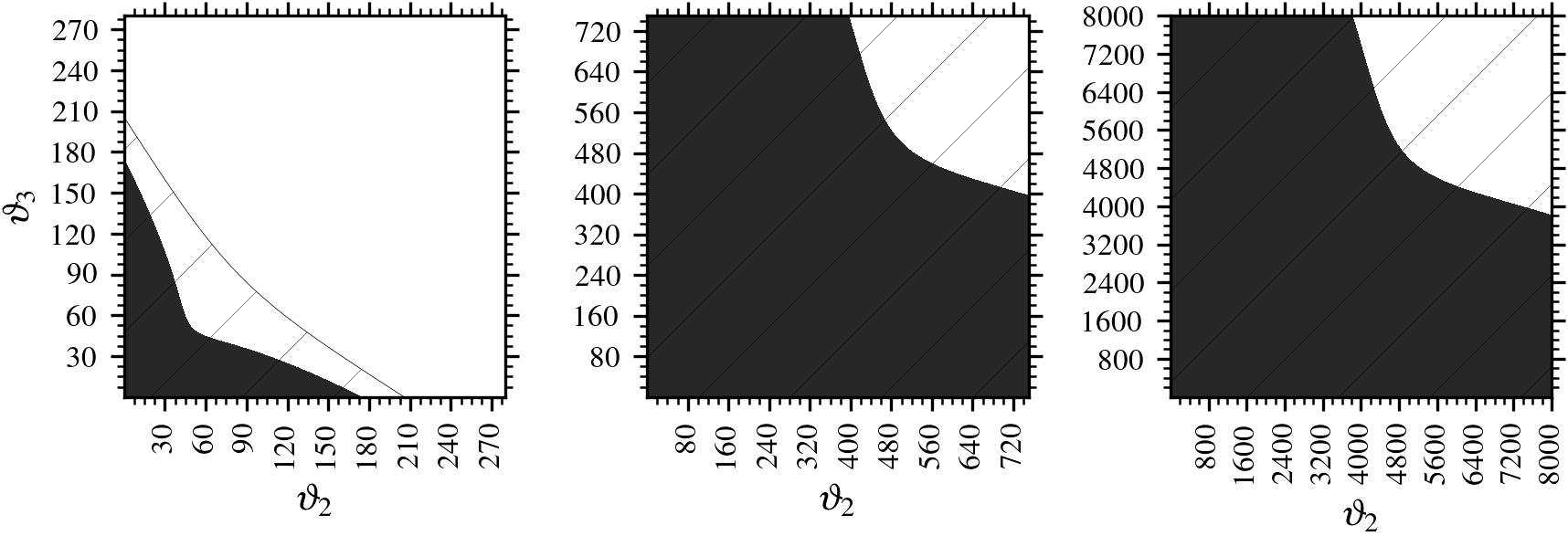}
  \caption{Stability domains as a function of $\vartheta_2$ and $\vartheta_3$, indicated by the dark, shaded area, and evaluated using Eqs.~\eqref{eq:first+hurwitz} and \eqref{eq:quartic+stability+condition} for different values of $\lvert q_1 \rvert$.
  Specifically, $\lvert q_1 \rvert$ is set equal to $0.1$ (left top panel), $1$ (middle top panel), $10$ (right top panel), $10^2$ (left bottom panel), $10^3$ (middle bottom panel) and $10^4$ (right bottom panel).
  The stability domain determined by the stability conditions defined in \citet{1982AcA....32..147D} is larger and is also indicated in these panels by the hatched areas.
  Unhatched white areas of the figure panels indicate unstable domains.
  Note the different axis scales for the different panels, which indicates the importance of the value of $\lvert q_1 \rvert$ in determining the stability of stationary solutions.}
  \label{fig:stability+domains}
\end{figure*}

In addition to Eq.~\eqref{eq:hurwitz+determinant+criteria}, all $w_u$ also need to be positive to obtain stable parametric stationary solutions (see e.g., \citealt{Hahn1967_book} and Appendix~B.4 in \citealt{PhDThesisVanBeeck}).
Combining these additional conditions with Eq.~\eqref{eq:hurwitz+determinant+criteria}, the stability domain of a parametric three-mode resonant coupling with real amplitudes can be described by only three conditions: Eq.~\eqref{eq:first+hurwitz}, the hyperbolicity check, and the quartic condition
\begin{equation}\label{eq:quartic+stability+condition}
    \dfrac{-\gamma_1^3}{\gamma_\boxplus}\left[\mathfrak{d}_4 + \mathfrak{d}_2\, q^2 + \mathfrak{d}_0\, q^4\right] > 0\,\, \Leftrightarrow\,\, \mathfrak{d}_4 + \mathfrak{d}_2\, q^2 + \mathfrak{d}_0\, q^4 > 0\,.
\end{equation}
The factor $-\gamma_1^3 / \gamma_\boxplus$ in Eq.~\eqref{eq:quartic+stability+condition} is positive for parametric resonances that fulfill Eq.~\eqref{eq:first+hurwitz}. 
The dimensionless quartic coefficients $\mathfrak{d}_u$ ($u = 0, 2, 4$) in that equation are given by
\begin{subequations}\label{eq:coefficients+equivalent+stability+condition}
\begin{align}
    &\begin{aligned}\mathfrak{d}_0 =&\, -18\,\vartheta_2\vartheta_3 - 3 \left(1 -\vartheta_2 - \vartheta_3\right)\\&\left[\left(1 -\vartheta_2 - \vartheta_3\right)^2 + 4\left(\vartheta_2 + \vartheta_3 - \vartheta_2\vartheta_3\right)\right]\,,\end{aligned}\label{eq:delta+0+coefficient}\\
    &\begin{aligned}\mathfrak{d}_2 =&\, -12\,\vartheta_2\vartheta_3 - \left(1 -\vartheta_2 - \vartheta_3\right)\\&\left[2 + \left(\vartheta_2 - \vartheta_3\right)^2 + \left(\vartheta_2 + \vartheta_3\right)^2\right]\,,\end{aligned}\label{eq:delta+2+coefficient}\\
    &\mathfrak{d}_4 =\, -2\,\vartheta_2\vartheta_3 + \left(1 -\vartheta_2 - \vartheta_3\right)^3\,,\label{eq:delta+4+coefficient}
\end{align}
\end{subequations}
where we define $\vartheta_{2,3} \equiv -\gamma_{2,3}\,\displaystyle/\,\gamma_1$, which are positive for parametric resonances. 
The physical meaning of these dimensionless ratios is similar to that of the quality factor ratios discussed in Sect.~\ref{sect:stationary+AEs}, but inverse: they compare the damping and/or driving time scales.

The quartic stability condition~\eqref{eq:quartic+stability+condition} is symmetric in $\vartheta_2$ and $\vartheta_3$, in accordance with the symmetry of the coupling coefficient.
Coefficients $\mathfrak{d}_2$ and $\mathfrak{d}_4$ are the same as in Eq.~(6.14) of \citet{1982AcA....32..147D}. 
Although the coefficient $\mathfrak{d}_0$ differs from the one given in that equation, likely due to an error in \citet{1982AcA....32..147D}, the stability condition is derived from the same characteristic polynomial coefficients defined in Eq.~\eqref{eq:characteristic+equation+case+4D+coefficients}.
The coefficients $\mathfrak{d}_u$ ($u = 0, 2, 4$) are a function of $\vartheta_2$ and $\vartheta_3$ only, whereas $q$ can be expanded as a function of $\vartheta_2$, $\vartheta_3$, and the ratio of the linear frequency detuning to the parent's linear driving rate $q_1 \equiv \delta\omega / \gamma_1$. 
We therefore can explore the stability domain of the (hyperbolic) stationary solutions by varying only three dimensionless ratios of linear variables: $q_1 = \delta\omega / \gamma_1$, $\vartheta_2$ and $\vartheta_3$.
The dimensionless ratio $q_1$ can be written as
\begin{equation} \label{eq:alternative+def+q1}
    q_1 \equiv Q_1\,\left(\dfrac{\delta\omega}{\Omega_1}\right) =  \dfrac{Q_1\,\delta\omega\,}{2\,\pi}P_1\,,
\end{equation}
where the cyclic co-rotating frame period $P_1 = 2\,\pi\,\displaystyle/\,\Omega_1$.
It is therefore a period-weighted combined measure of the driving time scale of the linearly excited parent mode ($Q_1$) and the efficiency with which non-linear energy transfer occurs ($\delta \omega$).
Hence, when comparing values of $q_1$ for triads with linearly excited modes of similar period, one might expect that the triad with larger $\lvert q_1\rvert$ has larger stationary mode amplitudes because of decreased efficiency of non-linear energy transfer (larger $\delta\omega$) and/or smaller stationary mode energy ratios $(A_{2,\,3}^s / A_{1}^s)^2$ (larger $Q_1$, when values of $Q_{2,\,3}$ are comparable).

Figure~\ref{fig:stability+domains} displays the domains of stability of stationary solutions in the three-dimensional phase space of the parameters $q_1$, $\vartheta_2$ and $\vartheta_3$, covering commonly encountered parameter values when modeling mode interactions in SPB stars (see Sect.~\ref{subsect:parameter+ensembles}).
The necessary condition~\eqref{eq:first+hurwitz} for stability of the fixed point, which requires that $\vartheta_2 + \vartheta_3 > 1$, is clearly recognizable on the left and middle panels in the top row of Fig.~\ref{fig:stability+domains}.
Physically, this expresses that daughter modes must be sufficiently damped compared to the linear excitation of the parent mode for stability, otherwise, resonant energy transfer will increase all amplitudes without bound.

The quartic stability condition~\eqref{eq:quartic+stability+condition} is more difficult to interpret.
Term $(1 - \vartheta_2 - \vartheta_3)$ in that stability condition is always negative due to Eq.~\eqref{eq:first+hurwitz} and can be interpreted as an effective total-damping-to-linear-driving ratio $\gamma_\boxplus / \gamma_1$. 
If the absolute value of the ratio is large, the overall damping per unit of driving is large as well.
Other terms are less straightforward to interpret.
Hence, we base our interpretation of this stability condition primarily on the stability domains pictured in Fig.~\ref{fig:stability+domains}.
The quartic stability condition derived in this work is stricter than that of \citet{1982AcA....32..147D} (displayed as hatched areas in Fig.~\ref{fig:stability+domains}) due to the difference in value of the coefficient $\mathfrak{d}_0$.
Condition~\eqref{eq:quartic+stability+condition} can also be expressed in terms of $q_1$, $\vartheta_2$ and $\vartheta_3$. 
Hence, conclusions drawn from the visualized stability domains in the chosen phase space (shown in Fig.~\ref{fig:stability+domains}) can be related to the equivalent stability condition that is expressed in terms of these three variables.

For large values of $\vartheta_2$ and $\vartheta_3$ a regime of strong damping (relative to the linear excitation of the parent mode) is reached.
In such strong damping regimes the fixed point solutions are unstable because the transfer of energy (per cycle) is not enough to overcome linear damping.
The domain of stability at the strong damping end of the $\vartheta_2$-$\vartheta_3$ plane moves towards larger values of $\vartheta_2$ and $\vartheta_3$ for larger values of $q_1$, as shown in the different panels of Fig.~\ref{fig:stability+domains}.
Conversely, the stability domain decreases considerably in size for smaller values of $q_1$.
This can be explained by an increase (decrease) in energy transfer efficiency and/or an increase (decrease) in energy available for transfer, leading to faster (slower) rates of energy transfer and corresponding smaller (larger) domains of stability, based on the physical explanation of Eq.~\eqref{eq:alternative+def+q1}.
Specifically, for faster (slower) rates of energy transfer per cycle, the stationary solutions can endure smaller (larger) perturbations around the stationary solutions before energy transfer renders the fixed points unstable.
The stability of the stationary solutions is thus primarily determined by the speed of energy transfer.

\subsection{The onset of parametric instability}\label{sect:onset+parametric+instability}

In this section we derive the minimum amplitude conditions for the onset of the parametric resonance instability for the non-linear interaction among three distinct modes in a sum or difference frequency.
If this instability does not occur for these three distinct modes, the amplitudes of the modes in the triad can only be limited to stable stationary values by higher-order non-linear mode coupling terms, such as the cubic self-coupling terms that were described in \citet{1996A&A...308...66V}.
Alternatively, limit cycles characterized by non-stationary amplitudes may occur.
Similar conditions for the onset of parametric and direct resonant instability derived for harmonic resonances can be found in Appendix~B.2 of \citet{PhDThesisVanBeeck}.

The initial growth of the daughter modes can be described using the complex AEs~\eqref{eq:final+form+extended+AEs} for one of the daughter modes and its complex conjugate for the other daughter mode.
If we then set $a_k = S_k\,\exp(-i\,\delta\omega\,t_1/2)$ (for $k \in \left\{2\,,\,3\right\}$, following \citealt{1982AcA....32..147D}) the explicit time-dependence disappears.
Further assuming that the complex amplitude factor $a_1$ stays constant, a plausible assumption in the initial phase of energy transfer, yields
\begin{subequations}\label{eq:onset+first+set+eqs}
\begin{align}
\pdv{S_2}{t_1} &= \left(\gamma_2 + i\,\dfrac{\delta\omega}{2}\right)\,S_2 + 2\,i\,\eta^*\,a_1\,S_3^*\,\Omega_2\,,\\
\pdv{S_3^*}{t_1} &= \left(\gamma_3 - i\,\dfrac{\delta\omega}{2}\right)\,S_3^* - 2\,i\,\eta\,a_1^*\,S_2\,\Omega_3\,.
\end{align}
\end{subequations}
Under the assumption that $S_k \sim \exp(\sigma\,t_1)$ (for $k \in \left\{2,3\right\}$), Eq.~\eqref{eq:onset+first+set+eqs} can be solved for the growth parameter $\sigma$, yielding
\begin{equation}\label{eq:characteristic+equation+sigma}
    \sigma = \dfrac{\gamma_2 + \gamma_3 \pm \sqrt{(\gamma_3 - \gamma_2 - i\,\delta\omega)^2 + 16\lvert \eta_1\rvert^2 A_1^2\,\Omega_2\,\Omega_3}}{2}\,,
\end{equation}
equivalent to expressions given by \citet{1981SvAL....7..128V} and \citet{1982AcA....32..147D}.
Parametric instability will occur if $\rp{\sigma} > 0$, because this ensures growth of the daughter mode amplitudes $A_{2}$ and $A_3$.
At the onset of parametric instability, $\rp{\sigma} = 0$.
We therefore define the instability threshold amplitude for the parent mode $A_t$ as the value of $A_1$ for $\rp{\sigma} = 0$.
The growth parameter $\sigma$ is then imaginary and we can set $\sigma = p\,i$, with $p$ determined by solving Eq.~\eqref{eq:characteristic+equation+sigma}:
\begin{equation}\label{eq:parameter+onset+instability}
    p = \dfrac{\delta\omega}{2}\dfrac{\left(\gamma_3 - \gamma_2\right)}{\left(\gamma_3 + \gamma_2\right)}\,.
\end{equation}
Using this expression in the real part of the growth parameter~\eqref{eq:characteristic+equation+sigma} yields the parametric instability threshold amplitude
\begin{equation}\label{eq:instability+threshold+amplitude}
    A_t = \dfrac{1}{2\,\lvert \eta_1 \rvert\sqrt{Q_2\,Q_3}}\sqrt{1 + \left(\dfrac{\delta\omega}{\gamma_2 + \gamma_3}\right)^2}\,.
\end{equation}
Instability threshold amplitude~\eqref{eq:instability+threshold+amplitude} is equivalent to the ones derived in \citet{1982AcA....32..147D}, \citet{2001ApJ...546..469W} and \citet{2003ApJ...591.1129A}.

Parametric resonant mode triad interactions require that the parent mode amplitude $A_1 \geq A_t$, and $A_1^s$ is always larger than $A_t$.
In the limit of very small (non-zero) detuning $\delta\omega$, the threshold amplitude is solely dependent on the coupling coefficient and the quality factors.
In that case, $A_t$ increases with decreasing $\lvert\eta_1\rvert$ and faster damping of the daughter modes (expressed by the quality factors) because both terms limit the amplitude growth of the daughter modes due to non-linear energy transfer, thus requiring a larger parent mode energy for a visible non-linear effect.
A larger detuning increases the threshold amplitude because of the less efficient energy transfer.

%
%

\section{Theoretically predicted observables}\label{sect:theoretical+observables}

An important observable in linear $g$-mode asteroseismic modeling is a $g$-mode period spacing pattern (which are extensively described in the literature; see e.g., \citealt{2018ApJS..237...15A,2021A&A...650A.175M,2021A&A...656A.158B} for some recent examples of how they can be used to probe internal mixing).
In this section we derive additional observables based on the theoretical AE formalism described in Sect.~\ref{sect:theoretical+model} and outline how to compare them to observed quantities.

An inherent assumption of our models is that the modes are coherent.
That assumption is justified because the detected frequencies of variability in SPB stars have been observed to be stable with a frequency precision of order $10^{-7}$ d$^{-1}$, based on long-term ground-based photometric monitoring \citep{2002A&A...393..965D}.
This is not necessarily the case for other pulsators: $\delta$~Sct stars, for example, show frequency and amplitude modulation in the majority of detected signals \citep{2016MNRAS.460.1970B}.
Amplitude and frequency modulation also occurs among $g$ mode triplets in oscillating white dwarfs (see e.g., the pioneering study of the oscillating DB white dwarf star KIC08626021 by \citealt{2016A&A...585A..22Z}).
Moreover, the amplitudes of the oscillating hot B subdwarf star KIC10139564 reveal that the modulation of its observed $p$ modes is larger than that of its $g$ modes \citep{2016A&A...594A..46Z}.
Whether this trend is generic among oscillating stars remains to be verified.

\subsection{Amplitudes: model-generated luminosity fluctuations}

We cannot compare the theoretical stationary amplitudes derived in Sect.~\ref{sect:stationary+AEs} with the surface luminosity amplitudes $\Tilde{A}$ determined from observations.
These theoretical stationary amplitudes must first be converted to the corresponding observables, the theoretical luminosity fluctuations at the stellar surface, $\mathfrak{L}$.

In this section we follow the approach of \citet{2017MNRAS.472.1538F} to compute a conversion factor $o_{A}$ used to convert a theoretical amplitude into a theoretical luminosity fluctuation $\mathfrak{L}$ caused by an oscillation mode of a resonant triad.
Analogous to Eq.~(82) in \citet{2017MNRAS.472.1538F}, we estimate the disc-averaged luminosity fluctuation for a single SPB star due to $g$ mode $\varphi$ as
\begin{equation}\label{eq:visible+lum+fluctuation}
    \dfrac{\Delta L_\varphi}{L} = \left(\dfrac{\Delta L_{\varphi}}{L}\right)_{\rm mode}\,H_r(i_s; k)\,,
\end{equation}
within the TAR, where
\begin{equation}\label{eq:mode+luminosity+perturbation}
    \left(\dfrac{\Delta L_{\varphi}}{L}\right)_{\rm mode} = t_k\,\dfrac{\Delta L_{\varphi,\,R}(R)}{L(R)} - e_k\,\dfrac{\xi_{\varphi,\,r}(R)}{R}
\end{equation}
in which $\Delta L_{\varphi,\,R}(R)$ is the Lagrangian surface luminosity perturbation due to mode $\varphi$, 
$\xi_{\varphi,\,r}$ is the radial part of the adiabatic eigenfunction of that mode, $i_s$ is the angle between the rotation axis and the line of sight at time $t = 0$ s (also called the spin inclination angle, see e.g., \citealt{2012MNRAS.420.3126F}), and $t_k$ and $e_k$ are limb-darkening coefficients given by the overlap integrals (in analogy to e.g., \citealt{2012MNRAS.421..983B})
\begin{subequations}\label{eq:disc+integration+factors}
\begin{align}
    t_k &= \int_0^1\,\mu\,H_{r}(\mu; k)\,h(\mu)\,d\mu \,,\\
    e_k &= \int_0^1\left[2\,\mu^2\,\dv{H_r(\mu; k)}{\mu} - \left(\mu - \mu^3\right)\dv[2]{H_r(\mu; k)}{\mu}\right]\,h(\mu)\,d\mu\,,
\end{align}
\end{subequations}
where $\mu \equiv \cos\theta$ and $h(\mu)$ is a limb-darkening function.
To derive Eq.~\eqref{eq:visible+lum+fluctuation}, we assume that the Lagrangian flux perturbation of mode $\varphi$, which causes the luminosity perturbation, is equal to the radiative flux perturbation (see e.g., \citealt{1989nos..book.....U}).
We use a linear limb-darkening function $h(\mu) = 1 + (3\, \mu / 2)$ in our computations (as has been customary for decades; see e.g., \citealt{1971PASJ...23..485O} and \citealt{1992A&A...266..294A}, who set $\mu = 0.36$ for B stars).
One can however easily account for other, more sophisticated limb-darkening laws by changing the limb-darkening function.
Numerical evaluations of $t_k$ and $e_k$ are necessary, because no analytic closed forms of the classical Hough functions exist.

Equation~\eqref{eq:visible+lum+fluctuation} thus separates contributions to the disc-averaged luminosity fluctuation $(\Delta L_{\varphi}\,\displaystyle/\,L)$ into two multiplicative factors: a factor attributed to the properties of the mode $\varphi$ under consideration, $(\Delta L_\varphi\,\displaystyle/\,L)_{\rm mode}$, and an angular factor that describes the observer's orientation in the rotating frame.
The theoretical flux fluctuation of that mode at the surface, $\mathfrak{L}_\varphi$, can then be computed by multiplying the (complex) amplitude $c_\varphi$ obtained from solving the AEs with the factor $(\Delta L_{\varphi}\,\displaystyle/\,L)$. 
Its modulus $\lvert\mathfrak{L}_\varphi\rvert$ can directly be compared with observed luminosity amplitudes $\Tilde{A}$.
The amplitude conversion factor $o_{A,\,\varphi}$ for a mode $\varphi$ is then defined as
\begin{equation}\label{eq:amp+conversion+factor}
    o_{A,\,\varphi} = \left|\left(\dfrac{\Delta L_\varphi}{L}\right)_{\rm mode}\right|\,H_r(i_s; k)\,.
\end{equation}
We compute the expected theoretical threshold surface luminosity fluctuations $\lvert \mathfrak{L}_t \rvert$, which is the minimum observed luminosity fluctuation that a parent mode in a mode triads needs for (parametric) resonant mode coupling to occur, and the stationary surface luminosity fluctuation $\lvert\mathfrak{L}_\varphi^s\rvert$ of mode $\varphi$, as
\begin{equation}\label{eq:surface+luminosity+fluctuations+stationary+threshold}
    \lvert \mathfrak{L}_t \rvert =  o_{A,1}\,A_t\,,\hspace{0.3cm} \lvert\mathfrak{L}_\varphi^s\rvert = o_{A,\,\varphi}\,A_\varphi^s\,.
\end{equation}
The theoretical $g$ mode stationary amplitudes $A_\varphi^s$ and the theoretical $g$ mode (parametric) threshold amplitudes $A_t$ in Eq.~\eqref{eq:surface+luminosity+fluctuations+stationary+threshold} are computed by setting the bookkeeping parameter $\mathfrak{J} = 1$ (similar to e.g., \citealt{1984ApJ...279..394B}), so that $\delta\omega = \Delta\Omega^l$.

The conversion factors~\eqref{eq:amp+conversion+factor} are sensitive to the choice of a limb-darkening function (because this affects $t_k$ and $e_k$), as well as the normalization factors for the Hough functions and the radial parts of the mode eigenfunctions defined in Eq.~\eqref{eq:normalization+radial+hough} and Sect.~\ref{sect:coupled+mode+equations}, respectively.
The stationary daughter-parent surface luminosity ratios 
$\absfraktxt{\mathfrak{L}^s_2}{\mathfrak{L}^s_1}$, $\absfraktxt{\mathfrak{L}^s_3}{\mathfrak{L}^s_1}$ minimize the influence of the choice of limb-darkening function and normalization factors.
We determine these ratios as
\begin{equation}\label{eq:stat+surf+amp+ratios}
    \dfrac{\lvert\mathfrak{L}^s_2\rvert}{\lvert\mathfrak{L}^s_1\rvert} = \left(\dfrac{o_{A,\,2}}{o_{A,\,1}}\right)\sqrt{\dfrac{|Q_2|}{|Q_1|}}\,,\hspace{0.3cm}
    \dfrac{\lvert\mathfrak{L}^s_3\rvert}{\lvert\mathfrak{L}^s_1\rvert} = \left(\dfrac{o_{A,\,3}}{o_{A,\,1}}\right)\sqrt{\dfrac{|Q_3|}{|Q_1|}}\,.
\end{equation}
The daughter-parent surface luminosity fluctuation ratios~\eqref{eq:stat+surf+amp+ratios} are the most robust amplitude-based theoretically predicted observables that can be used in resonant non-linear asteroseismic modeling.
To derive the expressions for these ratios, we assume a parametric resonant mode triad (for a three-mode sum-frequency coupling or its difference frequency analogue), and use the definition of the quality factor $Q_\varphi$, in addition to Eq.~\eqref{eq:theoretical+amplitude+ratios}.
Stationary surface luminosity ratios can thus be computed in terms of linear non-adiabatic parameters.
The equivalent expression for the daughter-parent surface luminosity fluctuation ratio of a harmonic dyad is given in Appendix~B.2 of \citet{PhDThesisVanBeeck}.

In this work, we only envision a rough comparison between theoretical predictions and observables by limiting ourselves to monochromatic predictions.
As highlighted by \citet{2023arXiv231108453A}, future studies of measured amplitude ratios from multi-color space photometry 
by combining Gaia \citep{2016A&A...595A...1G}, {\it Kepler\/} \citep{2010ApJ...713L..79K} or PLATO \citep{2014ExA....38..249R} data offer additional opportunities to characterize stellar atmospheric properties.
Concrete applications of our theory require integrations over particular passbands instead of the monochromatic predictions for the daughter-parent ratios considered here.
Such integrations will not be considered in this work but will be considered in follow-up application papers.

\subsection{Frequencies and phases: frequency detuning and combination phase for $\Omega_1 \approx \Omega_2 + \Omega_3$}\label{sect:observed+freq+detunings+combination+phases}

The inherent assumption made in linear asteroseismic inference is that any non-linear frequency shifts (e.g., those determined by Eq.~\eqref{eq:explicit+frequency+shift}) are negligible, so that theoretical frequencies computed within a linear formalism can directly be compared to their observed counterparts.
A non-linear formalism, such as the one we derive in Sect.~\ref{sect:theoretical+model}, allows one to verify that assumption.

We define the observed frequency detuning $\Delta\Tilde{\Omega}$ as
\begin{equation}\label{eq:observed+detuning}
    \Delta\Tilde{\Omega} = \Tilde{\Omega}_{1,\,\mathfrak{i}} - \Tilde{\Omega}_{2,\,\mathfrak{i}} - \Tilde{\Omega}_{3,\,\mathfrak{i}}\,,
\end{equation}
where the observed frequencies of modes $\varphi$ are defined as $\Tilde{\Omega}_{\varphi,\,\mathfrak{i}}$ with subscript $\mathfrak{i}$ indicating that observed frequencies are measured in the inertial frame.
Because of the azimuthal selection rule~\eqref{eq:azimuthal+selection+rule}, $\Delta\Tilde{\Omega}$ is also equal to its co-rotating frame equivalent $\Delta\Tilde{\Omega}_{\mathfrak{c}}$, that is, $\Delta\Tilde{\Omega} = \Delta\Tilde{\Omega}_{\mathfrak{c}} \equiv \Tilde{\Omega}_{1} - \Tilde{\Omega}_{2} - \Tilde{\Omega}_{3}$.
This equivalence, along with Eq.~\eqref{eq:locked+condition} and the stationary equivalent of Eq.~\eqref{eq:observed+detuning}, then determine that
\begin{equation}\label{eq:necessary+detuning+condition}
    \Delta\Tilde{\Omega}^s = \Delta\Tilde{\Omega}^{nl} = 0.0\,,
\end{equation}
needs to be fulfilled for a isolated (resonantly locked) mode triad.
In identifying such couplings observationally, one should therefore search for combinations of observed modes with stationary amplitudes, for which
\begin{equation}\label{eq:approx+resonance+freq+condition}
    \lvert\Delta\Tilde{\Omega}^s\rvert + \lvert\sigma_{\Delta\Tilde{\Omega}^s}\rvert \lesssim 2\,\pi\,\mathfrak{R}_\nu\,,
\end{equation}
where $\sigma_{\Delta\Tilde{\Omega}^s}$ is the propagated uncertainty of $\Delta\Tilde{\Omega}^s$, and $\mathfrak{R}_\nu \equiv \tfrac{1}{T}$ denotes the Rayleigh limit, with $T$ being the total time span of the time series of the SPB star that needs to be modeled.

The observable that can directly be compared with observed frequencies of modes in an inferred candidate resonance is
\begin{equation}\label{eq:observational+nonlinear+freq}
    \Omega_{\varphi,\,\mathfrak{i}}^{\rm NL,s} \equiv \Omega_\varphi^s + m_\varphi\,\Omega + \delta\Omega_\varphi^s = \Omega_{\varphi,\,\mathfrak{i}}^s + \delta\Omega_\varphi^s\,,
\end{equation}
which includes the quadratic stationary non-linear frequency shift~\eqref{eq:explicit+frequency+shift}.
By summing $\mathfrak{L}_\varphi = o_{A,\,\varphi}\,A_\varphi$ with its complex conjugate, we determine the individual stationary phase observables (see Sect. 4.3.2 in \citealt{PhDThesisVanBeeck} for the explicit manipulations)
\begin{equation}\label{eq:observed+mode+phases+locked}
    \begingroup
    \renewcommand*{\arraystretch}{1.2}\begin{bmatrix}
    \Tilde{\phi}_{1,s}^s\\
    \Tilde{\phi}_{2,s}^s\\
    \Tilde{\phi}_{3,s}^s\\
    \end{bmatrix}\endgroup = \dfrac{\pi}{2} - \begingroup\renewcommand*{\arraystretch}{1.2}\begin{bmatrix}
    \left(\phi^s_1\right)_0 \\
    \left(\phi^s_2\right)_0 \\
    \left(\phi^s_3\right)_0 \\
    \end{bmatrix}\endgroup\ - \begingroup\renewcommand*{\arraystretch}{1.2}\begin{bmatrix}
    \phi_{L\,1} \\
    \phi_{L\,2} \\
    \phi_{L\,3} \\
    \end{bmatrix}\endgroup\,,
\end{equation}
where we use the stationary equivalent of Eq.~\eqref{eq:AE+P}, as well as Eqs.~\eqref{eq:frequency+detuning}, \eqref{eq:stat+amp+relations}, and \eqref{eq:definition+q}.
In Eq.~\eqref{eq:observed+mode+phases+locked}, $\phi_{L\,\varphi}$ is the phase of the complex quantity $\left(\Delta L_\varphi\,\displaystyle/\, L\right)_{\rm mode}$ defined in Eq.~\eqref{eq:mode+luminosity+perturbation}.
The observed individual mode phases~\eqref{eq:observed+mode+phases+locked} therefore are independent of time if the modes are part of a locked mode triad.

In analogy with the definition of the stationary equivalent of the theoretical generic phase coordinate $\Upsilon$ in Eq.~\eqref{eq:upsilon+parameter}, we define the stationary combination phase observable $\Tilde{\Phi}^s_s$ as
\begin{equation}\label{eq:observable+combination+phase}
    \Tilde{\Phi}^s_s = \Tilde{\phi}_{1,s}^s - \Tilde{\phi}_{2,s}^s - \Tilde{\phi}_{3,s}^s = -\dfrac{\pi}{2} - \left(\Delta\phi^s\right)_0 - \Delta\phi_L^s\,,
\end{equation}
where the last equality holds because of Eq.~\eqref{eq:observed+mode+phases+locked}, and in which $\left(\Delta\phi^s\right)_0 =  \left(\phi^s_1\right)_0 -  \left(\phi^s_2\right)_0 -  \left(\phi^s_3\right)_0$ and $\Delta\phi_L^s = \phi_{L\,1} - \phi_{L\,2} - \phi_{L\,3}$.
The coupling coefficient $\eta_1$ is real-valued and therefore has no contribution to Eq.~\eqref{eq:observable+combination+phase}.
Stationary combination phase observable~\eqref{eq:observable+combination+phase} is to be compared with a relative phase computed from the phases of observed candidate resonance signals.

%
%

\section{Numerical results for SPB models}\label{sect:num+results}
We simulate stationary resonant parametric three-$g$-mode coupling processes by computing a grid of models representative for the SPB oscillators.
Numerical stellar evolution models are generated by the stellar evolution code MESA (version $15140$; \citealt{2011ApJS..192....3P,2013ApJS..208....4P,2015ApJS..220...15P,2018ApJS..234...34P,2019ApJS..243...10P}). 
Linear stellar oscillation models are generated by the stellar oscillation code GYRE (version $6.0.1$; \citealt{2013MNRAS.435.3406T,2018MNRAS.475..879T,2020ApJ...899..116G}), and use the MESA models as input.
Numerical mode coupling models use both MESA and GYRE models as input.

We discuss the MESA model grid setup in Sect.~\ref{subsect:MESA+grid+setup}, the GYRE model grid setup in Sect.~\ref{subsect:GYRE+grid+setup}, and display and discuss the numerical results for triad ensembles in Sect.~\ref{subsect:parameter+ensembles}.
The link to our inlists for these codes, as well as the link to our mode coupling code repository, can be found in Appendix~\ref{app:MESA+GYRE+code+link}.

\subsection{MESA model grid setup for SPB stars}\label{subsect:MESA+grid+setup}

We compute MESA stellar evolution models with the parameters given in Table~\ref{tab:mesa+parameters}.
These parameters cover the ranges of inferred initial mass (${\rm M}_{\rm ini}$) and core hydrogen mass fraction (${\rm X}_{\rm c}$) values for SPB stars given in Table~2 of \citet{2022ApJ...930...94P}.
The MESA models have the `standard' initial chemical mixture of nearby B-type stars derived by \citet{2012A&A...539A.143N} and \citet{2013EAS....63...13P}, and an Eddington gray atmosphere.
Following \citet{2021NatAs...5..715P}, we adjust the initial hydrogen and helium mass fractions ${\rm X}_{\rm ini}$ and ${\rm Y}_{\rm ini}$ so that the ratio $\tfrac{{\rm X}_{\rm ini}}{{\rm Y}_{\rm ini}} = \tfrac{{\rm X}_{*}}{{\rm Y}_{*}}$, with ${\rm X}_{*}$ and ${\rm Y}_{*}$ equal to the Galactic standard values for B-type stars in the solar neighborhood \citep{2013EAS....63...13P}.
We use Opacity Project (OP) opacity tables \citep{2005MNRAS.362L...1S} that were computed by \citet{2015A&A...580A..27M} for this elemental mixture.
The full proton-proton chain and CNO cycle nuclear reaction networks are used to describe core hydrogen fusion on the main sequence.
Beyond the zero age main sequence (ZAMS), we use the \citet{2001A&A...369..574V} hot wind scheme with a wind scaling factor fixed to a value of $0.3$ (see \citealt{2021A&A...648A..36B}).

Diffusive isotope mixing processes within the stellar interior are assumed to be described by the simplified transport Eq.~(1) of \citet{2021A&A...650A.175M} and \citet{2021NatAs...5..715P}.
Mixing processes in the radiative envelope are described by a diffusive mixing profile for internal gravity wave mixing deduced by \citet{2017ApJ...848L...1R} and used by \citet{2018A&A...614A.128P} in the context of asteroseismic modeling. This profile scales the radiative envelope mixing level ${\rm D}_{\rm env}$ with a factor inversely proportional to the (local) mass density.
To model core-boundary mixing (CBM) processes, we employ an approach similar to the diffusive exponential overshooting model with efficiency parameters $f_{\rm CBM}$ and $f_0$ fixed to $0.02$ and $0.005$ (see e.g., \citealt{2019A&A...628A..76M,2021A&A...650A.175M}).
The inner boundary of the overshooting zone (i.e., the convective core mass) is determined by the Ledoux criterion for mixing length parameter $\alpha_{\rm MLT} = 2.0$ within the \citet{1968pss..book.....C} formalism for mixing length theory.
In the implementation, we set the minimal level of diffusive isotope mixing 
$D_{\rm env,min}$ equal to $100$ cm$^2$ s$^{-1}$.
If the mixing level drops below that boundary at a certain location within the model, diffusive mixing is halted locally.

Our baseline (fiducial) model is a near terminal age main sequence (TAMS) model with ${\rm X}_{\rm c} = 0.09$, ${\rm M}_{\rm ini} = 4\,{\rm M}_\sun$, solar initial metallicity (${\rm Z}_{\rm ini} = 0.014$), and $f_{\rm CBM}$ equal to $0.02$. 
Models $\Delta {\rm X_{\rm c,\,1}}$ and $\Delta {\rm X_{\rm c,\,2}}$ have core hydrogen mass fractions ${\rm X}_{\rm c}$ equal to $0.29$ and $0.59$, respectively, and are representative for a mid-MS and near-ZAMS SPB star, because ${\rm X}_{\rm c}$ is a proxy for the main sequence age of the star (hydrogen is depleted in the core during the main sequence).
Model $\Delta {\rm X_{\rm c}}|\rm M_{\rm ini}$ is representative of a ${\rm M}_{\rm ini} = 6\,{\rm M}_\sun$ mid-MS SPB star.
The fixed values of the parameters ${\rm D}_{\rm env,min}$, $f_{\rm CBM}$ and ${\rm Z}_{\rm ini}$ are representative of the median values of these parameters in Table~2 of \citet{2022ApJ...930...94P}.

To compute the non-linear quadratic mode coupling coefficients~\eqref{eq:coupling+coefficient}, we need the adiabatic derivative of $\Gamma_1$ with respect to the stellar density, $\left(\pdv{\Gamma_1}{\ln \rho}\right)_S$.
We compute this quantity in the MESA models with a custom function defined in Appendix~\ref{app:ad+der}.

\begin{table}
	\centering
	\caption{Model parameters of the MESA model grid: initial mass ${\rm M}_{\rm ini}$, core hydrogen mass fraction ${\rm X}_{\rm c}$, initial metallicity ${\rm Z}_{\rm ini}$, exponential overshoot efficiency $f_{\rm CBM}$, and the logarithm of the minimal radiative envelope mixing level ${\rm D}_{\rm env, min}$.}
	\label{tab:mesa+parameters}
	\begin{tabular}{llllll} 
		\hline\hline\vspace{-9pt}\\
                            Model  & ${\rm M}_{\rm ini}$ (${\rm M}_\sun$)    & ${\rm X}_{\rm c}$ & ${\rm Z}_{\rm ini}$ & $f_{\rm CBM}$ & $\log(\tfrac{{\rm D}_{\rm env, min}}{{\rm cm}^2\,{\rm s}^{-1}})$\\[3pt]
		\hline\vspace{-7.5pt}\\
    fiducial                         & 4.0       & 0.09 & 0.014 & 0.02 & 2.00  \\
    $\Delta {\rm X_{\rm c,\,1}}$                         & 4.0       & 0.29 & 0.014 & 0.02 & 2.00    \\
    $\Delta {\rm X_{\rm c,\,2}}$                        & 4.0       & 0.59 & 0.014 & 0.02 & 2.00    \\
    $\Delta {\rm M_{\rm ini,\, 1}}$                        & 6.0       & 0.09 & 0.014 & 0.02 & 2.00    \\
    $\Delta {\rm X_{\rm c}}|\rm M_{\rm ini}$          & 6.0       & 0.29 & 0.014 & 0.02 & 2.00    \\
    $\Delta {\rm M_{\rm ini,\, 2}}$                       & 8.0       & 0.09 & 0.014 & 0.02 & 2.00    \\
 		\hline
	\end{tabular}
\end{table}

\subsection{GYRE model grid setup for SPB stars}\label{subsect:GYRE+grid+setup}

\begin{table*}
	\centering
	\caption{Radial order ranges $\left\{{\rm n}_u\right\}$ ($u \in \myintset{3}$) of the $(k,m) = (0,2)$ parent and $(k,m) = (0,1)$ daughter modes computed with GYRE and based on the MESA models listed in Table~\ref{tab:mesa+parameters}, angular rotation rates $\Omega$ expressed in percentages of the Roche critical rotation rate $\Omega_{\rm Roche}$ and in d$^{-1}$, model radius $R$, and the order of magnitude of the bounds for the ranges of $\vartheta_{2}$ and $\vartheta_{3}$ (denoted by $\vartheta_{2,3}$), $q_1$ and the quality factors $Q_\varphi$ ($\varphi \in \myintset{3}$).}
	\label{tab:radial+order+ranges}
	\begin{tabular}{lllllllll} 
		\hline\hline\vspace{-9pt}\\
                            Model  &  $\left\{{\rm n}_1\right\}$   & $\left\{{\rm n}_{2,3}\right\}$ & $\tfrac{\Omega}{\Omega_{\rm Roche}}$ ($\%$) & $\lceil{\vartheta_{2,3}}\rfloor$ &
                            $\lceil{q_1}\rfloor$ &
                            $\lceil{Q_\varphi}\rfloor$ & $\Omega$ (d$^{-1}$) & $R$ ($R_\sun$)
                            \\[3pt]
		\hline\vspace{-7.5pt}\\
    fiducial               & $\myintsetrange{70}{50}$       & $\myintsetrange{120}{70} \cup \myintsetrange{38}{10}$ & 20 & [$10^2$, $10^{-2}$] & [$10^{6}$, $10^{0}$] & [$10^7$, $10^2$] & $0.9995$& $5.1874$\\
    $\Delta {\rm X_{\rm c,\,1}}$                   & $\myintsetrange{40}{20}$       & $\myintsetrange{70}{40} \cup \myintsetrange{15}{3}$     & 20 & [$10^3$, $10^{-4}$] & [$10^7$, $10^{1}$] & [$10^8$, $10^2$] & $1.6006$& $3.7900$\\
    $\Delta {\rm X_{\rm c,\,2}}_{(a)}$                  & $\myintsetrange{20}{10}$       & $\myintsetrange{47}{17} \cup \myintsetrange{10}{3}$     & 20 & [$10^4$, $10^{-4}$] & [$10^7$, $10^{2}$] & [$10^8$, $10^2$] & $2.8419$& $2.5848$\\    
    $\Delta {\rm X_{\rm c,\,2}}_{(b)}$                  & $\myintsetrange{19}{10}$       & $\myintsetrange{48}{3}$     & 60 & [$10^4$, $10^{-4}$] & [$10^7$, $10^{1}$] & [$10^8$, $10^2$] & $8.5256$ & $2.5848$\\
    $\Delta {\rm M_{\rm ini,\, 1}}$                 & $\myintsetrange{55}{33}$       & $\myintsetrange{120}{45} \cup \myintsetrange{30}{5}$    & 20 & [$10^2$, $10^{-3}$] & [$10^6$, $10^{-1}$] & [$10^7$, $10^2$] & $0.8628$& $6.5483$\\
    $\Delta {\rm X_{\rm c}}|\rm M_{\rm ini}$                  & $\myintsetrange{26}{16}$       & $\myintsetrange{55}{5}$   & 20 & [$10^4$, $10^{-3}$] & [$10^8$, $10^{0}$] & [$10^7$, $10^2$] & $1.3904$& $4.7645$\\
    $\Delta {\rm M_{\rm ini,\, 2}}$               & $\myintsetrange{39}{30}$       & $\myintsetrange{85}{5}$    & 20 & [$10^3$, $10^{-2}$] & [$10^6$, $10^{0}$] & [$10^6$, $10^2$] & $0.7611$& $7.8322$ \\
 		\hline
	\end{tabular}
	\tablefoot{
We define the radial order ranges using $\myintsetrange{u}{h} \equiv \left\{\, x \in \mathbb{N}_0\,|\, -u \leq x \leq -h\,\right\}$, and we define the range of the dimensionless parameter $a$ as $\lceil a \rfloor \equiv \left[\mathcal{O}(\sup |a|)\,,\, \mathcal{O}(\inf |a|)\right]$.
Mode triads that consist of modes whose radial orders lie within the radial order ranges quoted in this table satisfy the necessary conditions for parametric resonance: $\gamma_1 > 0\,,\, \gamma_2 < 0$ and $\gamma_3 < 0$.
We consider the $\Delta {\rm X_{\rm c,\,2}}$ model at two rotation rates, labeled with the subscripts $(a)$ and $(b)$.
Gaps in the radial order ranges $\left\{{\rm n}_{2,3}\right\}$ indicate the presence of linearly excited $(k,m)=(0,1)$ modes.
Such linearly excited modes cannot act as potential daughter modes in parametric resonances with the $(k,m)=(0,2)$ parent modes and are therefore not considered in this work.
Note that the $\kappa$ mechanism excites a strongly different set of potential $(k,m)=(0,2)$ parent modes for each of the different models.}

\end{table*}

The most commonly observed $g$ modes in SPB stars are prograde $(k,m)=(0,1)$ and $(k,m)=(0,2)$ modes (see e.g., \citealt{2021NatAs...5..715P,2021MNRAS.503.5894S}).
For each of the computed MESA models, we therefore compute the GYRE linear adiabatic and non-adiabatic eigenfrequencies and eigenfunctions of these commonly detected $g$ modes, within the Cowling approximation \citep{1941MNRAS.101..367C} and the TAR.
We assume uniform rotation at $20\%$ of the Roche critical rotation rate $\Omega_{\rm Roche}$ for most models, except for the $\Delta {\rm X_{\rm c,\,2}}$ model, for which we consider rotation at rates equal to $20\%$ and (the excessively high) $60\%$ of $\Omega_{\rm Roche}$.
The positive (negative) linear driving (damping) rates of the modes are estimated as the imaginary part of the non-adiabatic linear eigenfrequencies computed by GYRE.

The radial orders of the $g$ modes computed for the different MESA stellar evolution models are listed in Table~\ref{tab:radial+order+ranges}.
These are similar to those of \citet{2021NatAs...5..715P}.
We ensure that the linear non-adiabatic GYRE computations -- which use trial frequencies based on the adiabatic GYRE computation results -- do not show evidence of missing non-adiabatic mode radial orders (see e.g., \citealt{2020ApJ...899..116G}), so that there is a one-to-one mapping between the adiabatic and non-adiabatic mode radial orders.
The ranges for the quality factors in Table~\ref{tab:radial+order+ranges} show that the condition $|\gamma_\varphi / \omega_\varphi| \equiv |1/Q_\varphi| \ll 1$ mentioned in Sect.~\ref{sect:AEs} is fulfilled for all modes (i.e., all modes are on the slow manifold).

\subsection{Search for isolated mode couplings: triad parameter ensembles for SPB stars}\label{subsect:parameter+ensembles}

Following the stability analysis in Sect.~\ref{sect:AE+stability}, only parametric resonance scenarios are able to produce stable stationary solutions for resonances $\Omega_1 \approx \Omega_2 + \Omega_3$ (or their difference frequency analogues).
We therefore restrict the GYRE computations of potential parent modes to those $(k,m) = (0,2)$ modes with linear driving rates $\gamma_1 > 0$, and restrict the computations of potential daughter modes to those $(k,m) = (0,1)$ modes with linear damping rates $\gamma_2 < 0$ and $\gamma_3 < 0$.
Gaps in the daughter mode radial order ranges $\{{\rm n}_{2,3}\}$ listed for some of the models in Table~\ref{tab:radial+order+ranges} indicate the presence of linearly excited $(k,m) = (0,1)$ modes that cannot partake in a parametric resonance.
For the $6$ M$_\sun$ near-ZAMS and $8$ M$_\sun$ mid-MS and near-ZAMS models with a rotation rate of $20$\% of $\Omega_{\rm Roche}$, we obtain no linearly excited potential parent modes.

We compute mode coupling coefficients $\eta_1$ for mode triads whose stationary AE solutions satisfy the (linear) stability conditions derived in Sect.~\ref{sect:AE+stability}.
To guarantee non-linear stability of the stationary solutions, these mode triads also need to fulfill the hyperbolicity check.
This hyperbolicity check can only be done after performing the non-linear mode coupling computations, because the computed stationary Jacobian matrix contains the coupling coefficient $|\eta_1|$ (see Eq.~(B.39) in \citealt{PhDThesisVanBeeck}).
The additional model validity constraints discussed in the next subsection are implemented to ensure that the solutions are away from the edge of the domains of applicability.

\subsubsection{Model validity criteria}\label{sect:model_validity_boundaries}

\citet{1997A&A...321..159B} state that resonant AEs are valid not only near a resonance center, but also far from resonance, with the solutions from resonant AEs slowly approaching the non-resonant AE solutions when moving away from that resonance center.
This point however only holds under the assumption that all modes participating in the modal coupling are linearly unstable, so that a stable multi-mode fixed point is reached with the same modes for resonant and non-resonant AEs \citep{1997A&A...321..159B}.
If that were not the case, the non-resonant solutions would predict negligibly small amplitudes for linearly damped modes.
For the cubic couplings considered by \citet{1997A&A...321..159B}, stable stationary points might exist for the driven resonance scenarios required for this type of behavior.
However, for the quadratic non-linear couplings considered in this work, stable multi-mode fixed points can only be found for parametric or direct resonant scenarios (with the latter only yielding stable stationary solutions for harmonic resonances; see Sect.~\ref{sect:AE+stability}).
Linearly damped modes participate in such coupling scenarios, invalidating the \citet{1997A&A...321..159B} assumption.

The resonant AEs derived in Sect.~\ref{sect:AEs} are therefore valid only near the resonance center, that is, if the absolute value of the linear triad frequency detuning $\Delta\Omega^l$ defined in Eq.~\eqref{eq:frequency+detuning}, is small in comparison to the absolute values of the individual co-rotating-frame mode frequencies: $\lvert\Delta\Omega^l\rvert \ll \lvert\Omega_\varphi\rvert \,\,\forall \varphi \in \myintset{3}$.
If it is not small, there are no comparatively slow amplitude variations and additional mode interactions need to be considered to describe the dynamics.
We therefore estimate the validity domain of the AEs in terms of the parameter:
\begin{equation}\label{eq:omega+AE}
    \Psi_{\rm AE} \equiv \dfrac{\left|\Delta\Omega^l\right|}{\left|\min\left(\Omega_1,\, \Omega_2,\, \Omega_3\right)\right|}\,,
\end{equation}
which compares $\left|\Delta\Omega^l\right|$ with the minimum absolute value of the co-rotating-frame angular mode frequencies of the triad.
We require that mode triads considered for modeling have $\Psi_{\rm AE} \leq 0.1$, so that $\left|\Delta\Omega^l\right| \ll \left|\min\left(\Omega_1,\, \Omega_2,\, \Omega_3\right)\right|$.
Many mode triads satisfy this model validity criterion, as is shown in Table~\ref{tab:selection+results}.

It is furthermore important to consider the threshold surface luminosity fluctuations $\lvert\mathfrak{L}_t\rvert$ of all possible mode triads, because these determine the onset of mode interaction.
To obtain isolated mode couplings, we determine the lowest and second-lowest $\lvert\mathfrak{L}_t\rvert$ for each of the possible mode triads in which a given parent mode participates.
Doing so allows us to label the individual modes of those triads with the computed value of $\lvert\mathfrak{L}_t\rvert$, which can be compared for each mode individually.
Identified isolated mode couplings must be (1) triads for which the lowest $\lvert\mathfrak{L}_t\rvert$ labels of all three modes (across all mode triads in which each mode is involved) are the same; and (2) triads for which $\lvert\mathfrak{L}_1^s\rvert$ of the lowest-$\lvert\mathfrak{L}_t\rvert$ mode triad is smaller than the second-lowest $\lvert\mathfrak{L}_t\rvert$ associated with the parent mode.
If at least one of those conditions is not fulfilled for a specific mode triad, the amplitude of at least one of the members of that mode triad is set by parametric energy transfer in a different mode triad.
The AEs derived in Sect.~\ref{sect:AEs} would then be invalid to describe the multi-mode coupling that occurs among such mode triads.
We collectively refer to these two conditions as the isolation criterion, and call mode triads that satisfy this criterion isolated mode triads.

\begin{figure}
	\centering
	\resizebox{\hsize}{!}{\includegraphics{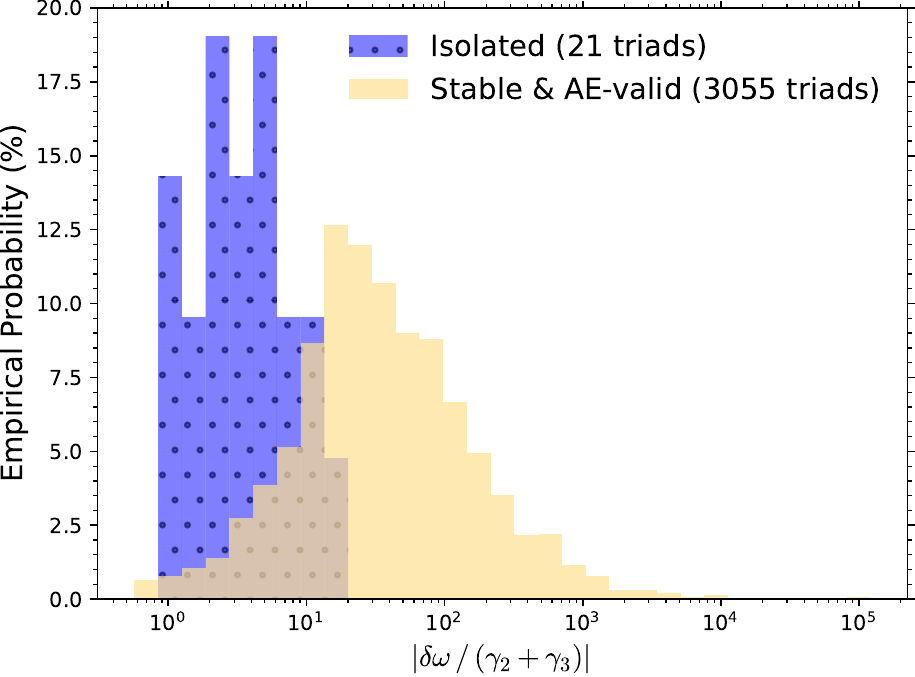}}
	\caption{Empirical probability distributions of $\lvert \delta\omega\,/\,(\gamma_2 + \gamma_3)\rvert$ for the $21$ identified (stable and AE$\,$-valid) isolated mode triads (blue and dotted) and their AE$\,$-valid and stable counterparts (yellow) that do not fulfill the isolation criterion. We choose the bin width in a similar way to what is done in \citet{2008+Hogg}.}
	\label{fig:arras+isolation+check}
\end{figure}

\citet{2003ApJ...591.1129A} have a different implicit `isolation criterion': they stated that if the resonance is sharp, it is plausible that only the parent and two daughter modes are relevant for the mode amplitude dynamics.
For sharp resonances, one expects $\lvert\delta\omega\,\displaystyle/\,\left(\gamma_2 + \gamma_3\right)\rvert$ to be small.
Figure~\ref{fig:arras+isolation+check} shows the empirical probability distribution of $\lvert\delta\omega\,\displaystyle/\,\left(\gamma_2 + \gamma_3\right)\rvert$ for the $21$ identified (stable, AE-valid) isolated mode triads, and their AE-valid and stable counterparts that do not fulfill the isolation criterion.
While it is certainly true that $\lvert\delta\omega\,\displaystyle/\,\left(\gamma_2 + \gamma_3\right)\rvert$ is small for the isolated mode triads, there are non-isolated mode triads that have similarly low values of $\lvert\delta\omega\,\displaystyle/\,\left(\gamma_2 + \gamma_3\right)\rvert$, which re-iterates the important role threshold surface luminosity fluctuations play in identifying such isolated mode triads.

Table~\ref{tab:selection+results} gives an overview of how restrictive the individual criteria are, and shows the effect of solely enforcing the validity or stability criteria.
Many stable mode triads are identified, and fixed point stability is essentially defined by the hyperbolicity check.
Both validity criteria are more restrictive than the stability criteria.
It does not seem obvious for a $g$ mode to participate in an isolated mode coupling scenario, as is reflected by the small number of identified isolated mode triads in Table~\ref{tab:selection+results}.
Moreover, strict enforcement of the TAR validity frequency hierarchies listed in for example \citet{2021A&A...652A.154D} would further reduce the number of mode triads considered for asteroseismic modeling.
This would however restrict the quadratic mode couplings identified for the $4$ M$_\sun$ near-ZAMS models from being used, contradicting observational evidence (e.g., \citetalias{2021A&A...655A..59V}).
We therefore do not strictly enforce these frequency hierarchies.

\begin{table*}
	\centering
	\caption{Overview of the number of mode triads that satisfy different model validity and stability criteria.}
	\label{tab:selection+results}
	\begin{tabular}{llllllllll} 
		\hline\hline\vspace{-9pt}\\
                            Model & Total &  $\gamma_\boxplus < 0$   & Quartic & Hyperbolic & AE & Is. & Stab. & Val. & Stab. \& Val. \\[3pt]
		\hline\vspace{-7.5pt}\\
    fiducial                   &   $68040$   &   $61040$   & $59491$ & $40517$ & $398$ & $6$ & $40517$ & $1$ &$1$ \\
    $\Delta {\rm X_{\rm c,\,1}}$                   &   $20790$   & $18920$       & $18883$ & $13173$ & $86$ & $8$  & $13173$ & $4$ & $4$  \\
    $\Delta {\rm X_{\rm c,\,2}}_{(a)}$               &    $8580$    & $8205$       & $8192$ & $5736$ & $121$ & $9$  & $5736$ & $3$ & $2$  \\
    $\Delta {\rm X_{\rm c,\,2}}_{(b)}$                &   $10810$    & $10008$      & $9975$ & $6940$ & $285$ & $12$ & $6940$ & $5$ & $4$   \\
    $\Delta {\rm M_{\rm ini,\, 1}}$                    &  $120819$  & $111568$      & $111124$ & $75289$ & $1906$ & $15$  & $75289$ & $6$ & $2$  \\
    $\Delta {\rm X_{\rm c}}|\rm M_{\rm ini}$        & $14586$ & $13691$    & $13565$ & $9421$ & $546$ & $17$  & $9421$ & $5$ & $3$  \\
    $\Delta {\rm M_{\rm ini,\, 2}}$                    &  $33210$ & $31568$       & $31014$ & $21062$ & $1235$ & $14$ & $21062$ & $5$ & $5$  \\
 		\hline
	\end{tabular}
 \tablefoot{
The different model stability criteria are denoted by the columns `$\gamma_\boxplus < 0$', `Quartic' and `Hyperbolic', which specify the number of modes that satisfy Eq.~\eqref{eq:first+hurwitz}, the quartic linear stability criterion \eqref{eq:quartic+stability+condition} and the hyperbolicity check (see first paragraph of Sect.~\ref{sect:AE+stability}), respectively.
How many of the mode triads fulfill all model stability criteria is given in the column `Stab.'.
Similarly, the different model validity criteria are denoted by the columns `AE' and `Is.', which specify the number of modes that satisfy the AE validity criterion ($\Psi_{\rm AE} \leq 0.1$, with $\Psi_{\rm AE}$ defined by Eq.~\eqref{eq:omega+AE}) and the isolation criterion.
The number of mode triads satisfying both validity criteria is given in the column `Val.', whereas the number of mode triads satisfying both the model stability and validity criteria is given in the column `Stab. \& Val.'.
The column `Total' contains the total number of mode triads considered for the different models listed in Table~\ref{tab:radial+order+ranges}.
}
\end{table*}

\subsubsection{Radial coupling contributions}\label{sect:radial+coupling+contributions}

For isolated mode triads that fulfill the stability and validity conditions, we compute a radial profile of the coupling coefficient $\eta_1(r)$ by integrating the overlap integrals in Eq.~\eqref{eq:coupling+coefficient} up to an internal radial coordinate $r$ instead of the stellar model surface radius $R$ (i.e., $r \leq R$ and $\eta_1 \equiv \eta_1(R)$).
The profile $\eta_1(r)$ then indicates the zones in the stellar interior for which contributions to the coupling coefficient are significant for these mode triads.

In the near-core regions the squared Brunt-V\"ais\"al\"a frequency ($N^2$) profile has a (local) maximum due to the presence of a $\mu$-gradient left behind by the receding convective core during the main sequence.
This maximum introduces a sharp transition in $N^2$ which changes the eigenfunctions: their modal inertia becomes more confined to the near-core region \citep{2008MNRAS.386.1487M,2021A&A...650A.175M}, at the location of the (local) maximum.
This can be rationalized by the increasing confinement of modal inertia to the TAR-valid near-core mixing zone with model age.
The width of that maximum depends on the evolutionary stage: more evolved SPB models have wider $N^2$ maxima, and some modes become trapped in the near-core regions \citep{2016ApJ...823..130M,2021A&A...650A.175M}.
For these more evolved models, the instability strip for $(k,m) = (0,2)$ parent $g$ modes also moves towards higher radial orders, further confining modal inertia to the oscillatory near-core regions.
Both these effects likely explain why the contributions to coupling coefficients are more concentrated in the near-core regions for more evolved models.

\begin{figure}
  \resizebox{\hsize}{!}{\includegraphics{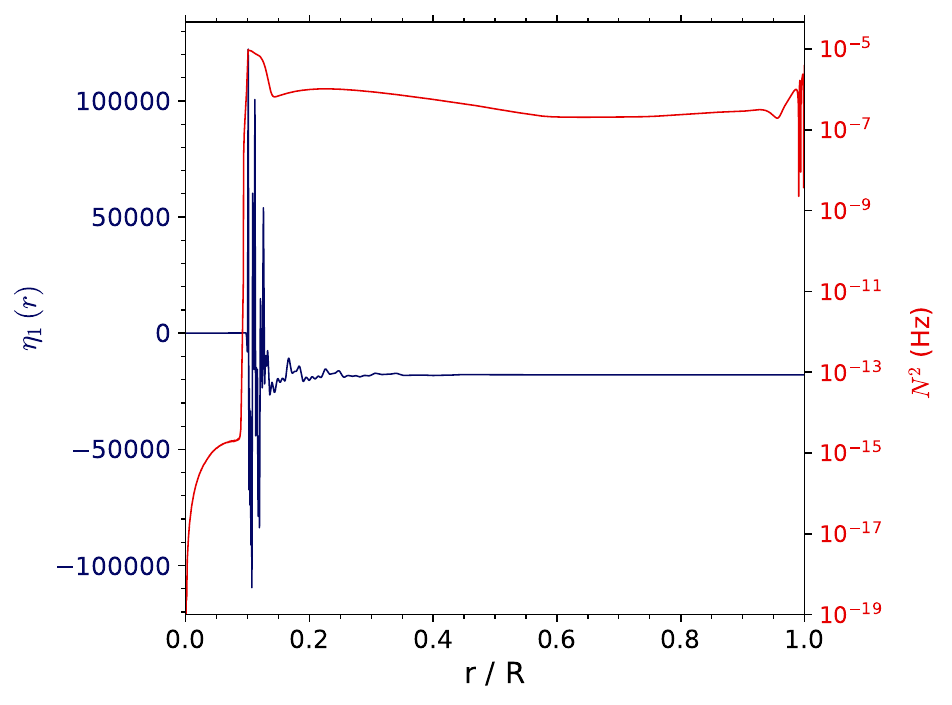}}
  \caption{Coupling coefficient and squared Brunt-V\"ais\"al\"a frequency ($N^2$) profiles as a function of the fractional radius for the isolated mode triad with $g$ mode radial orders $\left(n_1,\,n_2,\,n_3\right) = \left(-22,\,-15,\,-53\right)$ in the mid-MS $\Delta {\rm X_{\rm c,\,1}}$ model of Table~\ref{tab:isolated+mode+triad+properties+1}.}
  \label{fig:coupling+coefficient+profile+example+good}
\end{figure}

\begin{figure}
  \resizebox{\hsize}{!}{\includegraphics{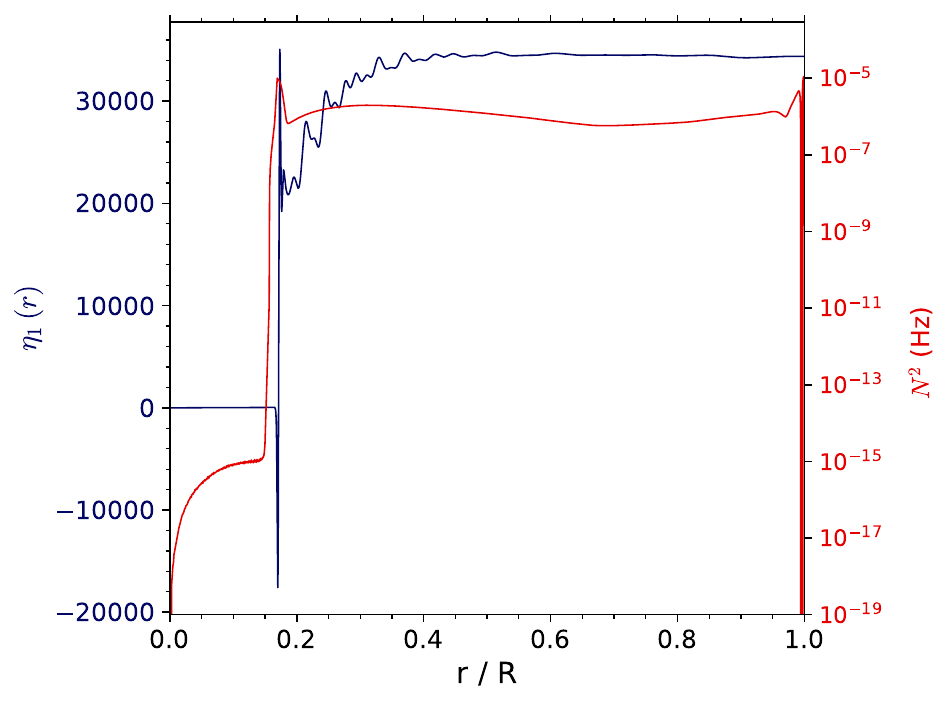}}
  \caption{Same as Fig.~\ref{fig:coupling+coefficient+profile+example+good}, but for the isolated mode triad with $g$ mode radial orders $\left(n_1,\,n_2,\,n_3\right) = \left(-14,\,-26,\,-10\right)$ in the $\Delta {\rm X_{\rm c,\,2}}_{(a)}$ model of Table~\ref{tab:isolated+mode+triad+properties+1}.}
  \label{fig:coupling+coefficient+profile+example+bad}
\end{figure}

Examples of such radial coupling coefficient profiles are given in Figs.~\ref{fig:coupling+coefficient+profile+example+good} and \ref{fig:coupling+coefficient+profile+example+bad}, which display the $N^2$ profile along with the coupling coefficient profile $\eta_1(r)$.
The profile in Fig.~\ref{fig:coupling+coefficient+profile+example+good} describes the coupling contributions for the mode triad characterized by $g$ mode radial orders $\left(n_1,\,n_2,\,n_3\right) = \left(-22,\,-15,\,-53\right)$ in the mid-MS $\Delta {\rm X_{\rm c,\,1}}$ model in Table~\ref{tab:isolated+mode+triad+properties+1}.
The near-ZAMS example displayed in Fig.~\ref{fig:coupling+coefficient+profile+example+bad} describes the coupling contributions for the mode triad characterized by $g$ mode radial orders $\left(n_1,\,n_2,\,n_3\right) = \left(-14,\,-10,\,-26\right)$ in the $\Delta {\rm X_{\rm c,\,2}}_{(a)}$ model of Table~\ref{tab:isolated+mode+triad+properties+1}.
These figures confirm that contributions to the coupling coefficient of the mode coupling in the mid-MS model are typically more confined to the near-core region in comparison to those contributions in the near-ZAMS model.

\subsubsection{Properties of isolated $g$ mode triads in SPB stars}

\begin{table*}
	\centering
	\caption{
 Linear co-rotating-frame frequencies $\Omega_\varphi$, largest absolute spin parameters $\vert s_\varphi\rvert$ (of the triad modes), radial orders $n_\varphi$ and linear driving or linear damping rates $\gamma_\varphi$ of $g$ modes $\varphi$ in all identified isolated ($g$) mode triads with ordering numbers $\left(k_1,\,k_2,\,k_3\right) = \left(0,\,0,\,0\right)$ and azimuthal orders $\left(m_1,\,m_2,\,m_3\right) = \left(2,\,1,\,1\right)$ for the models in Table~\ref{tab:radial+order+ranges} that yield stable and valid stationary solutions.
 The effective triad damping $\gamma_\boxplus$ and the triad detuning $\delta\omega$ are also listed. }
	\label{tab:isolated+mode+triad+properties+1}
	\begin{tabular}{lllllll} 
	\hline\hline\vspace{-9pt}\\
    Model  &  $\triadquant{n_1}{n_2}{n_3}$ & $\delta\omega$ (d$^{-1}$) & $\gamma_\boxplus$ ($10^{-3}$ d$^{-1}$) & $\triadquant{\Omega_1}{\Omega_2}{\Omega_3}$ (d$^{-1}$) & $\triadquant{\gamma_1}{\gamma_2}{\gamma_3}$ ($10^{-5}$ d$^{-1}$) & $\lvert s\rvert_{\rm max}$ \\[3pt]
	\hline\vspace{-7.5pt}\\
    fiducial       &    $\triadquantrad{51}{38}{90}$        &   $0.00145$    &   $-0.32249$   & $\triadquant{3.23}{2.32}{0.91}$ &  $\triadquant{0.39}{-0.30}{-32.3}$   & $2.20$   \\
    $\Delta {\rm X_{\rm c,\,1}}$       &    $\triadquantrad{20}{13}{58}$        &   $-0.00197$    &   $-2.0248$   & $\triadquant{6.92}{5.74}{1.18}$ &  $\triadquant{0.18}{-0.16}{-202.5}$   & $2.71$   \\
    &    $\triadquantrad{21}{14}{54}$        &   $-0.00528$    &   $-1.657$   & $\triadquant{6.60}{5.33}{1.28}$ &  $\triadquant{0.57}{-0.17}{-166.1}$   & $2.51$   \\
    &    $\triadquantrad{22}{15}{53}$        &   $0.00370$    &   $-1.536$  & $\triadquant{6.29}{4.99}{1.30}$ &  $\triadquant{1.06}{-0.19}{-154.4}$  & $2.46$   \\
    &    $\triadquantrad{39}{40}{40}$        &   $0.00303$    &   $-0.350$  & $\triadquant{3.53}{1.76}{1.76}$ &  $\triadquant{11.2}{-23.1}{-23.1}$  & $1.81$   \\
    $\Delta {\rm X_{\rm c,\,2}}_{(a)}$     &        $\triadquantrad{10}{7}{22}$      &    $-0.01510$   &    $-1.025$  & $\triadquant{12.0}{9.16}{2.88}$ &  $\triadquant{0.06}{-0.08}{-102.4}$ & $1.98$ \\    
    &        $\triadquantrad{14}{10}{26}$      &   $0.03121$    &    $-3.252$  & $\triadquant{8.85}{6.41}{2.41}$ &  $\triadquant{4.19}{-0.08}{-329.3}$ & $2.36$ \\ 
    $\Delta {\rm X_{\rm c,\,2}}_{(b)}$      &      $\triadquantrad{10}{6}{36}$      &    $-0.01614$    &   $-7.765$   & $\triadquant{11.26}{9.65}{1.63}$ & $\triadquant{0.07}{-0.05}{-776.5}$ & $10.5$ \\
    &      $\triadquantrad{12}{8}{23}$      &    $-0.00844$    &   $-1.565$   & $\triadquant{9.62}{7.08}{2.54}$ & $\triadquant{0.94}{-0.12}{-157.3}$ & $6.71$ \\
    &      $\triadquantrad{17}{12}{30}$      &    $-0.00723$    &   $-5.144$   & $\triadquant{6.80}{4.86}{1.95}$ & $\triadquant{8.45}{-0.08}{-522.8}$ & $8.74$ \\
    &      $\triadquantrad{18}{13}{31}$      &    $0.03334$    &   $-5.825$   & $\triadquant{6.42}{4.49}{1.89}$ & $\triadquant{7.03}{-0.19}{-589.3}$ & $9.02$ \\
    $\Delta {\rm M_{\rm ini,\, 1}}$      &  $\triadquantrad{50}{48}{56}$       &    $-0.00083$   &  $-0.507$ & $\triadquant{2.37}{1.29}{1.09}$ & $\triadquant{15.5}{-13.3}{-52.9}$ & $1.59$  \\
    &   $\triadquantrad{40}{30}{70}$       &    $0.00836$   &  $-1.429$  & $\triadquant{2.98}{2.12}{0.85}$ & $\triadquant{6.65}{-0.62}{-148.9}$ & $2.03$ \\
    $\Delta {\rm X_{\rm c}}|\rm M_{\rm ini}$         &      $\triadquantrad{16}{14}{22}$     &    $0.00130$  &  $-0.120$  & $\triadquant{6.19}{3.81}{2.38}$ &  $\triadquant{0.01}{-0.56}{-11.4}$ & $1.17$ \\
    &     $\triadquantrad{17}{13}{29}$     &    $0.00830$  &  $-0.915$  & $\triadquant{5.88}{4.08}{1.79}$ &  $\triadquant{0.16}{-0.57}{-91.1}$ & $1.56$ \\
    &     $\triadquantrad{23}{17}{41}$     &   $0.01453$   &  $-4.495$  & $\triadquant{4.37}{3.12}{1.23}$ &  $\triadquant{5.33}{-1.15}{-453.6}$ & $2.26$ \\
    $\Delta {\rm M_{\rm ini,\, 2}}$      &   $\triadquantrad{31}{23}{56}$     &   $0.00283$     & $-2.018$  & $\triadquant{3.03}{2.18}{0.85}$ &  $\triadquant{2.30}{-2.14}{-202.0}$ & $1.80$ \\
    &    $\triadquantrad{32}{24}{58}$     &    $0.00631$    &  $-2.176$ & $\triadquant{2.92}{2.10}{0.82}$ &  $\triadquant{2.43}{-2.42}{-217.6}$ & $1.86$ \\
    &    $\triadquantrad{35}{26}{62}$     &    $-0.00274$    &  $-2.915$ & $\triadquant{2.68}{1.93}{0.76}$ &  $\triadquant{7.54}{-3.13}{-295.9}$ & $2.01$ \\
    &   $\triadquantrad{36}{27}{63}$     &     $0.00546$   &  $-2.734$  & $\triadquant{2.61}{1.86}{0.75}$ &  $\triadquant{2.02}{-3.20}{-272.2}$ & $2.04$ \\
    &    $\triadquantrad{38}{28}{67}$     &    $-0.00575$    &  $-2.724$  & $\triadquant{2.47}{1.78}{0.70}$ &  $\triadquant{0.24}{-4.27}{-268.3}$ & $2.18$ \\
 		\hline
	\end{tabular}
	\tablefoot{
The mode triad detuning $\delta\omega$ is equal to $\Delta\Omega^l$ because we set $\mathfrak{J} = 1$ in Eq.~(\ref{eq:frequency+detuning}). 
We compute the largest absolute spin parameter as $\lvert s\rvert_{\rm max} = \max\left(\lvert s_\varphi\rvert\right) \equiv \max\left(\lvert 2\,\Omega / \Omega_\varphi\rvert\right)\,\hspace{0.1cm}\forall\varphi\in\myintset{3}$.
}
\end{table*}

\begin{table*}
	\centering
	\caption{
 Dimensionless quantities for the mode triads listed in Table~\ref{tab:isolated+mode+triad+properties+1}: dimensionless AE validity parameter $\Psi_{\rm AE}$, absolute detuning-damping ratio $\lvert q\rvert$ and absolute detuning-driving ratio $\lvert q_1\rvert$, the driving-damping rate ratios $\vartheta_2$ and $\vartheta_3$, as well as the order of magnitude of the energy-normalized coupling coefficient, $\mathcal{O}_{\lvert\eta_1\rvert}$. 
 These mode triads are identified by their $g$ mode radial orders $n_\varphi$.
 }
	\label{tab:isolated+mode+triad+properties+2}
	\begin{tabular}{lllllllllll} 
	\hline\hline\vspace{-9pt}\\
    Model  &  $(n_1,\,n_2,\,n_3)$ & $\Psi_{\rm AE}$ &  $\lvert q\rvert$ & $\lvert q_1\rvert$ & $\vartheta_2$ & $\vartheta_3$ & $\mathcal{O}_{\lvert\eta_1\rvert}$\\[3pt]
	\hline\vspace{-7.5pt}\\
    fiducial       &    $\triadquantrad{51}{38}{90}$    &    $0.00160$   & $4.51022$ &  $373.680$   & $0.76947$ & $83.0825$ & $10^6$ \\
    $\Delta {\rm X_{\rm c,\,1}}$       &    $\triadquantrad{20}{13}{58}$    &    $0.00167$   & $0.97315$ &  $1122.39$   & $0.92560$ & $1153.43$ & $10^3$ \\
    &    $\triadquantrad{21}{14}{54}$    &    $0.00414$   & $3.18616$ &  $919.730$   & $0.30289$ & $289.361$ & $10^4$ \\
    &    $\triadquantrad{22}{15}{53}$    &    $0.00284$  &    $2.41124$  &  $350.601$  & $0.17994$ & $146.223$  & $10^4$ \\
    &    $\triadquantrad{39}{40}{40}$    &    $0.00172$  &    $8.64606$  &  $27.0505$  & $2.06433$ & $2.06433$  & $10^4$ \\
    $\Delta {\rm X_{\rm c,\,2}}_{(a)}$     &       $\triadquantrad{10}{7}{22}$     &   $0.00525$    & $14.7345$ &  $23416.0$   & $1.27865$ & $1588.92$ & $10^4$ \\    
    &        $\triadquantrad{14}{10}{26}$        &   $0.01297$  &   $9.59642$   & $744.512$  & $0.01816$ & $78.5641$ & $10^4$ \\ 
    $\Delta {\rm X_{\rm c,\,2}}_{(b)}$    &     $\triadquantrad{10}{6}{36}$   &   $0.00993$     & $2.07856$ &   $24822.5$    & $0.76510$ & $11942.4$ & $10^3$ \\  
    &     $\triadquantrad{12}{8}{23}$   &   $0.00332$     & $5.39600$ &   $897.596$    & $0.13080$ & $167.214$ & $10^4$ \\
    &     $\triadquantrad{17}{12}{30}$   &   $0.00371$     & $1.40528$ &   $85.5389$    & $0.00907$ & $61.8605$ & $10^4$ \\
    &     $\triadquantrad{18}{13}{31}$   &   $0.01764$     & $5.72364$ &   $474.296$    & $0.02771$ & $83.8384$ & $10^4$ \\
    $\Delta {\rm M_{\rm ini,\, 1}}$      &  $\triadquantrad{50}{48}{56}$    &   $0.00076$    & $1.63256$ &  $5.34811$   & $0.85748$ & $3.41842$ & $10^4$ \\
    &   $\triadquantrad{40}{30}{70}$        &   $0.00983$  & $5.85327$  &  $125.797$  & $0.09346$ & $22.3982$  & $10^6$ \\
    $\Delta {\rm X_{\rm c}}|\rm M_{\rm ini}$         &  $\triadquantrad{16}{14}{22}$     &  $0.00055$   & $10.8770$  & $8987.04$  & $38.7770$ & $788.467$ & $10^4$ \\
    &    $\triadquantrad{17}{13}{29}$      & $0.00464$   &  $9.06636$ &  $5197.85$  & $3.58916$ & $570.722$ & $10^4$  \\
    &   $\triadquantrad{23}{17}{41}$     &   $0.01180$  & $3.23295$ &  $272.498$   & $0.21492$ & $85.0729$ & $10^5$ \\
    $\Delta {\rm M_{\rm ini,\, 2}}$      &   $\triadquantrad{31}{23}{56}$    &   $0.00334$      & $1.40262$ & $122.926$ & $0.92741$ & $87.7126$ & $10^6$ \\
    &    $\triadquantrad{32}{24}{58}$      &   $0.00772$   & $2.89766$  &  $259.527$  & $0.99670$ & $89.5678$ & $10^6$ \\
    &    $\triadquantrad{35}{26}{62}$      &   $0.00362$   & $0.94161$  &  $36.4148$  & $0.41545$ & $39.2574$ & $10^6$ \\
    &    $\triadquantrad{36}{27}{63}$      &   $0.00733$  & $1.99803$ & $270.435$  & $1.58532$ & $134.765$ & $10^6$ \\
    &    $\triadquantrad{38}{28}{67}$     &  $0.00825$   &  $2.11147$ & $2423.35$  & $18.0122$ & $1130.69$ & $10^6$ \\
 		\hline
	\end{tabular}
\end{table*}

Isolated mode triads can be identified for each model listed in Table~\ref{tab:radial+order+ranges}.
Their characteristic mode and triad properties are listed in Table~\ref{tab:isolated+mode+triad+properties+1}, and their dimensionless model validity and stability estimators, including the order of magnitude of the coupling coefficient $\lvert \eta_1\rvert$, are displayed in Table~\ref{tab:isolated+mode+triad+properties+2}.
It is clear from the largest absolute values of the spin parameters that at least one of the modes in the identified isolated mode triads is sub-inertial (i.e., $\lvert s_\varphi\rvert > 1$), requiring the non-perturbative (TAR) description of the Coriolis force.

The Doppler shift causes a logical trend to appear when correlating the average values of $\Omega_\varphi$ with a change in the rotation frequency (i.e., upon comparison of mode frequencies for the models $\Delta {\rm X_{\rm c,\,2}}_{(a)}$ and $\Delta {\rm X_{\rm c,\,2}}_{(b)}$): $\Omega_\varphi$ decreases for increasing rotation rates.
This trend is only weakly influenced by the effect of the Coriolis force for these sectoral prograde $g$ or equatorial Kelvin waves, which is not the case for non-sectoral modes (due to geostrophic balance; see e.g., \citealt{1982GillBook,2005MNRAS.360..465T,2015MNRAS.446.1438D,2017MNRAS.469...13S}).

The co-rotating-frame mode frequencies also (on average) decrease for higher-mass models, and decrease with model age.
This change with mass can be attributed to the more massive SPB models having a larger model radius, which enlarges the pulsation cavity, although it is to a degree counteracted by the decrease in $\Omega$.

Mode frequency in the co-rotating frame is instrumental for determining linear, heat-driven mode excitation: it determines the geometrically optimal driving region for a mode, which needs to coincide with the Z bump for the mode to be linearly excited in an SPB star.
In general, higher radial order $g$ modes\footnote{A $g$ mode $\alpha$ is considered to have a higher radial order than $g$ mode $\beta$ when $n_\alpha < n_\beta$ (i.e., when the absolute value of its radial order is larger).} -- which have lower frequencies and are therefore more susceptible to rotational effects -- are excited in more evolved models (on the main sequence).
The expected higher mode densities for more evolved stars might explain why only one isolated mode triad was identified for the fiducial model.
The linear (vibrational) stability sensitively depends on the opacity profile of any particular stellar model (e.g., \citealt{2017MNRAS.466.2284D}).
A change in opacity profile can therefore lead to the identification of different isolated mode triads.

Stabilization of the $(k,m)=(0,1)$ $g$ modes in more rapidly rotating stars (e.g., \citealt{2005MNRAS.360..465T}) can lead to opportunities for non-linear excitation by a parametric resonance, as it enlarges the pool of potential daughter modes with which combination can be made.
A physical explanation of the effect of rotation within the TAR on mode excitation by the $\kappa$ mechanism can be found in \citet{1998ApJ...497L.101U} and \citet{2005MNRAS.360..465T}. 
That increasing potential for mode combination with larger rotation rate can be observed in the `Total' column of Table~\ref{tab:selection+results}, where more combinations are considered for the more rapidly rotating model $\Delta {\rm X_{\rm c,\,2}}_{(b)}$, if compared to the number of combinations considered for model $\Delta {\rm X_{\rm c,\,2}}_{(a)}$.
The stabilizing effect for the parent $(k,m)=(0,2)$ $g$ modes comparatively seems weaker (as was also observed by \citealt{2005MNRAS.360..465T}): only one of the parent modes is stabilized, in comparison to seven of the daughter modes, when we compare the $\Delta {\rm X_{\rm c,\,2}}_{(a)}$ and $\Delta {\rm X_{\rm c,\,2}}_{(b)}$ models.
In this case, we identify four isolated triads for model $\Delta {\rm X_{\rm c,\,2}}_{(b)}$, and two isolated triads for the slower-rotating model $\Delta {\rm X_{\rm c,\,2}}_{(a)}$.
However, the stabilizing effect does not necessarily mean that we can identify more isolated mode triads for rapid rotators, because those triads need to fulfill the stability and validity criteria.

The range of $\lvert q_1\rvert$ for the identified isolated mode triads is $\left[10^4,\,10^0\right]$ (using the notation of Table~\ref{tab:radial+order+ranges}).
This indicates that the two rightmost panels on the top of Fig.~\ref{fig:stability+domains}, as well as the panels on the bottom row of Fig.~\ref{fig:stability+domains}, describe the linear stability of the isolated stationary solutions.
These solutions' resonant nature is important, as is indicated by the values of $\lvert q\rvert$ being of the order of unity to ten, affecting the stationary amplitude~\eqref{eq:stat+amps+q}.

We can identify only one isolated harmonic resonant triad (for the $\Delta {\rm X_{\rm c,\,1}}$ model), which is a small number in comparison to the $20$ identified resonances of the combination type $\Omega_1 \approx \Omega_2 + \Omega_3$.
It therefore seems comparatively harder to satisfy all stability and validity conditions for harmonic resonances, which provides a reason for why only a few of such resonances are identified in the observational data of \citetalias{2021A&A...655A..59V}.

We only consider combinations of a small number of modes in this proof-of-concept study and do not study the dependence of non-linear parameters -- such as the coupling coefficient -- on stellar parameters in detail.

\section{Impact on asteroseismic modeling}\label{sect:asteroseismic+modeling+impact}

\begin{table*}
	\centering
	\caption{
 Maximal absolute non-linear frequency shifts as a fraction of the linear inertial-frame angular frequency $\lvert\delta\Omega^s_{\varphi,{\rm m}} \,\displaystyle/\, \Omega_{\varphi,\,i}\rvert$ (in parts-per-thousand), zero-point-corrected non-linear combination phases $\Tilde{\Phi}^s_{s,0}$, and expected stationary surface luminosity fluctuation ratios $\absfraktxt{\mathfrak{L}^s_2}{\mathfrak{L}^s_1}$ and $\absfraktxt{\mathfrak{L}^s_3}{\mathfrak{L}^s_1}$ for the mode triads listed in Table~\ref{tab:isolated+mode+triad+properties+1}.
 These mode triads are identified by their $g$ mode radial orders $n_\varphi$.
 }
	\label{tab:isolated+triad+observables}
	\begin{tabular}{llllll} 
	\hline\hline\vspace{-9pt}\\
    Model  &  $(n_1,\,n_2,\,n_3)$ & $\left|\tfrac{\delta\Omega^s_{\varphi,{\rm m}}}{\Omega_{\varphi,\,i}}\right|$ (ppt) & $\Tilde{\Phi}^s_{s,0}$ (rad) & $\absfraktxt{\mathfrak{L}^s_2}{\mathfrak{L}^s_1}$ & $\absfraktxt{\mathfrak{L}^s_3}{\mathfrak{L}^s_1}$\\[3pt]
	\hline\vspace{-7.5pt}\\
    fiducial     &    $\triadquantrad{51}{38}{90}$  &  $0.764733$  &   $0.21819$  &  $0.55605$   & $0.06168$  \\
    $\Delta {\rm X_{\rm c,\,1}}$       &    $\triadquantrad{20}{13}{58}$   &  $0.708263$ &  $-0.79901$   &    $0.42858$   & $0.03261$  \\
    &    $\triadquantrad{21}{14}{54}$   &  $1.839526$ &  $-0.30412$   &    $0.65585$   & $0.06390$  \\
    &    $\triadquantrad{22}{15}{53}$       & $1.283029$ & $0.39313$ &    $0.91831$  &    $0.08677$  \\
    &    $\triadquantrad{39}{40}{40}$       & $0.593527$ & $0.11515$ &    $0.48298$  &    $0.48298$  \\
    $\Delta {\rm X_{\rm c,\,2}}_{(a)}$     &       $\triadquantrad{10}{7}{22}$   &  $2.639689$ & $-0.06776$ &   $0.32000$    & $0.05700$  \\    
    &        $\triadquantrad{14}{10}{26}$      & $6.021050$ & $0.10383$ &   $2.53183$  &   $0.14379$     \\ 
    $\Delta {\rm X_{\rm c,\,2}}_{(b)}$      &      $\triadquantrad{10}{6}{36}$  &  $1.589962$ &  $-0.44842$  &   $0.32190$     & $0.00750$   \\
    &      $\triadquantrad{12}{8}{23}$  &  $0.766883$ &  $-0.18324$  &   $0.84057$     & $0.08189$   \\
    &      $\triadquantrad{17}{12}{30}$  &  $0.701324$ &  $-0.61847$  &   $3.08804$     & $0.06054$   \\
    &      $\triadquantrad{18}{13}{31}$  &  $3.238349$ &  $0.17297$  &   $2.02044$     & $0.04897$   \\
    $\Delta {\rm M_{\rm ini,\, 1}}$      &   $\triadquantrad{50}{48}{56}$   &  $0.443377$  &  $-0.54958$  &   $0.79629$    & $0.41179$   \\
    &      $\triadquantrad{40}{30}{70}$       & $5.086921$ & $0.16921$ &   $2.05217$  & $0.20288$  \\
    $\Delta {\rm X_{\rm c}}|\rm M_{\rm ini}$         &    $\triadquantrad{16}{14}{22}$    & $0.330309$ & $0.09168$ &  $0.07627$   & $0.05190$   \\
    &    $\triadquantrad{17}{13}{29}$     & $2.600074$ & $0.10985$ & $0.29260$   &  $0.08237$ \\
    &    $\triadquantrad{23}{17}{41}$     & $5.594778$ & $0.29998$ & $0.72069$   &  $0.08163$ \\
    $\Delta {\rm M_{\rm ini,\, 2}}$      &   $\triadquantrad{31}{23}{56}$   & $1.760714$ & $0.61936$ &   $0.34623$     & $0.07864$  \\
    &    $\triadquantrad{32}{24}{58}$     & $3.997386$ & $0.33231$ &   $0.48095$   & $0.07250$  \\
    &    $\triadquantrad{35}{26}{62}$     & $1.832881$ & $-0.81546$ &   $0.61223$   & $0.07525$  \\
    &    $\triadquantrad{36}{27}{63}$     & $3.610779$ & $0.46404$ &   $0.30289$   & $0.04729$  \\
    &    $\triadquantrad{38}{28}{67}$     & $3.885665$ & $-0.44231$ &   $0.12459$   & $0.01283$  \\
 		\hline
	\end{tabular}
	\tablefoot{
We define $\Tilde{\Phi}^s_{s,0}$ as $\Tilde{\Phi}^s_{s} + \left(\Delta\phi^s\right)_0$.
The observables $\absfraktxt{\mathfrak{L}^s_2}{\mathfrak{L}^s_1}$ and $\absfraktxt{\mathfrak{L}^s_3}{\mathfrak{L}^s_1}$, as well as $\Tilde{\Phi}^s_s$, are defined in Eqs.~\eqref{eq:stat+surf+amp+ratios} and \eqref{eq:observable+combination+phase}, respectively. 
}
\end{table*}

In this section, we discuss how the non-linear theoretically predicted observables defined in Sect.~\ref{sect:theoretical+observables} can aid or complement current frequency-based asteroseismic modeling of SPB stars.
The various non-linear observables of interest, or derived quantities thereof, are listed in Table~\ref{tab:isolated+triad+observables}.

\subsection{Oscillation frequency \& combination phase observables}\label{subsect:asteroseismic+modeling+impact+frequencies+phases}

We find that the non-linear frequency shifts of the isolated mode triads are small in comparison to the (linear) inertial mode frequencies, as is illustrated by the values in the column displaying values of $\lvert\delta\Omega^s_{\varphi,{\rm m}} \,\displaystyle/\, \Omega_{\varphi,\,i}\rvert$ in Table~\ref{tab:isolated+triad+observables}, which shows the largest value of the frequency shift for any of the three modes in the isolated triad, expressed in units of the inertial mode frequency.
These dimensionless frequency shifts are typically of order $10^{-3}$ to $10^{-4}$, and we find no clear correlations with any of the model parameters (potentially due to the small number of modes considered).

Translated to units of d$^{-1}$, this means that there are several frequencies whose (non-linear) frequency shifts are of order $10^{-5}$ d$^{-1}$ or smaller, rendering them of similar order as the typical errors for the frequencies derived from $4$-year {\em Kepler} light curves (e.g., \citealt{2021A&A...656A.158B}).
The largest shifts are obtained for the modes in the faster-rotating models, and are of order $10^{-2}$ or $10^{-3}$ d$^{-1}$, which is approximately two orders of magnitude larger than typical uncertainties.
In practice, the resonant non-linear frequency shifts are hard to distill from frequency lists deduced from prewhitening analyses similar to what was done in \citetalias{2021A&A...655A..59V}.
The non-resonant frequency shifts are even smaller, justifying the approximation of using linear frequencies in asteroseismic modeling.

We show the zero-point-corrected non-linear combination phases $\Tilde{\Phi}^s_{s,0}$ for all identified isolated mode triads in Table~\ref{tab:isolated+triad+observables}, where $\Tilde{\Phi}^s_{s,0} = \Tilde{\Phi}^s_s + \left(\Delta\phi^s\right)_0$.
We find no specific trends for these combination phases when correlated with other model parameters (potentially due to the small number of modes considered).
In principle, such phases can be compared with observed relative phases, if the relative phase at the zero point is accounted for.
The latter is however dependent on the initial values of the individual mode phases, $\vec{\phi}_0$, which are unconstrained.

\subsection{Amplitude ratio observables}\label{subsect:asteroseismic+modeling+impact+amplitude+ratios}

The stationary surface luminosity fluctuation ratios $\absfraktxt{\mathfrak{L}^s_2}{\mathfrak{L}^s_1}$ and $\absfraktxt{\mathfrak{L}^s_3}{\mathfrak{L}^s_1}$ offer the best constraints for asteroseismic modeling.
We find that most of these ratios are smaller than unity, indicating that most of the stationary-state (or saturation) energies of the linearly driven $(k,m) = (0,2)$ parent modes are larger than those of the $(k,m) = (0,1)$ daughter modes that are linearly damped and parametrically excited.
Daughter modes have larger stationary mode amplitude ratios $\absfraktxt{\mathfrak{L}^s_2}{\mathfrak{L}^s_1}$ only for certain mode triads in the $4$ M$_\sun$ near-ZAMS models $\Delta {\rm X_{\rm c,\,2}}_{(a)}$ and $\Delta {\rm X_{\rm c,\,2}}_{(b)}$, as well as the $6$ M$_\sun$ near-TAMS models $\Delta M_{\rm ini,\,1}$.
For these specific mode triads, the other daughter mode's amplitude is significantly smaller than that of the parent mode.

The stationary amplitude ratios~\eqref{eq:stat+surf+amp+ratios} depend on the ratios of the mode quality factors that express the number of periods necessary to linearly increase or decrease modal energy by a factor of $e$.
Such ratios are inversely proportional to $\vartheta_{2}$ or $\vartheta_3$ and proportional to the frequency ratio $\Omega_{2}\,\displaystyle/\,\Omega_1$ or $\Omega_{3}\,\displaystyle/\,\Omega_1$.
We find that the frequency ratios for these specific isolated mode triads that have larger-amplitude daughter modes are not particularly different from those in other identified isolated mode triads, as can be rationalized from the co-rotating mode frequency values in Table~\ref{tab:isolated+mode+triad+properties+1}.
The observed difference in amplitude ratio is thus is mostly due to different values of $\vartheta_2$.
The smaller $\vartheta_2$, the longer the linear damping timescale of daughter mode $2$ is, in comparison to the linear driving timescale of parent mode $1$.
Based on the listed values of $\vartheta_2$ in Table~\ref{tab:isolated+mode+triad+properties+2}, this effect seems to become important when $\vartheta_2 \lesssim 0.13 \equiv \vartheta_{2,{\rm b}}$, that is, when the linear damping rate of one of the daughter modes is less than $13$\% of the linear driving rate of the parent mode.
We also estimate the value of $\vartheta_2$ at which this effect becomes important by performing a linear regression between the $\vartheta_2$ and $\absfraktxt{\mathfrak{L}^s_2}{\mathfrak{L}^s_1}$ values of four isolated mode triads in Table~\ref{tab:isolated+mode+triad+properties+2}.
Specifically, we construct a regression model using the $\vartheta_2$ and $\absfraktxt{\mathfrak{L}^s_2}{\mathfrak{L}^s_1}$ values obtained from the two mode triads with parent-daughter amplitude ratios greater than unity but approaching the unity limit the closest, and the values of these quantities obtained from those mode triads with parent-daughter amplitude ratios smaller than unity that approach the unity limit the closest.
The linear regression yields a linear slope of $-0.08 \pm 0.04$ ($\vartheta_2$ per unit of $\absfraktxt{\mathfrak{L}^s_2}{\mathfrak{L}^s_1}$) and an intercept $0.22 \pm 0.06$, when we assume normally distributed residuals.
This leads to an estimated boundary $\hat{\vartheta}_{2,{\rm b}} = 0.14 \pm 0.07$ that is consistent with the earlier defined boundary $\vartheta_{2,{\rm b}}$.

In general, energy transfer occurs from the parent to the daughter mode only if the latter's energy is smaller than the former's (e.g., \citealt{2003ApJ...591.1129A}).
However, if one of the daughter modes participating in the mode triad is strongly damped (linearly), mode $\beta$, when the other daughter mode, mode $\alpha$, is only weakly damped (linearly), we find that daughter mode $\alpha$ can saturate at larger energies than the parent mode.
In such situations the parent mode needs to transfer a lot of energy to the daughter modes per cycle in order to overcome the strong linear damping of daughter mode $\beta$ and reach a stationary state.

Isolated parametric couplings in which both daughter modes attain amplitudes larger than the parent mode are not observed in our calculations.
Of the $276\,835$ potential triads considered in this work, only $32$ non-AE valid triads have both daughter-parent luminosity fluctuation energy ratios~\eqref{eq:stat+surf+amp+ratios} greater than unity, which all originate from the $\Delta {\rm M_{\rm ini,\, 1}}$ model.
Of those $32$ solutions, $26$ satisfy $q > Q_1$, the necessary but not sufficient condition for both stationary daughter-parent rotating-frame mode energy ratios~\eqref{eq:theoretical+amplitude+ratios} to be greater than unity, because of Eqs.~\eqref{eq:definition+q} and \eqref{eq:necessary+detuning+condition}.
The other $6$ solutions are relatively close to satisfying that criterion ($q$ and $Q_1$ are of the same order of magnitude) and likely obtain daughter-parent mode luminosity fluctuation energy ratios~\eqref{eq:stat+surf+amp+ratios} greater than unity because of the differing ratios of amplitude conversion factors $o_{\rm A, \varphi}$.
Given the typical values of $\Omega_\varphi$, $\gamma_\varphi$ and $q$ listed in Tables~\ref{tab:isolated+mode+triad+properties+1} and \ref{tab:isolated+mode+triad+properties+2}, as well as the quality factor ranges listed in Table~\ref{tab:radial+order+ranges}, we deem it unlikely to encounter isolated solutions of this kind.

\subsection{Comparison with observations of SPB stars}\label{subsect:asteroseismic+modeling+impact+comparison+observations}

To assess the impact of mode amplitude ratios on period spacing pattern modeling, we compare daughter-parent surface luminosity fluctuation ratios predicted by our formalism with observed equivalents derived for the ensemble of $38$ SPB stars of \citetalias{2021A&A...655A..59V}.
We limit ourselves to comparison among theoretically predicted amplitude ratios and observed equivalents of sum frequencies, because these make up the majority of the isolated mode triads listed in Table~\ref{tab:isolated+mode+triad+properties+2}.
Our comparison is also valid for difference frequencies, because the AEs and stationary solutions for these difference frequencies are of the same form as those of sum frequencies (see Appendix~\ref{app:diff+freq}).

\begin{figure*}
    \centering
    \begin{minipage}{9cm}
        \includegraphics[width=\textwidth]{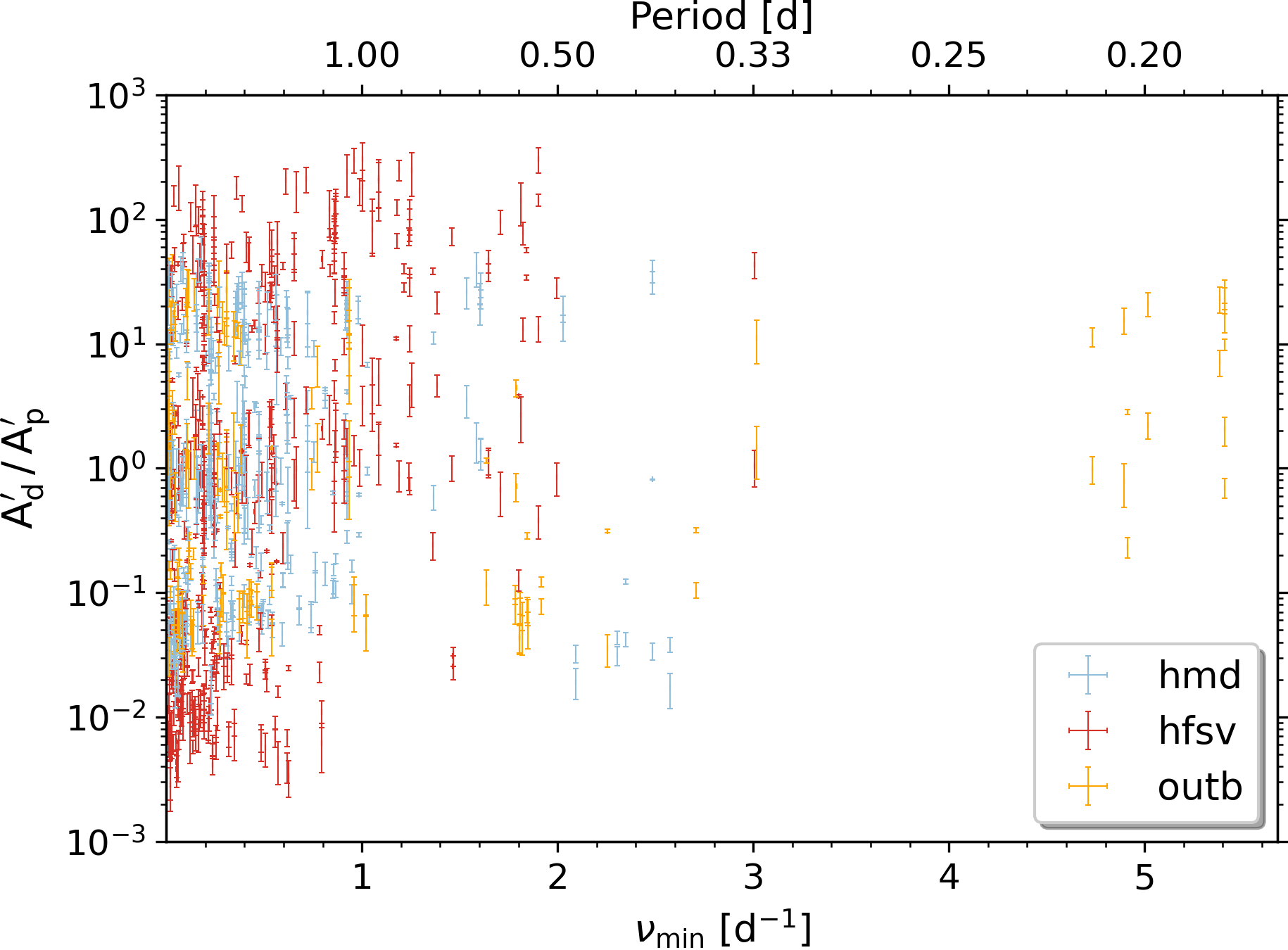}
    \end{minipage}
    \begin{minipage}{9cm}
        \vspace*{0.08cm}
        \includegraphics[width=1.019\textwidth]{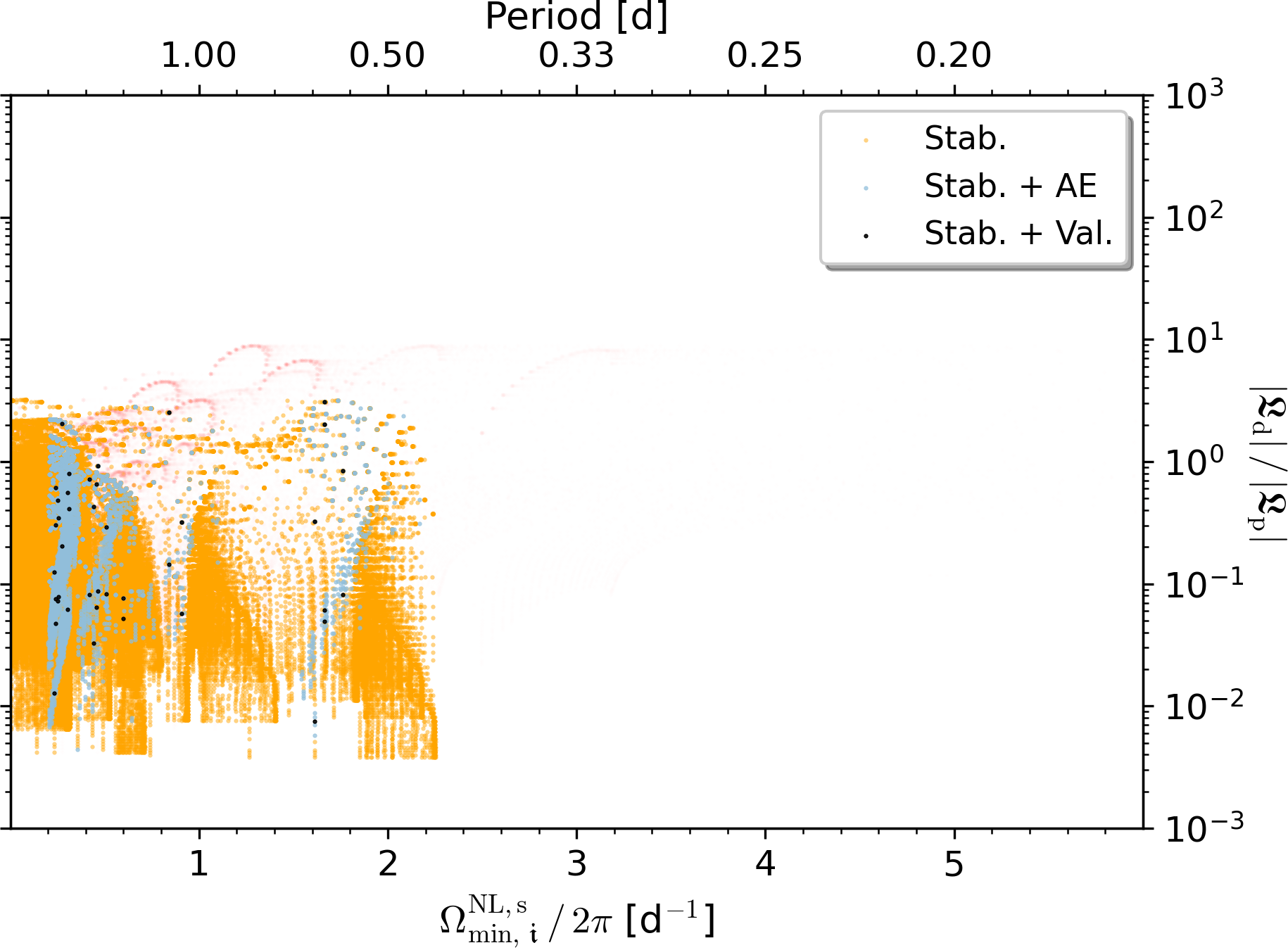}
    \end{minipage}
  \caption{Observed daughter-parent amplitude ratios $A_d'\,/\,A_p'$ of candidate resonances of SPB stars detected in models of light curves measured by {\em Kepler}, compared to their theoretically predicted equivalent observables, the monochromatic stationary daughter-parent surface luminosity perturbation ratios $\absfraktxt{\mathfrak{L}_d}{\mathfrak{L}_p}$, which are computed with the theoretical oscillation model described in Sect.~\ref{sect:theoretical+model}.
  These quantities are shown as a function of the minimal inertial-frame frequency $\nu_{\rm min}$ of the three signals in the considered candidate resonance, or, the minimal mode frequency observable $\Omega_{{\rm min},\,\mathfrak{i}}^{{\rm NL},\,s}\,/\,2\pi$ among the three mode frequency observables of the considered triad, respectively.
  \textit{Left panel:} $A_d'\,/\,A_p'$ as a function of $\nu_{\rm min}$. The light curve models were generated using strategy 3 of \citetalias{2021A&A...655A..59V}, which uses the signal-to-noise ratio of the detected signals to determine their significance. Colors indicate the pseudo-class defined in \citetalias{2021A&A...655A..59V}: high-$f_{\rm sv}$ (`hfsv'), high-mode-density (`hmd'), or outbursting (`outb'). \textit{Right panel:} $\absfraktxt{\mathfrak{L}_d}{\mathfrak{L}_p}$ as a function of $\Omega_{{\rm min},\,\mathfrak{i}}^{{\rm NL},\,s}\,/\,2\pi$. Colors indicate whether all stability criteria are fulfilled (`Stab.'), or if additionally the AE validity criterion (`Stab. + AE') or all validity criteria (`Stab. + Val.'; i.e., the isolated triads listed in Table~\ref{tab:isolated+triad+observables}) are fulfilled. We show alias signals in the frequency range of $0$ to $24.4512$ d$^{-1}$ as fainter orange-red (`Stab.') symbols, because we do not account for the splitting of the alias signals (which will affect the amplitude ratios).}
  \label{fig:observed+vs+computed+amplitude+ratios+SPBs+strategy+3}
\end{figure*}

\citetalias{2021A&A...655A..59V} generated variability models for the light curves, hereafter referred to as (harmonic) light curve models, of $38$ SPB stars using five different harmonic analysis strategies. 
In this work, we use the models generated by their strategy $3$, which uses the signal-to-noise ratios to determine the significance of detected variability signals, because this resembles commonly used strategies in literature (see Table~1 and the corresponding discussion in Sect.~2.2 of \citetalias{2021A&A...655A..59V} for additional details).
They assigned pseudo-classes to the different members of the ensemble of analyzed SPB stars to denote the difficulties encountered during the analysis process: high-$f_{\rm sv}$ stars had light curves that were described adequately by the harmonic light curve models; high mode density stars had many close-spaced frequencies in their Fourier transforms or Lomb-Scargle periodograms; and outbursting stars were found to have distinct variability `outbursts' in their light curves, resulting in many close-spaced and high-amplitude frequency groups (see \citetalias{2021A&A...655A..59V}). 
To select the relevant observed signals for our comparison, we look for sum-frequency resonance combinations of three signals that are part of the light curve models and fulfill the resonance criterion $\lvert\nu_1 - \nu_2 - \nu_3\rvert \leq \mathfrak{R}_\nu + \sigma_{\nu,\,{\rm prop}}$ (with $\sigma_{\nu,\,{\rm prop}}$ the propagated uncertainty of the observed frequency difference $\nu_1 - \nu_2 - \nu_3$ and $\nu_1$, $\nu_2$ and $\nu_3$ the observed frequencies). 
These combinations also need to contain (i) at least one of the two highest-amplitude signals in the light curve model, and (ii) lower-amplitude components (i.e., not one of the two highest-amplitude signals) that are not present in other considered combinations.
The low-over-highest frequency amplitude ratios of these combinations, $A_d'\,/\,A_p'$ (i.e., two ratios per combination), are displayed in the left panel of Fig.~\ref{fig:observed+vs+computed+amplitude+ratios+SPBs+strategy+3} as a function of the minimal frequency $\nu_{\rm min}$ of the three modes in the combination, as measured within the inertial reference frame.
Under the assumption that the non-linear parametric resonant coupling process dominates the energy transfer among the considered modes, we hereafter refer to $A_d'\,/\,A_p'$ as the observed daughter-parent amplitude ratios; under that assumption, $\nu_{\rm min}$ is the minimal daughter mode frequency.
Because the stationary combination phase observable $\Tilde{\Phi}^s_s$ defined in Eq.~\eqref{eq:observable+combination+phase} depends on the unconstrained initial zero points of the individual mode phases and because no trends are observed in the zero-point-corrected non-linear combination phases $\Tilde{\Phi}^s_{s,0}$ of isolated mode triads listed in Table~\ref{tab:isolated+triad+observables}, we do not enforce conditions on the relative phase of these observed combinations of signals, unlike \citetalias{2021A&A...655A..59V}.

One compares the observed daughter-parent amplitude ratios with the relevant computed daughter-parent (stationary) surface luminosity fluctuation ratios $\absfraktxt{\mathfrak{L}_d}{\mathfrak{L}_p}$ of triads found among the $(k, m) = (0, 1)$ daughter $g$ modes and of the $(k, m) = (0, 2)$ parent $g$ modes computed by GYRE for the models listed in Table~\ref{tab:radial+order+ranges} (shown in the right panel of Fig.~\ref{fig:observed+vs+computed+amplitude+ratios+SPBs+strategy+3}).
These mode triads have small frequency detunings~\eqref{eq:frequency+detuning} and satisfy all stability criteria, and we further distinguish between triads with modes that satisfy (i) only the stability criteria, or (ii) the stability criteria and the AE validity criterion, or (iii) all stability and validity criteria (i.e., these are the isolated mode triads).
The minimal mode frequency observable $\Omega_{{\rm min},\,\mathfrak{i}}^{{\rm NL},\,s}$ (used as the x-axis variable in the right panel of Fig.~\ref{fig:observed+vs+computed+amplitude+ratios+SPBs+strategy+3}) is smaller than the non-linear frequency shift~\eqref{eq:explicit+frequency+shift} for some of the considered modes in mode triads with linearly stable solutions (i.e., triads with modes that only satisfy the stability criteria and none of the validity criteria).
In that case, we compute their alias frequencies in the frequency range $[24.4512, 0]$ d$^{-1}$ (using the notation in Table~\ref{tab:radial+order+ranges}; i.e., up to the notional {\em Kepler} long-cadence Nyquist frequency; \citealt{2014MNRAS.445..946C}).
Such aliased signals would however have their amplitude ratios affected by their splitting, which we do not account for in this work.
The predicted surface luminosity fluctuation ratios of these alias frequencies are therefore shown on the right panel of Fig.~\ref{fig:observed+vs+computed+amplitude+ratios+SPBs+strategy+3} as faint (orange-red) symbols.

The ensemble of stationary surface luminosity fluctuation ratios of computed triads seems to be able to explain the low-amplitude-ratio part of the ensemble of observed daughter-parent amplitude ratios, as can be derived from the panels in Fig.~\ref{fig:observed+vs+computed+amplitude+ratios+SPBs+strategy+3}.
Some SPB modeling targets, for example KIC008714886, show promise for matching a part of their observed amplitude ratios with their theoretically predicted equivalent.
Other targets (notably the slow rotator KIC0010526994) have observed daughter-parent amplitude ratios that are all much greater than unity, which we do not have in the calculated ratios.

The absence of high theoretically predicted surface luminosity fluctuation ratios~\eqref{eq:theoretical+amplitude+ratios} might be caused by our choice of opacity tables, because they determine the linear growth or linear damping rates of the modes.
For example, one of the analyzed SPB stars, KIC0010536147, is on the massive end of the SPB instability region (according to the stellar parameters determined in \citealt{2022ApJ...930...94P}), where, according to our simulations that account for the classical $\kappa$ mechanism driving mechanism, no $(k,m) = (0,2)$ parent modes are excited (see Table~\ref{tab:radial+order+ranges}).
The required changes in opacity to reach the highest observed amplitude ratios are, however, unlikely to be realistic.
The modified opacities of \citet{2017MNRAS.466.2284D} used in modeling the linear stability of the hybrid main-sequence early B-type pulsator $\nu$ Eridani for example lead to normalized driving rates that were only two to four times larger than the unmodified driving rates.
Asteroseismic modeling of dedicated targets is needed to further assess the effect of opacity on theoretical surface luminosity fluctuation ratios.

Additionally, some of the observed modes might have different nonradial geometries (i.e., different values of $m$, $k$) than the ones considered in the limited number of simulations carried out for this work.
For example, the period spacing patterns considered by \citet{2021NatAs...5..715P} are made up of zonal $(k,m) = (1,0)$ dipole $g$ modes for $9$ of their considered $26$ SPB stars, and one star, KIC008714886, has an identified retrograde $(k,m) = (0,-1)$ dipole $g$ mode period spacing pattern (see Supplementary Table 1 of \citealt{2021NatAs...5..715P}).
When couplings between $g$ modes other than the ones considered in this work have linear driving or linear damping rates, frequencies, as well as amplitude conversion factors $o_{A,\varphi}$ of the same order as those obtained for the $g$ modes considered in this work, we expect to obtain similar theoretically predicted luminosity fluctuation ratios.
We note, however, that for such different combinations the parent mode can be a lower frequency mode.
An example of such a combination that satisfies the coupling coefficient selection rules is the triad that consists of a driven $(k,m) = (0,1)$ parent mode $\alpha$, a damped $(k,m) = (0,2)$ daughter mode $\beta$, and a damped $(k,m) = (0, -1)$ daughter mode $\gamma$.
The damped $(k,m) = (0,2)$ daughter mode $\beta$ can have a lower frequency than the parent mode $\alpha$ in the co-rotating frame, while having a higher frequency in the inertial frame.
Such a damped $(k,m) = (0,2)$ mode $\beta$ would therefore be labeled as the parent when following the rules used to generate the observed daughter-parent amplitude ratios shown in Fig.~\ref{fig:observed+vs+computed+amplitude+ratios+SPBs+strategy+3}. 
For comparison with the theoretically predicted daughter-parent surface luminosity fluctuation ratios, the relevant observed amplitude ratios then need to either be computed (as the ratio of the current amplitude ratios) or inverted.
In our example, we need to compute the amplitude ratio of modes $\beta$ and $\gamma$ ($A_\gamma' / A_\beta' = (A_\gamma' / A_\alpha') / (A_\beta' / A_\alpha')$), and we need to invert the currently computed amplitude ratio of modes $\alpha$ and $\beta$.

We also deem it likely that some of the high observed amplitude ratios can be attributed to other amplitude saturation mechanisms, such as multi-mode coupling or higher-order couplings.
Correlation of the mean values of the 10 largest mode amplitudes (determined from the light curve models) with the mean values of the 10 largest identified amplitude ratios for each of the different SPB stars considered by \citetalias{2021A&A...655A..59V} for example yields a positive Spearman correlation coefficient. 
This suggests that higher-order couplings might be needed to explain the high amplitude ratios of high-amplitude SPB pulsators.

%
%

\section{Conclusions}\label{sect:conclusions}

We derive a theoretical oscillation modeling framework that describes non-linear three-mode coupling among gravito-inertial ($g$) modes of Slowly Pulsating B (SPB) stars within the Traditional Approximation for Rotation (TAR; e.g., \citealt{1968RSPTA.262..511L,1997ApJ...491..839L,2003MNRAS.340.1020T,2013LNP...865...23M}), extending and correcting terms in the formalism of \citetalias{2012MNRAS.420.2387L} (see Appendix~\ref{app:expressions+qcc}).
This framework relates the $g$ mode adiabatic eigenfunctions to potential non-linear energy exchange between the modes, by computing three-mode (energy-scaled) coupling coefficients $\lvert \eta_1\rvert$ based on a phase space mode decomposition inserted in the relevant coupled equations of motion.
To describe three-mode resonant couplings, we derive amplitude equations (AEs) from the coupled equations of motion for a sum-frequency resonance $\Omega_1 \simeq \Omega_2 + \Omega_3$, using the multiple time scales perturbation method (e.g., \citealt{1973pm.book.....N,1981itpt.book.....N,1979noos.book.....N}).
These coupling coefficients need to satisfy angular selection rules~\eqref{eq:azimuthal+selection+rule} and \eqref{eq:meridional+selection+rule} if energy is to be exchanged among the modes.
The isolated stable stationary solutions of these AEs (i.e., their stable fixed points) then describe locked three-mode resonant couplings among $g$ modes in rapidly rotating $g$-mode pulsating stars, such as SPB stars.

We use this framework to compute examples of isolated mode couplings in stellar structure and pulsation models that represent SPB stars analyzed in \citetalias{2021A&A...655A..59V}.
We limit ourselves to computing couplings among $g$ modes with ordering numbers $\left(k_1,\,k_2,\,k_3\right) = \left(0,\,0,\,0\right)$ and azimuthal orders $\left(m_1,\,m_2,\,m_3\right) = \left(2,\,1,\,1\right)$; the most frequently observed modes in SPB stars (see e.g., \citealt{2021NatAs...5..715P,2021MNRAS.503.5894S,2022ApJ...930...94P}).
The locked mode solutions have to fulfill multiple stability and validity criteria to describe physical three-mode coupling scenarios in such stars.
To ensure stability of the three-mode stationary solution, it needs to be hyperbolic, the resonance must be parametric, and the solution must fulfill the condition in Eq.~\eqref{eq:first+hurwitz} and the linear quartic stability criterion~\eqref{eq:quartic+stability+condition}.
The validity of the solution is guaranteed if it fulfills the AE validity condition $\Psi_{\rm AE} \leq 0.1$ (with $\Psi_{\rm AE}$ defined in Eq.~\eqref{eq:omega+AE}), and the isolation criterion.
The most restricting of all these conditions is the isolation criterion, which ensures that no multi-mode coupling scenarios take place.
We find that the restriction to sharp resonances as done by \citet{2003ApJ...591.1129A} overestimates the number of isolated mode triads.

By performing coupling computations up to a certain inner radius, we map the regions that contribute significantly to the mode coupling.
Typically, we obtain strong contributions to the resonant coupling among the three $g$ modes in the near-core zones of the SPB models, where $N^2$ peaks and modes can become (partially) trapped \citep{2008MNRAS.386.1487M,2021A&A...650A.175M}.
For more evolved models, this $N^2$ maximum is wider and the $\kappa$ mechanism typically excites higher radial order modes, which are increasingly confined to the near-core zones.
Conversely, for less evolved models, significant contributions to $\lvert\eta_1\rvert$ can be found outside of the near-core zones.

Linear heat-driven vibrational instability defines which modes are available for the non-linear (parametric) couplings we study in this work.
The individual mode frequencies depend on a variety of factors, including rotation rate, evolutionary stage, and mass, among others.
These frequencies are important, because they define the optimal linear driving regions of the corresponding modes.
The linear driving of $g$ modes in SPB stars, however, is ultimately reliant upon the opacity profiles inside the \ce{Fe} bump.
Opacity tables used during stellar modeling define the opacity gradients in the driving zones, and therefore influence (i) the number of possible mode triads that can be formed for a specific model, but also (ii) the linear driving and linear damping rates themselves.
These linear driving and linear damping rates affect the non-linear observables, such as the computed stationary mode luminosity fluctuations and their ratios.

For the few models considered in this work, we find no obvious correlations for changes in the coupling coefficients~\eqref{eq:coupling+coefficient} with changes in model parameters.
It is not clear whether such trends can be detected if the number of models and the number of considered modes is increased while varying model parameters such as the opacity tables.
Because the absolute values of the detuning-damping ratio, $\lvert q \rvert$, are larger than unity, it is clear however that the resonant nature of the coupling strongly determines the stationary state mode amplitudes~\eqref{eq:stat+amps+q}.

The non-linear frequency shifts~\eqref{eq:explicit+frequency+shift} for isolated triads are small enough to neglect when compared with the typical errors for the frequencies derived from $4$-year {\em Kepler} light curves, and the combination phases~\eqref{eq:observable+combination+phase} contain an unconstrained initial mode phase parameter.
The best constraints for asteroseismic modeling are therefore obtained from the predicted luminosity fluctuation ratios~\eqref{eq:stat+surf+amp+ratios} of coupled modes.
Most of these ratios indicate that the $\left(k,\,m\right) = \left(0,\,2\right)$ parent $g$ modes considered in this work have higher apparent mode amplitudes than their coupled $\left(k,\,m\right) = \left(0,\,1\right)$ daughter $g$ modes.
We obtain disparate linear daughter mode damping rates for each of the isolated mode triads.
If one of the linearly damped daughter modes in the isolated mode triad is weakly damped (linearly; at a rate $\lesssim 13$\% that of the parent linear driving rate), we find that that daughter mode can saturate at a higher amplitude than its coupled parent mode.
We deem it unlikely that isolated solutions exist in which both daughter-parent amplitude ratios are greater than unity, because of the typical values for the quality factors and detuning-damping ratios associated with $g$ modes in SPB star models.

The monochromatic stationary luminosity fluctuation ratios are consistent with some of the lowest amplitude ratios $A_d'\,\displaystyle/\,A_p'$ observed in the {\em Kepler} space photometric time series of some of the target SPB stars (see Sect.~\ref{subsect:asteroseismic+modeling+impact+comparison+observations}).
Even the lowest order model of resonant non-linear mode coupling developed in this work can thus aid future asteroseismic modeling of rapidly rotating SPB stars based on $g$ modes, because it offers additional constraints on some of the observed combination frequencies based on stationary amplitude ratios of specific, resonantly coupled modes.
This can in principle also serve as an additional source of mode identification.
The logical next step in understanding non-linear amplitude saturation is to apply this framework to the modeling of some of the SPB stars analyzed in \citetalias{2021A&A...655A..59V} by integrating our monochromatic predictions over the {\it Kepler\/} passband.

It should be investigated whether larger resonantly coupled networks of modes or higher-order and $r$-$g$ mode coupling can saturate some of the unexplained observed candidate couplings in SPB stars.
Higher-order coupling will definitely already include self-saturation effects that are neglected in the current formalism (see e.g., \citealt{1998MNRAS.297..536V} for an example of a study that included these effects for radial modes while neglecting rotation, and \citealt{2008A&A...484...29G,2008A&A...490..743G} for direct numerical simulations of the $\kappa$ mechanism and non-linear saturation for radial modes in Cepheids).
Mode triads containing $g$ modes not considered in this work can also explain the presence of some of the highest observed amplitude ratios obtained in this work.

Other authors, such as \citet{2001ApJ...557..311L} and \citet{2011MNRAS.412.2265A}, used a mode expansion formalism that did not invoke the TAR and accounted for linear mode couplings to describe the linear eigenmodes.
They found that their formalism linearly stabilizes some of the $g$ modes that were excited when the TAR was assumed (i.e., constituting a systematical error).
That method, which truncates the series expansion at some manageable expansion order, is computationally more intensive, and might not be accurate \citep{2007MNRAS.374..248D}.
Perhaps the hybrid expansion method discussed in Chapter $7$ of \citet{2013LNP...865.....G} that describes linear mode coupling using expansions of modes whose angular eigenfunctions are described by Hough functions is a good compromise that softens the computational load with increasing expansion order, compared to the original expansion formalism.
Because we find that the non-linear saturation is crucially dependent on the linear excitation properties, in-depth investigations of the linear vibrational instability of $g$ modes in SPB stars using the latest improvements in opacity computations are crucial for future non-linear excitation studies (e.g., \citealt{2017MNRAS.466.2284D}).
Such investigations will allow us to estimate the systematical error we make in the theoretical predictions of surface luminosity fluctuation ratios due to opacity.

%
%

\begin{acknowledgements}
    The research leading to these results received funding from the KU Leuven Research Council (grant C16/18/005: PARADISE, with PIs CA and TVH). The research leading to part of these results has also received funding from the European Research Council (ERC) under the Horizon Europe programme (Synergy Grant agreement N$^\circ$101071505: 4D-STAR).
    Funded by the European Union.
    Views and opinions expressed are however those of the author(s) only and do not necessarily reflect those of the European Union or the European Research Council.
    Neither the European Union nor the granting authority can be held responsible for them.
    JVB and CA acknowledge support from the Research Foundation Flanders (FWO) under grant agreements N°V421221N (Travel Grant) and N°K802922N (Sabbatical leave).
    This research was also supported in part by the National Science Foundation under Grant No. PHY-1748958. 
    JVB is grateful to Dominic Bowman for the useful discussion and comments on the manuscript.
    JVB is also grateful to Michel Rieutord, Rony Keppens and Timothy Van Reeth for the useful comments on his PhD manuscript, which improved this work.
    CA is grateful for the kind hospitality offered by the staff of the Center for Computational Astrophysics at the Flatiron Institute of the Simons Foundation in New York during her work visits in the fall of 2022 and spring of 2023. 
    Throughout this work we have made use of the following Python packages: {\sc lmfit} \citep{matt_newville_2020_3814709}, {\sc SciPy} \citep{2020SciPy-NMeth}, {\sc NumPy} \citep{2020NumPy-Array}, {\sc Pandas} \citep{mckinney-proc-scipy-2010}, and {\sc Matplotlib} \citep{4160265}.
    We thank their authors for making these great software packages open source.
\end{acknowledgements}

%
%

\bibliographystyle{aa}
\bibliography{aanda}

%
%
\begin{appendix}


\section{Phase-space representation of the oscillation model within the TAR}\label{app:alternative+representation+oscillation+model}

Equation~\eqref{eq:general+linear+eqs+motion} can be written as a dynamical system in phase space that satisfies (following \citetalias{2001PhRvD..65b4001S})
\begin{equation}\label{eq:tensor+equations+of+motion}
    \dot{\vec{\zeta}} = \tens{T}\vec{\cdot}\vec{\zeta} + \tens{F}\,,
\end{equation}
in which the non-Hermitian operator $\tens{T}$ is defined as
\begin{equation}
    \tens{T} = \begin{bmatrix}
    \,-\dfrac{1}{2}\vec{B} & \vec{1}\,\\
    \,-\vec{C} + \dfrac{1}{4}\vec{B}^2 & -\dfrac{1}{2}\vec{B}\,\\
    \end{bmatrix}\,,
\end{equation}
where $\vec{\zeta}$ and $\tens{F}$ are defined as
\begin{equation}
    \vec{\zeta}\left(\vec{x},\,t\right) \equiv \begin{bmatrix}
    \vec{\xi}\left(\vec{x},\,t\right)\\
    \vec{\pi}\left(\vec{x},\,t\right)
    \end{bmatrix}\,,\hspace{5mm}\tens{F} = \begin{bmatrix}
    0\\
    \vec{a}_{\rm ext}
    \end{bmatrix}\,,
\end{equation}
with $\vec{x}$ the position vector, $t$ the time coordinate, and
\begin{equation}\label{eq:conjugate+momentum+schenk}
    \vec{\pi} = \dot{\vec{\xi}} + \dfrac{1}{2}\vec{B}\left(\vec{\xi}\right)\,.
\end{equation}
By making a time dependence {\em Ansatz} similar to Eq.~\eqref{eq:time+Ansatz},
\begin{equation}\label{eq:time+dependence+zeta}
    \vec{\zeta}\left(\vec{x},t\right) = \vec{\zeta}\left(\vec{x}\right)\,e^{-i\omega t}\,,
\end{equation}
the equations of motion for free linear oscillations yield the eigenvalue problem
\begin{equation}\label{eq:phase+space+eigenvalue+problem}
    \left[\tens{T} + i\omega\right]\vec{\cdot}\,\vec{\zeta}\left(\vec{x}\right) = 0\,.
\end{equation}

Because $\tens{T}$ is non-Hermitian, one should distinguish between its right and left eigenvectors.
The right eigenvectors $\vec{\zeta}_\varphi$ of operator $\tens{T}$ are the solutions of
\begin{equation}\label{eq:right+eigenmode}
    \left[\tens{T} + i\omega_\varphi\right]\vec{\cdot}\,\vec{\zeta}_\varphi\left(\vec{x}\right) = 0\,.
\end{equation}
These right eigenvectors defined by Eq.~\eqref{eq:right+eigenmode} are the eigenvectors of non-conjugated eigenmodes satisfying Eq.~\eqref{eq:general+linear+eqs+motion}.
We refer the reader to \citetalias{2001PhRvD..65b4001S} for details on the left eigenvector problem.


\section{Proof of the orthogonality condition}\label{app:orthogonality}

We give an explicit proof of Eq.~\eqref{eq:complex+orthogonality+relation} that employs the rotating-frame symplectic product $W\left(\vec{\xi}_\varphi,\vec{\xi}_\beta\right)\,$, which was defined by \citet{1978ApJ...221..937F} as
\begin{equation}\label{eq:symplectic+product}
\begin{aligned}
    W\left(\vec{\xi}_\varphi,\vec{\xi}_\beta\right) =&\, \left<\vec{\xi}_\varphi,\, \dot{\vec{\xi}_\beta} + \dfrac{1}{2}\vec{B}(\vec{\xi}_\beta)\right> - \left<\dot{\vec{\xi}_\varphi} + \dfrac{1}{2}\vec{B}(\vec{\xi}_\varphi), \vec{\xi}_\beta\right>\,,
\end{aligned}
\end{equation}
where the definition of the inner product in Eq.~\eqref{eq:inner+product} was used.
Because of the time dependence {\em Ansatz}~\eqref{eq:time+Ansatz}, Eq.~\eqref{eq:symplectic+product} becomes
\begin{equation}\label{eq:scaled+orthogonality+condition}
    iW\left(\vec{\xi}_\varphi,\vec{\xi}_\beta\right) = \left(\omega_\beta + \omega_\varphi^*\right)\left<\vec{\xi}_\varphi, \vec{\xi}_\beta\right> + \left<\vec{\xi}_\varphi, i\vec{B}(\vec{\xi}_\beta)\right>\,,
\end{equation}
after multiplication with $i$, and using the anti-Hermiticity of $\vec{B}$.
The symplectic product $W$ is a conserved quantity \citep[e.g.,][]{1978ApJ...222..281F}, hence, we have that
\begin{equation}\label{eq:time+dependence+symplectic+product}
    \pdv{W\left(\vec{\xi}_\varphi,\vec{\xi}_\beta\right)}{t} = 0 = i\left(\omega_\varphi^* - \omega_\beta\right)W\left(\vec{\xi}_\varphi,\vec{\xi}_\beta\right)\,,
\end{equation}
because of {\em Ansatz}~\eqref{eq:time+Ansatz} and the assumed time-independence of $\vec{\Omega}$.
For $\omega_\beta \neq \omega_\varphi^*$, we obtain from Eq.~\eqref{eq:time+dependence+symplectic+product} that $iW\left(\vec{\xi}_\varphi,\vec{\xi}_\beta\right)=0$, which proves that in that case Eq.~\eqref{eq:complex+orthogonality+relation} holds.
For modes with real eigenfrequencies $\Omega_\beta \neq \Omega_\varphi$, Eqs.~\eqref{eq:scaled+orthogonality+condition} and \eqref{eq:time+dependence+symplectic+product} imply that
\begin{equation}\label{eq:time+dependence+symplectic+product_real}
    \left(\Omega_\beta + \Omega_\varphi\right)\left<\vec{\xi}_\varphi, \vec{\xi}_\beta\right> + \left<\vec{\xi}_\varphi,  i\vec{B}(\vec{\xi}_\beta)\right> = 0\,,
\end{equation}
which is equal to Eq.~\eqref{eq:real+orthogonality+relation} for $\varphi \neq \beta$.
In the case of complex conjugate degeneracy (i.e., $\omega_\beta = \omega_\varphi^*$), or, in the case of degeneracy for real-valued eigenfrequencies (i.e., $\Omega_\beta = \Omega_\varphi$), Eqs.~\eqref{eq:time+dependence+symplectic+product} and \eqref{eq:time+dependence+symplectic+product_real} yield no orthogonality conditions.
We do not consider degenerate modes in this work, and instead refer the reader to \citetalias{2001PhRvD..65b4001S} for information on the degenerate eigenvalue problem.

For the mode $\vec{\xi}_\varphi$, we obtain from Eq.~\eqref{eq:symplectic+product} that
\begin{equation}\label{eq:same+modes+symplectic}
    W\left(\vec{\xi}_\varphi,\vec{\xi}_\varphi\right) = -i \left(\omega_\varphi + \omega_\varphi^*\right)\left<\vec{\xi}_\varphi, \vec{\xi}_\varphi\right> + \left<\vec{\xi}_\varphi, \vec{B}(\vec{\xi}_\varphi)\right>\,.
\end{equation}
Because of the conserved nature of $W\left(\vec{\xi}_\varphi,\vec{\xi}_\varphi\right)$ and the assumed time-independence of $\vec{\Omega}$, we write the time derivative of the symplectic product~\eqref{eq:same+modes+symplectic} as
\begin{equation}\label{eq:time+dependence+symplectic+product_same_modes}
    \pdv{W\left(\vec{\xi}_\varphi,\vec{\xi}_\varphi\right)}{t} = i\left(\omega_\varphi^* - \omega_\varphi\right)W\left(\vec{\xi}_\varphi,\vec{\xi}_\varphi\right) = 0\,,
\end{equation}
indicating that $\omega_\varphi^* = \omega_\varphi$ (i.e., $\omega_\varphi$ is real) or $iW\left(\vec{\xi}_\varphi,\vec{\xi}_\varphi\right) = 0$.
For real eigenfrequencies $\Omega_\varphi$, Eq.~\eqref{eq:time+dependence+symplectic+product_same_modes} is always fulfilled.
This proves Eq.~\eqref{eq:real+orthogonality+relation} for $\varphi = \beta$, because Eq.~\eqref{eq:same+modes+symplectic} multiplied by $i$ is equal to the real-valued constant $b_\varphi$ defined in Eq.~\eqref{eq:b+A} for a real eigenfrequency $\Omega_\varphi$.
For complex non-degenerate eigenfrequencies $\omega_\varphi$, Eq.~\eqref{eq:time+dependence+symplectic+product_same_modes} is fulfilled only if $b_\varphi = 0$, because $\omega_\varphi \neq \omega_\varphi^*$.

An entirely equivalent proof can be constructed using left contractions of Eq.~\eqref{eq:general+linear+eqs+motion+time+dependence} and its complex conjugate.
\citet{2022MNRAS.513.2522L} provided an example of such a proof, relying on a different time dependence {\em Ansatz}.
Their derived Eq.~(11) is similar to Eq.~\eqref{eq:real+orthogonality+relation}, and is valid for non-degenerate real-valued eigenfrequencies.
The lack of an orthogonality relation for the non-adiabatic complex-valued eigenfunctions of oscillations in rotating stars (with associated complex-valued eigenfrequencies) that does not involve Jordan chain modes (see Appendix A of \citealt{2001PhRvD..65b4001S}) motivates the use of adiabatic real-valued eigenfunctions (with associated real-valued eigenfrequencies) in the computations for the mode coupling coefficient~\eqref{eq:coupling+coefficient}.


\section{Expressions for the explicit terms of the three-mode coupling coefficients} \label{app:expressions+qcc}

We base ourselves on explicit expressions derived by \citetalias{2012MNRAS.420.2387L} for the terms that make up the three-mode coupling coefficients $\kappa_{ABC}$ defined in their Eq.~(26), and correct some of these expressions before using them to compute the coupling coefficient $\kappa^{\beta\gamma}_{\varphi}$ defined in Eq.~\eqref{eq:coupling+coefficient}.
Additionally, we show how we compute the adiabatic derivative of the adiabatic exponent $\Gamma_1$, $\left(\pdv{\Gamma_1}{\ln \rho}\right)_S$, which appears in one of the terms of $\kappa^{\beta\gamma}_{\varphi}$.

\subsection{Information on the explicit terms of $\kappa^{\beta\gamma}_{\varphi}$}\label{sect:explicit+terms+cc}

Equations~(\ref{eq:a+G+expressions}), (\ref{eq:tensors+nonlinear+acceleration}) and (\ref{eq:traces+scalar+nonlinear+acceleration}) contain covariant derivatives of the displacements $\vec{\xi}$ and the gravitational potential $\Phi$, which are used to compute the non-linear coupling term in the Cowling approximation.
These covariant derivatives are computed in spherical coordinates using Eqs. (B6) to (B16) in appendix B of \citetalias{2012MNRAS.420.2387L}, in which we replace the Hough functions defined in \citetalias{2012MNRAS.420.2387L}, $\Tilde{\Theta}$, $\Tilde{\Theta}^\theta$ and $\Tilde{\Theta}^\phi$, by their equivalents $H_r(\theta)\,e^{i\phi_\varphi}$, $H_\theta(\theta)\,e^{i\phi_\varphi}$ and $H_\phi(\theta)\,e^{i\phi_\varphi}$ defined for a mode $\varphi$ in this work.
We further note that equation (B12) of \citetalias{2012MNRAS.420.2387L} should read
\begin{equation}\label{eq:B12+lee}
    \nabla_r\,\xi^\phi = \dfrac{1}{r\sin\theta}\pdv{\xi^\phi}{r} = \dfrac{i}{r}\pdv{}{r}\left(r\,\dfrac{z_2}{c_1\,\Bar{\omega}^2}\right)\dfrac{H_\phi}{\sin\theta}\,,
\end{equation}
in which $z_2$ denotes the dimensionless horizontal displacement vector of a mode in the Cowling approximation, $\Bar{\omega}$ is the dimensionless mode frequency of that mode, and $c_1$ is a dimensionless ratio; all defined in \citetalias{2012MNRAS.420.2387L}.

Following \citetalias{2001PhRvD..65b4001S} and \citetalias{2012MNRAS.420.2387L}, the terms in the coupling coefficient integral can be split up in four different integrals $\kappa_{\varphi}^{\beta\gamma\,(u)}$ ($u \in \myintset{4}$), similar to the definitions of $\kappa^{(u)}_{ABC}$ in Eqs.~(B1) to (B5) of \citetalias{2012MNRAS.420.2387L}, or their equivalents $\eta^{(u)}_1 \equiv \kappa_{\varphi}^{\beta\gamma\,(u)}\,/\,\epsilon_\varphi$.
Equations (B2) and (B5) of \citetalias{2012MNRAS.420.2387L} describe how to compute $\kappa_{ABC}^{(1)}$ and $\kappa_{ABC}^{(4)}$, and are employed in this work to compute the coupling integrals $\kappa_{\varphi}^{\beta\gamma\,(1)}$ and $\eta_1^{(1)}$, and $\kappa_{\varphi}^{\beta\gamma\,(4)}$ and $\eta_1^{(4)}$, respectively.
We note that in Eq.~(B3) of \citetalias{2012MNRAS.420.2387L} for $\kappa_{ABC}^{(2)}$, the derivative $\pdv{\Gamma_1}{\ln\rho}$ should be replaced by its adiabatic equivalent, $\left(\pdv{\Gamma_1}{\ln \rho}\right)_S$, before using that equation to compute the coupling integrals $\kappa_{\varphi}^{\beta\gamma\,(2)}$ and $\eta_1^{(2)}$.
We compute this adiabatic derivative as described in Appendix~\ref{app:ad+der}.
Each of the integrands of the coupling integrals $\kappa_{\varphi}^{\beta\gamma\,(u)}$ or $\eta_1^{(u)}$ consists of a factor composed only of quantities derived from the stellar equilibrium structure, multiplied with some expression that either involves covariant derivatives of the eigenfunctions, or covariant derivatives of the gravitational potential $\Phi$ and the eigenfunctions themselves (for $\kappa_{\varphi}^{\beta\gamma\,(4)}$ and $\eta_1^{(4)}$).
The explicit expressions of these latter factors of the integrands are constructed from the expressions in Eqs.~(B17) to (B22) of \citetalias{2012MNRAS.420.2387L}.
We note that the last four terms of Eq.~(B19) in \citetalias{2012MNRAS.420.2387L} need to be multiplied with a factor $2$ before using them to compute the coupling integrals $\kappa_{\varphi}^{\beta\gamma\,(3)}$ and $\eta_1^{(3)}$, and that the first two terms of their Eq.~(B19) can be simplified, because
\begin{equation}
    \begin{aligned}
        &S\left(\pdv{(r\,z_1)}{r}\left(z_1 - \dfrac{z_2}{c_1\,\Bar{\omega}^2}\right)\pdv{}{r}\left(r\,\dfrac{z_2}{c_1\,\Bar{\omega}^2}\right)\,:\,H_r\left[H_\theta H_\theta - H_\phi H_\phi\right]\right)\, +\,\\
        &S\left(\pdv{(r\,z_1)}{r}\,z_1\,\pdv{}{r}\left(r\,\dfrac{z_2}{c_1\,\Bar{\omega}^2}\right)\,:H_r\left[\delta H_r H_\theta - \dfrac{m\,H_r\,H_\phi}{\sin\theta}\right] \,\right)\, -\\
        &S\left(\pdv{(r\,z_1)}{r}\,z_1\,\pdv{}{r}\left(r\,\dfrac{z_2}{c_1\,\Bar{\omega}^2}\right)\,:\, H_r\left[H_\theta H_\theta - H_\phi H_\phi\right]\right) =\\
        &S\left(\pdv{(r\,z_1)}{r}\,z_1\,\pdv{}{r}\left(r\,\dfrac{z_2}{c_1\,\Bar{\omega}^2}\right)\,:H_r\left[\delta H_r H_\theta - \dfrac{m\,H_r\,H_\phi}{\sin\theta}\right] \,\right)\, -\\
        &S\left(\pdv{(r\,z_1)}{r} \dfrac{z_2}{c_1\,\Bar{\omega}^2}\pdv{}{r}\left(r\,\dfrac{z_2}{c_1\,\Bar{\omega}^2}\right)\,:\,H_r\left[H_\theta H_\theta - H_\phi H_\phi\right]\right)\,,
    \end{aligned}
\end{equation}
in which we separated the factors containing the azimuthal angle $\phi$, and where $\delta H_r$ replaces $\delta\Tilde{\Theta}$, which was defined together with the symmetrical operator $S$ in \citetalias{2012MNRAS.420.2387L}.

Finally, as noted by for example \citetalias{2012MNRAS.420.2387L}, the coupling coefficient integrals over the spherical star can be transformed into products of integrals over the radial part of the integrand with integrals over the angular part of the integrand.
The selection rules discussed in Sect.~\ref{sect:selection+rules} originate from the integral over the angular part of the integrand of these coupling integrals (see appendix~\ref{app:meridional+selection+rule} for the explicit dependence of the selection rules on the Hough functions).

\subsection{A computation-friendly form of $\left(\pdv{\Gamma_1}{\ln \rho}\right)_S$}\label{app:ad+der}

In this section, we show that the adiabatic derivative $\left(\pdv{\Gamma_1}{\ln \rho}\right)_S$, which should be used in Eq.~(B3) of \citetalias{2012MNRAS.420.2387L}, can be written as a function of non-adiabatic derivatives that are computed in stellar evolution codes such as MESA \citep{2011ApJS..192....3P,2013ApJS..208....4P,2015ApJS..220...15P,2018ApJS..234...34P,2019ApJS..243...10P}.
First, let us write a two-dimensional Jacobian as
\begin{equation}\label{eq:jacobian}
    \pdv{(u,\,v)}{(x,\,y)} \equiv \begingroup\renewcommand*{\arraystretch}{1.2}\begin{vmatrix}
    \left(\pdv{u}{x}\right)_y & \left(\pdv{u}{y}\right)_x\\
    \left(\pdv{v}{x}\right)_y & \left(\pdv{v}{y}\right)_x\\
    \end{vmatrix}\endgroup\,.
\end{equation}
Using Eq.~(\ref{eq:jacobian}), the expression for $\left(\pdv{\Gamma_1}{\ln \rho}\right)_S$ becomes
\begin{equation}\label{eq:jacobian+S}
    \left(\pdv{\Gamma_1}{\ln \rho}\right)_S  = \rho\,\begingroup\renewcommand*{\arraystretch}{1.2}\begin{vmatrix}
    \left(\pdv{\Gamma_1}{\rho}\right)_S & \left(\pdv{\Gamma_1}{S}\right)_\rho\\
    0 & 1\\
    \end{vmatrix}\endgroup =  \rho\,\pdv{(\Gamma_1,\,S)}{(\rho,\,S)}\,.
\end{equation}
If we then multiply Eq.~(\ref{eq:jacobian+S}) with a Jacobian equal to $1$, we can write
\begin{equation}
    \left(\pdv{\Gamma_1}{\ln \rho}\right)_S = \rho\,\pdv{(\Gamma_1,\,S)}{(\rho,\,S)}\pdv{(\rho,\,T)}{(\rho,\,T)} = \rho\,\dfrac{\pdv{(\Gamma_1,\,S)}{(\rho,\,T)}}{\pdv{(\rho,\,S)}{(\rho,\,T)}}\,,
\end{equation}
which after using the definition of the Jacobian in Eq.~(\ref{eq:jacobian}) becomes
\begin{equation}\label{eq:computation+friendly+adiabatic+derivative}
    \left(\pdv{\Gamma_1}{\ln \rho}\right)_S =\rho\,\left[\left(\pdv{\Gamma_1}{\rho}\right)_T -\left\{\dfrac{\left(\pdv{S}{\rho}\right)_T}{\left(\pdv{S}{T}\right)_\rho}\right\}\,\left(\pdv{\Gamma_1}{T}\right)_\rho\right]\,,
\end{equation}
whose right hand side only contains thermodynamic quantities that are computed in stellar evolution codes such as MESA.

An adiabatic second-order derivative of adiabatic exponent $\Gamma_1$ with respect to mass density $\rho$ appears in coupling coefficients that describe four-mode interactions (or higher-order derivatives when higher-order coupling terms are considered).
Higher-order derivatives of the thermodynamic quantities used in Eq.~(\ref{eq:computation+friendly+adiabatic+derivative}) are not readily computed in stellar evolution codes such as MESA, but would be necessary to obtain an analytical expression for the adiabatic higher-order derivatives of $\Gamma_1$.
Such higher-order derivatives can alternatively be estimated from numerical derivatives of Eq.~(\ref{eq:computation+friendly+adiabatic+derivative}) when using the output of current state-of-the-art stellar evolution codes.


\section{Defining $c_\varphi(t)$ and $c_\varphi^*(t)$}\label{app:c_coefficients}

We prove the validity of the expressions~\eqref{eq:real+inverse+mode+expansions} for $c_\varphi(t)$ and $c_\varphi^*(t)$ by making use of the orthogonality relations among the modes.
First, let us derive the complex orthogonality relations containing complex conjugate modes using the approach in Appendix~\ref{app:orthogonality}.
For $W\left(\vec{\xi}_\varphi^*,\vec{\xi}_\beta\right)$, $W\left(\vec{\xi}_\varphi,\vec{\xi}_\beta^*\right)$ and $W\left(\vec{\xi}_\varphi^*,\vec{\xi}_\beta^*\right)$ we obtain from the definition of the symplectic product in Eq.~\eqref{eq:symplectic+product} that
\begin{subequations}\label{eq:complex+conjugated+orthog+symplectic+product}
\begin{align}
    W\left(\vec{\xi}_\varphi^*,\vec{\xi}_\beta\right) =&\, -i\left(\omega_\beta - \omega_\varphi\right)\left<\vec{\xi}_\varphi^*, \vec{\xi}_\beta\right> + \left<\vec{\xi}_\varphi^*,  \vec{B}(\vec{\xi}_\beta)\right>\,,\label{eq:symplectic+kconj}\\
    W\left(\vec{\xi}_\varphi,\vec{\xi}_\beta^*\right) =&\, -i\left(\omega_\varphi^* - \omega_\beta^*\right)\left<\vec{\xi}_\varphi, \vec{\xi}_\beta^*\right> + \left<\vec{\xi}_\varphi,  \vec{B}(\vec{\xi}_\beta^*)\right>\,,\label{eq:symplectic+jconj}\\
    W\left(\vec{\xi}_\varphi^*,\vec{\xi}_\beta^*\right) =&\,\, i\left(\omega_\varphi + \omega_\beta^*\right)\left<\vec{\xi}_\varphi^*, \vec{\xi}_\beta^*\right> + \left<\vec{\xi}_\varphi^*,  \vec{B}(\vec{\xi}_\beta^*)\right>\,.\label{eq:symplectic+kjconj}
\end{align}
\end{subequations}
The orthogonality relation derived from Eq.~\eqref{eq:symplectic+kconj} becomes
\begin{equation}\label{eq:orthog+kconj}
    \left(\omega_\beta - \omega_\varphi\right)\left<\vec{\xi}_\varphi^*, \vec{\xi}_\beta\right> + \left<\vec{\xi}_\varphi^*,  i\vec{B}(\vec{\xi}_\beta)\right> = 0\,,
\end{equation}
when $\omega_\varphi \neq  -\omega_\beta$, which has a real-valued equivalent
\begin{equation}\label{eq:orthog+jconj}
    \left(\Omega_\beta - \Omega_\varphi\right)\left<\vec{\xi}_\varphi^*, \vec{\xi}_\beta\right> + \left<\vec{\xi}_\varphi^*,  i\vec{B}(\vec{\xi}_\beta)\right> = 0\,.
\end{equation}
Orthogonality relation~\eqref{eq:orthog+jconj} is valid if $\Omega_\varphi \neq -\Omega_\beta$.
Similarly, we obtain for Eq.~\eqref{eq:symplectic+jconj} that
\begin{equation}\label{eq:orthog+kjconj}
    \left(\omega_\varphi^* - \omega_\beta^*\right)\left<\vec{\xi}_\varphi, \vec{\xi}_\beta^*\right> + \left<\vec{\xi}_\varphi,  i\vec{B}(\vec{\xi}_\beta^*)\right> = 0\,,
\end{equation}
when $\omega_\varphi^* \neq -\omega_\beta^*$. 
The real-valued equivalent of Eq.~\eqref{eq:orthog+kjconj} is 
\begin{equation}\label{eq:orthog+bconj}
    \left(\Omega_\varphi - \Omega_\beta\right)\left<\vec{\xi}_\varphi, \vec{\xi}_\beta^*\right> + \left<\vec{\xi}_\varphi,  i\vec{B}(\vec{\xi}_\beta^*)\right> = 0\,,    
\end{equation}
when $\Omega_\varphi \neq -\Omega_\beta$.
Finally, we have that
\begin{equation}
    \left(\omega_\beta^* + \omega_\varphi\right)\left<\vec{\xi}_\varphi^*, \vec{\xi}_\beta^*\right> - \left<\vec{\xi}_\varphi^*,  i\vec{B}(\vec{\xi}_\beta^*)\right> = 0\,,
\end{equation}
for Eq.~\eqref{eq:symplectic+kjconj} and $\omega_\varphi \neq \omega_\beta^*$.
Its real-valued equivalent
\begin{equation}\label{eq:symplectic+orthog+real+double+conj}
    \left(\Omega_\beta + \Omega_\varphi\right)\left<\vec{\xi}_\varphi^*, \vec{\xi}_\beta^*\right> - \left<\vec{\xi}_\varphi^*,  i\vec{B}(\vec{\xi}_\beta^*)\right> = \delta^{\varphi}_\beta\, b_\varphi^* = \delta^{\varphi}_\beta\, b_\varphi\,,
\end{equation}
is the complex conjugate of Eq.~\eqref{eq:real+orthogonality+relation}, and is valid for non-degenerate eigenfrequencies (i.e., $\Omega_\varphi\neq\Omega_\beta$).
The second equality in Eq.~\eqref{eq:symplectic+orthog+real+double+conj} is valid because $b_\beta$ is a real number for the adiabatic eigenfunctions used in this work.

Multiplying the term $ \Omega_\beta\,\vec{\xi}\left(\vec{x},\,t\right) + i\,\Dot{\vec{\xi}}\left(\vec{x},\,t\right) + i\vec{B}\left(\vec{\xi}\left(\vec{x},\,t\right)\right) $ with $\vec{\xi}_\beta^*$ and integrating over the stellar mass yields
\begin{equation}
    \left<\vec{\xi}_\beta,\,\Omega_\beta\,\vec{\xi}\left(\vec{x},\,t\right) + i\,\Dot{\vec{\xi}}\left(\vec{x},\,t\right) + i\vec{B}\left(\vec{\xi}\left(\vec{x},\,t\right)\right)\right> \equiv \mathcal{P}\,.
\end{equation}
By virtue of the phase space mode expansion~\eqref{eq:real+mode+expansion}, we write the quantity $\mathcal{P}$ as
\begin{equation}\label{eq:step+proof+c+alpha}
\begin{aligned}
    \mathcal{P} = &\sum_\varphi\,\left(\Omega_\beta + \Omega_\varphi\right)\,\ccoef{\varphi}\left<\spatmode{\beta},\,\spatmode{\varphi}\right>\\
    &\hspace{3.5mm}+ \left(\Omega_\beta - \Omega_\varphi\right)\,\ccoefconj{\varphi}\left<\spatmode{\beta},\,\spatmodeconj{\varphi}\right>\\
    &\hspace{3.5mm}+\ccoef{\varphi}\left<\spatmode{\beta},\,i\BOP{\spatmode{\varphi}}\right>\\
    &\hspace{3.5mm}+\ccoefconj{\varphi}\left<\spatmode{\beta},\,i\BOP{\spatmodeconj{\varphi}}\right>\,.
\end{aligned}
\end{equation}
If we then substitute the orthogonality relation~\eqref{eq:real+orthogonality+relation} and a relabeled version of orthogonality relation~\eqref{eq:orthog+bconj} in Eq.~\eqref{eq:step+proof+c+alpha}, we get for non-degenerate modes $\varphi$ and $\beta$ that
\begin{equation}\label{eq:proof+c+alpha}
    \mathcal{P} = \ccoef{\beta}\,b_\beta\,,
\end{equation}
assuming $\Omega_\varphi \neq -\Omega_\beta$, and using the Kronecker delta in Eq.~\eqref{eq:real+orthogonality+relation}.
Equation~\eqref{eq:proof+c+alpha} thus proves Eq.~\eqref{eq:real+mode+coefficient} for non-degenerate modes $\varphi$ and $\beta$ that have $\Omega_\beta \neq -\Omega_\varphi$, and a non-zero $b_\beta$.

Similarly, if we multiply $ \Omega_\beta\,\vec{\xi}\left(\vec{x},\,t\right) - i\,\Dot{\vec{\xi}}\left(\vec{x},\,t\right) - i\vec{B}\left(\vec{\xi}\left(\vec{x},\,t\right)\right) $ with $\vec{\xi}_\beta$ and integrate over the stellar mass, we obtain
\begin{equation}
    \left<\vec{\xi}_\beta^*,\,\Omega_\beta\,\vec{\xi}\left(\vec{x},\,t\right) - i\,\Dot{\vec{\xi}}\left(\vec{x},\,t\right) - i\vec{B}\left(\vec{\xi}\left(\vec{x},\,t\right)\right)\right> \equiv \mathcal{Q}\,,
\end{equation}
which, by virtue of the phase space mode expansion~(\ref{eq:real+mode+expansion}) can be written as
\begin{equation}\label{eq:step+proof+c+alpha_conj}
\begin{aligned}
    \mathcal{Q} =&\, \sum_\varphi\,\left(\Omega_\beta - \Omega_\varphi\right)\,\ccoef{\varphi}\left<\spatmodeconj{\beta},\,\spatmode{\varphi}\right>\\
    &\hspace{3.5mm}+ \left(\Omega_\beta + \Omega_\varphi\right)\,\ccoefconj{\varphi}\left<\spatmodeconj{\beta},\,\spatmodeconj{\varphi}\right>\\
    &\hspace{3.5mm}-\ccoef{\varphi}\left<\spatmodeconj{\beta},\,i\BOP{\spatmode{\varphi}}\right>\\
    &\hspace{3.5mm}-\ccoefconj{\varphi}\left<\spatmodeconj{\beta},\,i\BOP{\spatmodeconj{\varphi}}\right>\,.
\end{aligned}
\end{equation}
The following relations
\begin{subequations}\label{eq:relations+inner+products+c+alpha+conj}
\begin{align}
    &\left<\spatmodeconj{\beta},\,\spatmode{\varphi}\right> = \left<\spatmodeconj{\varphi},\,\spatmode{\beta}\right>,\\
    &\left<\spatmodeconj{\beta},\,i\BOP{\spatmode{\varphi}}\right>\begin{aligned}[t] &= -\left<\spatmodeconj{\varphi},\,i\BOP{\spatmode{\beta}}\right>,
    \end{aligned}
\end{align}
\end{subequations}
are valid because of the definition of the inner product and the anti-Hermiticity of the $\vec{B}$ operator.
Substituting Eq.~\eqref{eq:relations+inner+products+c+alpha+conj} in orthogonality relation~\eqref{eq:orthog+jconj} yields
\begin{equation}\label{eq:modified+orthog+jconj}
    \left(\Omega_\beta - \Omega_\varphi\right)\left<\vec{\xi}_\beta^*, \vec{\xi}_\varphi\right> - \left<\vec{\xi}_\beta^*,  i\vec{B}(\vec{\xi}_\varphi)\right> = 0\,.
\end{equation}
By substituting the orthogonality relations~\eqref{eq:symplectic+orthog+real+double+conj} and~\eqref{eq:modified+orthog+jconj} in Eq.~\eqref{eq:step+proof+c+alpha_conj}, we obtain for $\Omega_\beta \neq - \Omega_\varphi$ that
\begin{equation}
    \mathcal{Q} = \ccoefconj{\beta}\,b_\beta\,,
\end{equation}
where we use the Kronecker delta in Eq.~\eqref{eq:symplectic+orthog+real+double+conj}.
This proves Eq.~\eqref{eq:imaginary+mode+coefficient} for non-zero $b_\beta$ and  $\Omega_\beta \neq - \Omega_\varphi$.


\section{Deriving the meridional selection rule}\label{app:meridional+selection+rule}

We derive the meridional selection rule mentioned in Sect.~\ref{sect:selection+rules} based on the fact that a rotating star is invariant under the map $f\colon(r,\theta,\phi)\to(r,\pi-\theta,\phi)$, in analogy to what was done in \citetalias{2001PhRvD..65b4001S} for their coupling coefficient.
This invariance implies that modes are either even or odd under the pullback operator $f_*$ (e.g., \citetalias{2001PhRvD..65b4001S})
\begin{equation}\label{eq:pullback+lag}
    f_* \vec{\xi} = \xi^{\hat{r}}(r,\pi - \theta,\phi)\,\bm{e}_{\hat{r}} - \xi^{\hat{\theta}}(r,\pi - \theta,\phi)\,\bm{e}_{\hat{\theta}} + \xi^{\hat{\phi}}(r,\pi - \theta,\phi)\,\bm{e}_{\hat{\phi}}\,.
\end{equation}
We then have 
\begin{equation}\label{eq:z-parity}
    f_*\vec{\xi}_\varphi = \mathcal{Z}_\varphi \,\vec{\xi}_\varphi\,,
\end{equation}
with $\mathcal{Z}_\varphi$ (called the z-parity in \citetalias{2001PhRvD..65b4001S}) equal to $1$ if mode $\varphi$ is an even mode and $-1$ if it is an odd mode.
The radial Hough function $H_r\left(\theta\right)$, the latitudinal Hough function $H_\theta\left(\theta\right)$, and the azimuthal Hough function $H_\phi\left(\theta\right)$ defined in \citet{2019A&A...627A..64P} are real-valued.
Based on Eq.~(\ref{eq:lag+disp+TAR}) we then have that
\begin{equation}\label{eq:z-parity+conj}
    f_*\vec{\xi}_\varphi^* = \mathcal{Z}_\varphi \,\vec{\xi}_\varphi^*\,.
\end{equation}

By applying the pullback operator $f_*$ on the coupling coefficient, and taking into account the commutation of $f$ with geometrical operations such as the computation of covariant derivatives (e.g., \citetalias{2001PhRvD..65b4001S}), we can write
\begin{equation}\label{eq:pullback+symmetry+cc}
    f_* \kappa_{\varphi}^{\beta\gamma} = \kappa\left(f_*\vec{\xi}_\varphi^*,\, f_*\vec{\xi}_\beta,\, f_*\vec{\xi}_\gamma\right) = \mathcal{Z}_\varphi\,\mathcal{Z}_\beta\,\mathcal{Z}_\gamma\,\kappa\left(\vec{\xi}_\varphi^*,\, \vec{\xi}_\beta,\, \vec{\xi}_\gamma\right) \,.
\end{equation}
Because $\kappa_{\varphi}^{\beta\gamma}$ is a number and therefore invariant under $f_*$, it follows that
\begin{equation}\label{eq:z+parity+expression}
    \left(1 - \mathcal{Z}_\varphi\mathcal{Z}_\beta\mathcal{Z}_\gamma\right)\kappa_{\varphi}^{\beta\gamma} = 0\,,
\end{equation}
resulting in the meridional or z-parity selection rule \citepalias{2001PhRvD..65b4001S}
\begin{equation}\label{eq:z+parity+selection}
    i_{\rm odd} = \mymatmod{0}{2}\,,
\end{equation}
where $i_{\rm odd}$ denotes the number of odd modes.
Hence, for three-mode coupling, the z-parity selection rule or meridional selection rule implies that two or zero odd modes need to be involved in the coupling, as mentioned in Sect.~\ref{sect:selection+rules}, where this is cast into a condition on the sum of the ordering numbers of the interacting modes.


\section{Equivalence of the amplitude equations for $\Omega_{1_d} \approx \Omega_{2_d} - \Omega_{3_d}$}\label{app:diff+freq}

In this section, we show the equivalence of the AEs determined in Sect.~\ref{sect:AEs} for the sum-frequencies $\Omega_1 \approx \Omega_2 + \Omega_3$ with those derived for difference-frequencies $\Omega_{1_d} \approx \Omega_{2_d} - \Omega_{3_d}$ using the same methods employed in Sect.~\ref{sect:AEs}.
This equivalence indicates that the stationary properties of resonant isolated sum-frequency triads that we derive in the main body of this paper are the same as those properties that one would derive for the (relabeled) difference frequencies for quadratic non-linear mode couplings.

To prove this equivalence, we first derive the difference-frequency AEs using the same procedure as in Sect.~\ref{sect:AEs}. Here, we define the resonance condition $\Omega_{1_d} \approx \Omega_{2_d} - \Omega_{3_d}$ in terms of the difference-frequency detuning parameter $\delta\omega_d$:
\begin{equation}\label{eq:detuning+diff+freq}
    \mathfrak{J}\,\delta\omega_d = \Omega_{1_d} \approx \Omega_{2_d} - \Omega_{3_d}\,.
\end{equation}
In Eq.~\eqref{eq:detuning+diff+freq}, the subscript $d$ refers to the fact that we are considering a difference-frequency resonance.
If we then set the secular-term-generating terms in Eq.~\eqref{eq:second+order+subst} equal to zero, use the difference-frequency resonance condition~\eqref{eq:detuning+diff+freq}, and introduce the linear growth or linear damping rates (similar to what was done in Sect.~\ref{sect:AEs}), we obtain the extended complex AEs for the difference-frequency resonance $\Omega_{1_d} \approx \Omega_{2_d} - \Omega_{3_d}$:
\begin{equation}\label{eq:difference+freq+complex+aes}
    \pdv{\vec{a}_d}{t_1} = \vec{\gamma}_d\circ\vec{a}_d + 2\,i\,\left(\vec{\eta_d}\circ\vec{a}_{M_{d}}\circ\vec{\Omega}_{M_{d}}\circ\vec{e}_d\right)\,.
\end{equation}
In the extended complex AEs~\eqref{eq:difference+freq+complex+aes}, we use the vectors
\begin{equation}\label{eq:defined+vectors+1+diff+freq}
    \vec{a}_d = \begingroup\renewcommand*{\arraystretch}{1.2}\begin{bmatrix}
    a_{1_d} \\ a_{2_d} \\ a_{3_d}
    \end{bmatrix}\endgroup\,,\hspace{0.3cm}\begingroup\renewcommand*{\arraystretch}{1.2}\vec{a}_{M_d} = \begin{bmatrix}
    a_{2_d}\,a_{3_d}^* \\ a_{1_d}\,a_{3_d} \\ a_{1_d}^*\,a_{2_d}
    \end{bmatrix}\endgroup\,,\hspace{0.3cm}\begingroup
    \renewcommand*{\arraystretch}{1.2}\vec{e}_d =  \begin{bmatrix}
    \exp(i\,\delta\omega_d\,t_1) \\ \exp(-i\,\delta\omega_d\,t_1) \\ \exp(i\,\delta\omega_d\,t_1)
    \end{bmatrix}\endgroup\,,
\end{equation}
and
\begin{equation}\label{eq:defined+vectors+2+diff+freq}
    \vec{\gamma}_d = \begingroup\renewcommand*{\arraystretch}{1.2}\begin{bmatrix}
    \gamma_{1_d} \\ \gamma_{2_d} \\ \gamma_{3_d}
    \end{bmatrix}\endgroup\,,\hspace{0.3cm} \begingroup\renewcommand*{\arraystretch}{1.2} \vec{\eta}_d = \begin{bmatrix}
    \eta_{1_d}^* \\ \eta_{1_d} \\ \eta_{1_d}^*
    \end{bmatrix}\endgroup
    \,,\hspace{0.3cm} \vec{\Omega}_{M_d} = \begingroup\renewcommand*{\arraystretch}{1.2}\begin{bmatrix}
    \Omega_{1_d} \\ \Omega_{2_d} \\ \Omega_{3_d}
    \end{bmatrix}\endgroup\,,
\end{equation}
in which we use the (difference-frequency) coupling coefficient $\eta_{1_d}$ that is defined as:
\begin{equation}\label{eq:coupling+coefficient+diff+freq}
    \eta_{2_d}^{1_{d}3_{d}} = \eta_{2_{d}}^{3_{d}1_{d}} = \left(\eta_{1_{d}}^{2_{d}\overline{3}_{d}}\right)^* = \left(\eta_{1_{d}}^{\overline{3}_{d}2_{d}}\right)^* = \left(\eta_{3_{d}}^{2_{d}\overline{1}_{d}}\right)^* = \left(\eta_{3_{d}}^{\overline{1}_{d}2_{d}}\right)^* \equiv \eta_{1_d} \,.
\end{equation}

By introducing real amplitudes $A_{\varphi_d}$ and phases $\phi_{\varphi_d}$ as in Sect.~\ref{sect:AEs}, $a_{\varphi_d} = A_{\varphi_d}\,\exp\left(i\,\phi_{\varphi_d}\right)$ (with $\varphi \in \myintset{3}$), we obtain the (real-valued) AEs for difference frequencies,
\begin{subequations}\label{eq:AEs+diff+freqs}
\begin{align}
    \pdv{\vec{A}_d}{t_1} &= \vec{\gamma}_d \circ \vec{A}_d + 2\, \lvert \eta_{1_d}\rvert\, \sin\left(\Upsilon_d\right)\, \left(\vec{A}_{N_d}\circ \vec{\Omega}_{M_d}\right)\,, \label{eq:AE+A+diff+freq}\\
    \vec{A}_d\circ\pdv{\vec{\phi}_d}{t_1} &= 2\, \lvert \eta_{1_d}\rvert\, \cos\left(\Upsilon_d\right)\, \left(\vec{A}_{P_d}\circ \vec{\Omega}_{M_d}\right)\,,\label{eq:AE+P+diff+freq}
\end{align}
\end{subequations}
where
\begin{equation}\label{eq:amplitude+products+1+diff+freq}
    \eta_{1_d} = |\eta_{1_d}|\,e^{-i\,\delta_{1_d}}\,,\hspace{0.2cm}\vec{A}_d = \begingroup\renewcommand*{\arraystretch}{1.2}\begin{bmatrix}
    A_{1_d} \\ A_{2_d} \\ A_{3_d}
    \end{bmatrix}\endgroup\,,\hspace{0.2cm} \vec{\phi}_d = \begingroup\renewcommand*{\arraystretch}{1.2}\begin{bmatrix}
    \phi_{1_d} \\ \phi_{2_d} \\ \phi_{3_d}
    \end{bmatrix}\endgroup\,,
\end{equation}
as well as
\begin{equation}\label{eq:amplitude+products+2+diff+freq}
    \vec{A}_{N_d} = \begingroup\renewcommand*{\arraystretch}{1.2}\begin{bmatrix}
    A_{2_d}\, A_{3_d} \\ -A_{1_d}\, A_{3_d} \\ A_{1_d} \, A_{2_d}
    \end{bmatrix}\endgroup\,,\hspace{0.2cm} \vec{A}_{P_d} = \begingroup\renewcommand*{\arraystretch}{1.2}\begin{bmatrix}
    A_{2_d}\, A_{3_d} \\ A_{1_d}\, A_{3_d} \\ A_{1_d} \, A_{2_d}
    \end{bmatrix}\endgroup\,.
\end{equation}
In Eq.~\eqref{eq:AEs+diff+freqs}, we define the generic (frequency-difference) phase coordinate $\Upsilon_d$ as
\begin{equation} \label{eq:upsilon+parameter+diff+freq}
    \Upsilon_d \equiv -\delta\omega_d\,t_1 + \phi_{1_d} - \phi_{2_d} + \phi_{3_d} - \delta_{1_d}\,,
\end{equation}
which contains the combination phase $\Phi_d = \phi_{1_d} - \phi_{2_d} + \phi_{3_d}$ for difference frequencies.
The time-dependence of $\Upsilon_d$ is described by
\begin{equation}\label{eq:time+dependence+phase+coordinate+diff+freq}
    \pdv{\Upsilon_d}{t_1} = -\delta\omega_d + \cot\left(\Upsilon_d\right)\left(-\gamma_{\boxplus_d} + \pdv{\ln A_{1_d}}{t_1} + \pdv{\ln A_{2_d}}{t_1} + \pdv{\ln A_{3_d}}{t_1}\right)\,,
\end{equation}
because the coupling coefficient~$\eta_{1_d}$ does not depend on time.
In Eq.~\eqref{eq:time+dependence+phase+coordinate+diff+freq}, $\gamma_{\boxplus_d} = \gamma_{1_d} + \gamma_{2_d} + \gamma_{3_d}$.
The equation for the time dependence of $\Upsilon_d$ is of the exact same form as Eq.~\eqref{eq:AE+4+no+zero+A}, which describes the time dependence of $\Upsilon$.
This suggests that the phase coordinates $\Upsilon$ and $\Upsilon_d$ are connected by some direct relation.

Finally, we prove the equivalence of the AEs derived in Sect.~\ref{sect:AEs} and the AEs derived in this Appendix using a relabeling operation.
The difference frequency $\Omega_{1_d} \approx \Omega_{2_d} - \Omega_{3_d}$ can be written as a sum frequency $\Omega_{2_d} \approx \Omega_{1_d} + \Omega_{3_d}$. 
Swapping labels in the definition of the coupling coefficient $\eta_{1_d}$ defined in Eq.~\eqref{eq:coupling+coefficient+diff+freq} to describe the coupling among modes in this sum-frequency analogue (dropping the subscript $d$ in the process), leads to the equivalence of its expression with that of the sum-frequency coupling coefficient $\eta_1$ defined in Eq.~\eqref{eq:coupling+coefficient}.
Performing the same relabeling operation for the frequency-difference phase coordinate $\Upsilon_d$ defined in Eq.~\eqref{eq:upsilon+parameter+diff+freq}, we obtain that it is equal to $-\Upsilon$.
Substituting this relation between $\Upsilon_d$ and $\Upsilon$ in the difference-frequency AEs~\eqref{eq:AEs+diff+freqs} converts them into the sum-frequency AEs~\eqref{eq:AE+A}.
The temporal dependence of the individual mode phases is therefore the same as those described by the AEs~\eqref{eq:AE+P}.
Hence, the AEs~\eqref{eq:AEs} and \eqref{eq:AE+4+no+zero+A} describe the temporal evolution of sum and difference frequencies, and the properties of the modes in these resonances are derived in the main body of this article.


\section{Materials available online}\label{app:MESA+GYRE+code+link}
The MESA and GYRE inlists used to generate our models, as well as the final model data products themselves can be accessed using the MESA market place, \url{https://cococubed.com/mesa_market/inlists.html}\,, which contains a link to a Zenodo repository.
This repository\footnote{The repository can directly be accessed using the following link: \url{https://doi.org/10.5281/zenodo.10814654}.} contains the respective inlists and additional information on the modeling process.

We developed a Python code that computes the stationary mode amplitude ratios using the formalism described in this work.
This code generates the figures included in this work.
You may download the latest version of the code from the GitHub repository at \url{https://github.com/JVB11/AESolver}.
The online documentation may be found at \url{https://jvb11.github.io/AESolver/}, and contains examples on how to manipulate different parts of the code.
The inlists for this Python code that generate the coupling models discussed in this work, as well as the final data products generated by this code, can be accessed using the same Zenodo repository that stores the MESA and GYRE inlists.

\end{appendix}
\end{document}